\documentclass[twocolumn,secnumarabic,amssymb, nobibnotes, aps, prb]{revtex4-2}

\usepackage{epsfig}
\usepackage{graphicx}
\usepackage{amsmath}
\usepackage{amssymb}
\usepackage[colorlinks, allcolors=blue]{hyperref}
\usepackage{url}
\usepackage{float}
\usepackage{xcolor}
\usepackage{orcidlink}

\begin{document}

%\title{Anomalous and diode Josephson effect in junctions of singlet superconductors and inhomogeneous ferromagnetic barrier with interfacial Rashba spin-orbit interaction}%
%\title{Microscopic study of nonreciprocal charge transport in  Josephson junctions with inhomogeneous ferromagnetic barriers and interfacial Rashba spin-orbit coupling}%
%\title{Nonreciprocal charge transport in  Josephson junctions with inhomogeneous ferromagnetic barrier and interfacial Rashba spin-orbit coupling}%
\title{Anomalous and diode Josephson effect in junctions with inhomogeneous ferromagnetic barrier and interfacial Rashba spin-orbit coupling}%
%\title{Anomalous and diode Josephson effect in junctions of singlet superconductors and inhomogeneous ferromagnetic barrier with interfacial Rashba spin-orbit interaction}

\author{Stevan Djurdjevi\' c$^{1,2}$\orcidlink{0000-0002-6409-1103}}\email{stevandj@ucg.ac.me} \author{Zorica Popovi\' c$^2$\orcidlink{0000-0002-1487-4695}}\email{pzorica@ff.bg.ac.rs}%
\affiliation{$^1$University of Montenegro, Faculty of Natural Sciences and Mathematics, D\v zord\v za Va\v singtona bb, 81000 Podgorica, Montenegro}
\affiliation{$^2$University of Belgrade, Faculty of Physics, Studentski trg 12, 11001 Belgrade, Serbia}
\date{May 11, 2026}%
\begin{abstract}
We theoretically investigate the anomalous and diode Josephson effects in planar two-dimensional Josephson junctions with arbitrarily oriented exchange fields in two ferromagnets within the barrier, and spin-orbit coupling at the superconductor/ferromagnet interfaces, where the superconducting electrodes can have $s$-wave or arbitrarily oriented $d$-wave order parameter lobes. We perform a systematic symmetry analysis of the junction Hamiltonian and identify the minimal conditions for breaking time-reversal and space-inversion symmetries, which are required for the emergence of anomalous and diode Josephson effects. We classify the junctions into three classes, with particular attention to those between $d_{x^2-y^2}$ and $d_{xy}$ oriented superconductors. Our symmetry analysis is supported by numerical calculations of the current-phase relation (CPR) obtained  using a generalized Furusaki-Tsukada (F-T) approach. By tuning the directions of exchange fields in the ferromagnets, Rashba SOC at the interfaces and superconducting order parameter orientations, nonreciprocity can be enhanced by more than 40\%. We further analyze the phase-dependent Andreev bound states (ABS) spectrum and their contribution to charge transport, as well as their signatures in the nonreciprocal transport characteristics. By comparing the current carried by ABS with that obtained using the F-T technique, we find that the contribution from continuum states above the gap becomes pronounced in presence of zero energy crossings in the ABS spectrum, and in junctions with $d$-wave superconducting electrodes due to the narrower superconducting gap, which may become closed. In the nonreciprocal regime, the ABS spectra show an asymmetric profile with respect to phase inversion, indicating the presence of a finite current at zero phase difference and unequal critical currents in opposite directions. Our results establish symmetry-based criteria and microscopic mechanism for engineering nonreciprocal Josephson transport in hybrid superconducting structures.

\end{abstract}
\maketitle

\section{Introduction}

Heterojunctions composed of superconducting and spin-splitting materials have attracted researchers' attention for many years due to their significance for applications in spintronics \cite{Zutic2004,Eschrig2011,Linder2015, Eschrig2015, Melnikov2022,Zutic2023} and quantum electronics \cite{DeFranceschi2010,Devoret2013,Golod2022,Alam2023}, as well as for phenomena interesting from fundamental point of view \cite{Buzdin2005, Bergeret2005}. Particular interest is given to Josephson junctions \cite{Josephson1962}, where two superconducting condensates are weakly coupled through various barrier materials, allowing nondissipative electric flow due to  the phase difference bias between the condensates. In the simplest case, the supercurrent depends sinusoidally on the phase difference, $\varphi$, in  tunnel junctions \cite{Golubov2004}. In ferromagnetic Josephson junctions  the current-phase relation (CPR) is nonsinusoidal and contains higher harmonics. Additionally, the direction of current flow can be altered depending on the exchange field and the thickness of  the ferromagnetic barrier, due to the oscillatory nature of the superconducting order parameter in ferromagnets proximitized to superconductors. This is known as the $0$-$\pi$ phase transition \cite{Buzdin1982,Tanaka1997,Ryazanov2001,Kontos2002,Blum2002,Golubov2004,Buzdin2005,Bergeret2005}, where the Josephson free energy reaches a minimum at zero phase difference for positive currents and at phase difference of $\varphi=\pi$ for negative currents. Besides junctions with conventional superconductors, this effect can also be realized in junctions where the superconducting order parameter has $d$-wave symmetry, as in cuprates, where the CPR depends on the orientation of the superconducting electrodes and shows a wide variety of profiles \cite{Harlingen1995,Tanaka1996,Tanaka97,Asano2001,Lofwander2001,Ilichev2001,Testa2005,Nas21}.
%In Josephson junctions with inhomogeneous ferromagnets, triplet pair correlations emerge due to spin flip scattering processes. Here, higher harmonics in the CPR are responsible for a finite critical current at the $0$-$\pi$ crossover points.
In Josephson junctions with conserved spatial inversion and time reversal symmetries, the CPR is an odd function of the phase difference, i.e., $I(\varphi)=-I(-\varphi)$. Consequently, the current at zero phase difference is zero, and the critical current is equal in both forward and backward directions. However, when both spatial inversion and time-reversal symmetries are simultaneously broken, typically by ferromagnetism or a magnetic field combined with spin-orbit interaction, a nonzero anomalous Josephson current (known as the anomalous Josephson effect  (AJE)), $I(\varphi=0)\not=0$, can occur due to the presence of a cosine term in the Fourier expansion of the anomalous CPR.  Anomalous current can be realized in a variety of Josephson junction with exchange fields and SOC where the conventional CPR acquires an additional phase shift \cite{Yip1995,Sigrist1998,Kashivaya2000,Buzdin2008,Zazunov2008,Reynoso2008,Tanaka2009,Liu2010a,Reynoso2012,Brunetti2013,Yokoyama2014,Konschelle2015,Bergeret2015,Lu2015,Nesterov2016,Silaev2017,Minutillo2018,Alidoust2021,Meng2022}. Breaking both symmetries can also result in different magnitudes of critical currents in opposite directions, leading to the Josephson diode effect (JDE) \cite{Hu2007,Flensberg2016,Masaki2021,Davydova2022,Zhang2022,He2022,Jiang2022,Nadeem2023,Vakili2024,Osin2024}.
%A quantitative measure of this nonreciprocity is the so called diode efficiency, defined as $Q=(I_c^+ -I_c^-)/(I_c^+ +I_c^-)$, where $I_c^+$ and $I_c^-$ are the maximum and minimum values of the current in the CPR.
After the experimental discovery of the superconducting diode effect in superlattices with a few percent of nonreciprocity \cite{Ando2020}, subsequent experiment with Josephson junctions have achieved higher diode efficiencies defined as $Q=(I_c^+ -|I_c^-|)/(I_c^+ +|I_c^-|)$, with $I_c^+$ and $I_c^-$ representing the maximum and minimum values of the current in the CPR \cite{Baumgartner2022}. This is promising for the application of Josephson junctions in nondissipative logic circuits, spintronic memory devices, and as important building blocks for quantum computers. These experiments have motivated intensive both theoretical and experimental research into nonreciprocal responses in junctions where AJE and JDE can be achieve \cite{Tanaka2022,Kokkeler2022,Baumgartner2022c,Wu2022,Pal2022,Turini2022,Jeon2022,Costa2023nat,Lu2023,Costa2023,Kochan2023,Lotfizadeh2024,Wang2024,Ilic2024,Danilo2025,Costa2025,Nas2025,Zhang2025,Wang2025,Behner2026,Cayao2026}.

Nondissipative charge transport in short Josephson junctions is primarily governed by subgap bound states, known as Andreev bound states (ABSs), which appear in the barrier between superconductors as a result of Andreev reflection due to strong spatial variation of the superconducting order parameter and determine the profile of CPR \cite{Bardeen1972,Kulik1975,Furusaki1991,Beenakker1991,Tanaka1996a,Furusaki1999, Kashivaya2000,Lofwander2001,Yokoyama2014,Mizushima2015,Sauls2018}.
  %For a $d$-wave symmetry of the superconducting order parameter, zero-energy Andreev bound states  can form, leading to an enhanced contribution of the second harmonic in CPR due to the sign change of  the pair potential on the Fermi surface. When both time reversal and space inversion symmetries are broken, asymmetric phase dependent discrete energy bands can emerge, resulting in $\sin\varphi$, $\sin 2\varphi$ and $\cos\varphi$ harmonics in the CPR, which promote the AJE and JDE.
The explanations for the occurrence of the AJE and JDE are connected to the emergence of asymmetric Andreev bound states in  junctions where both time reversal and space inversion symmetries are broken, due to joint influence of SOC and Zeeman coupling \cite{Yokoyama2014,Costa2023,Cayao2024,Mondal2025,Wang2025a}. Since the ABS spectrum determines the current in Josephson devices, phase asymmetry leaves signatures on the Josephson current-phase relation, resulting in $\sin\varphi$, $\sin 2\varphi$ and $\cos\varphi$ harmonics in various transverse channels \cite{Tanaka2022,Nas2025}, which promote AJE  and JDE, as well as current reversing $0-\pi$ like phase transitions.

In this paper, we study nonreciprocal charge transport in planar two-dimensional Josephson junctions with arbitrarily oriented exchange fields in two ferromagnets in the barrier, and spin-orbit coupling at the superconductor/ferromagnet interfaces. The superconducting electrodes have $s$-wave or arbitrary orientated $d$-wave order parameters. To obtain the necessary conditions for the appearance of a finite anomalous current and nonreciprocal critical current ($I_c^+\neq |I_c^-|$) in these junctions, we perform a systematic symmetry analysis of the junction Hamiltonian and identify the minimal conditions for simultaneous breaking of both time-reversal and space-inversion symmetry necessary for realization of anomalous and Josephson diode effects. Based on these constraints, we classify the junctions into three classes and identify the minimal requirements for the realization of these effects. For $s$-wave or oppositely oriented $d$-wave electrodes, both effects require noncoplanar spin-spliting fields and unequal opposite exchange field components along the spin quantization axis at SOC interfaces. For identically oriented $d$-wave electrodes (excluding the $d_{x^2-y^2}$ orientation of both superconductors), symmetry forbids both effects for equal out-of-plane and equal or opposite in-plane magnetization orientations. For all remaining $d$-wave configurations, a finite out-of-plane magnetization component in at least one ferromagnetic layer is sufficient to generate both effects. Our symmetry analysis is supported by numerical calculations of the Josephson current within the Bogoliubov-de Gennes framework using a generalized Furusaki-Tsukada (F-T) approach for spin active interfaces and anisotropic pairing in the superconducting electrodes. Angle-resolved CPR analysis reveals the coexistence of transport channels with dominant $\sin\varphi$, $\sin 2\varphi$, and $\cos\varphi$ components, establishing a microscopic condition for the diode effect \cite{Tanaka2022,Nas2025}. We pay particular attention to the junction between $d_{x^2-y^2}$ and $d_{xy}$ oriented superconductors, where we find that additional symmetries suppress diode behavior for coplanar spin splitting fields, yielding a CPR with $\sin(2n\varphi)$ and $\cos((2n-1)\varphi)$ harmonics \cite{Lu2015} and degenerate ground states around $\varphi=\pm\pi/2$. The anomalous current vanishes at the transition between these states and changes sign across it. We further analyze the phase-dependent ABS spectrum and its contribution to the charge transport. By comparing the current carried by ABSs with that obtained using the F-T technique, we find that the contribution from continuum states above the gap becomes pronounced  near zero energy crossings in the ABS bands, resulting in a sawtooth CPR profile, and in junctions with $d$-wave superconducting electrodes due to the narrower superconducting gap, which may become closed \cite{Yokoyama2014,Cayao2024,Wang2025}. We investigate the effect of the orientation of spin splitting fields in the two ferromagnets on CPR and on the phase dependent energy spectrum of ABSs and their signatures in the nonreciprocal transport characteristics. The ABS spectrum exhibits asymmetry with respect to phase inversion, which is directly linked to the appearance of the AJE and JDE.

%\textcolor{red}{In this paper, we study the appearance of AJE and JDE in planar two-dimensional Josephson junctions with two ferromagnetic layers featuring arbitrarily oriented exchange fields and Rashba spin–orbit coupling localized at the S/F interfaces. We consider both $s$-wave and $d$-wave superconducting electrodes with arbitrary orientations of the order-parameter lobes. Within the Bogoliubov–de Gennes formalism, we  perform a systematic symmetry analysis of the junction Hamiltonian and derive general conditions for the appearance or suppression of a finite Josephson current at zero phase difference and for nonreciprocal critical current. Based on these constraints, we classify the junctions into three distinct classes and identify the minimal requirements for realizing the AJE and JDE. Our analytical results are supported by numerical calculations using a generalized Furusaki–Tsukada approach for spin-active fields and anisotropic superconducting order. We also analyze the phase-dependent Andreev bound state spectrum and its role in nonreciprocal transport.}

In Sec. \ref{sec2}, the model of the considered Josephson junction and the formalism used for calculating the Josephson current and Andreev bound state energies calculation are introduced. The conditions for the appearance of AJE and JDE, derived from symmetry considerations, along with the corresponding numerical results, are presented in Secs. \ref{sec3} and \ref{sec4}. Sec. \ref{sec5} is devoted to the analysis of phase dependent Andreev bound state spectra across different transverse channels. Finally, Sec. \ref{sec6} provides concluding remarks.

\section{Model and theoretical frame} \label{sec2}
\begin{figure}[h]
	\centering
	\includegraphics[width=8.5cm]{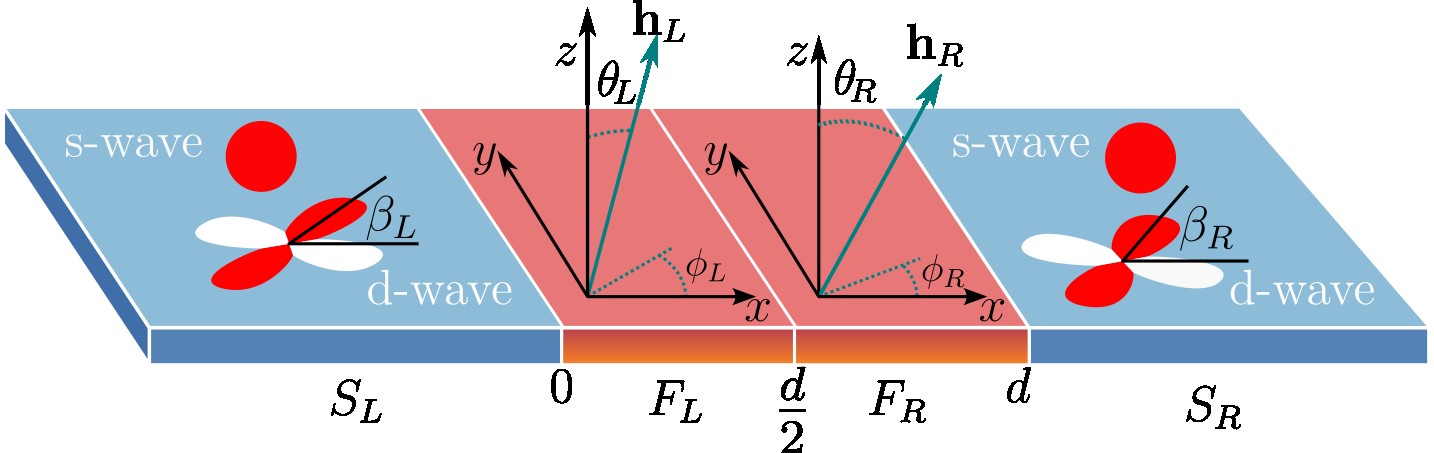}
	\caption{Schematic illustration of the $S_L F_L F_R S_R$ Josephson junction. The agnetization vectors in the left and right ferromagnets, $\mathbf{h}_L$ and $\mathbf{h}_R$, are arbitrarily oriented, with their orientations defined by the polar angles, $\theta_L$ and $\theta_R$, and the azimuthal angles, $\phi_L$ and $\phi_R$, respectively. The superconducting electrodes may have either an $s$-wave or $d$-wave order parameter, where $\beta_L$ and $\beta_R$ are the angles between the $\hat{a}$-axis of the crystal and the $x$-axes of the left and right superconductors, respectively.}\label{Fig1}
\end{figure}

We consider planar two dimensional superconductor ($S_L$)/ferromagnet ($F_L$)/ferromagnet ($F_R$)/superconductor ($S_R$) Josephson junction, placed in $xy$-plane with $x$-axis orthogonal to the layer interfaces. Left and right superconductors are located in half-planes $x<0$ and $x>d$, respectively. Two monodomain ferromagnetic layers of equal thickness are situated in $0<x<d/2$ for left ferromagnet, and in $d/2<x<d$ for right ferromagnet. In Fig. \ref{Fig1}, schematic illustration of the considered junction is shown.
Superconductors may have either isotropic ($s$-wave) or an anisotropic ($d$-wave) order parameter. For $d$-wave pairing symmetry, the orientations of the left and right superconducting crystals are defined by angles $\beta_L$ and $\beta_R$, respectively, which measure the angles between the direction of the order parameter lobe and the $x$-axis. In the  two ferromagnetic layers within the barrier, arbitrarily oriented magnetization directions are allowed. To make the model as realistic as possible, interface potentials at flat contacts between different materials ($S_L/F_L$ and $F_R/S_R$) are modeled by insulating barriers and, due to changing background electrostatic potentials across $S/F$ contacts, by interfacial Rashba spin-orbit coupling (SOC).

%We suppose that critical temperature of superconductors is much lower then critical temperature of ferromagnets, and using BCS description of superconductivity and Stoner model for ferromagnetism, we don't care about of temperature dependence of exchange field.
The superconductors are described within BCS model of superconductivity, while for ferromagnets we use the Stoner model. To calculate the Josephson current and ABS spectra, we model the junction using the Bogoliubov-de Gennes Hamiltonian \cite{deGennes1966,Nas2025}
\begin{equation}\label{HBdG}
	H_{BdG}=\left(\begin{array}{cc}
		H & \hat{\Delta}\\
		\hat{\Delta}^* & -\tilde{H}
	\end{array}\right),
\end{equation}
where $H=H_0+\sum_{j}H_j^F+\sum_{i}H_i$ represents the single particle Hamiltonian. Here, $H_0=\frac{\hbar^2k^2}{2m}-\mu$ is its kinetic part where $m$ is effective mass of quasiparticles and $\mu$ is chemical potential. For simplicity, we assume that chemical potential and effective mass are equal in every part of the junction. The second term in the single particle Hamiltonian refers to the exchange field, where $j\in\{L,R\}$ denotes the left and right ferromagnetic layers, so that $H_L^F=-\boldsymbol{h}_L\boldsymbol{\sigma}\Theta(x)\Theta(d/2-x)$ and $H_R^F=-\boldsymbol{h}_R\boldsymbol{\sigma}\Theta(x-d/2)\Theta(d-x)$, where  $\boldsymbol{\sigma}=(\sigma_x,\sigma_y,\sigma_z)^T$ is the vector of Pauli matrices and $\Theta(x)$ is Heaviside step function. We assume that the two ferromagnetic layers have arbitrarily oriented magnetization directions defined by the azimuthal $\phi_j$ and polar $\theta_j$ angles (see Fig. \ref{Fig1}), with equal magnitudes of exchange interactions in the ferromagnetic domains, $\boldsymbol{h}_j=h(\sin\theta_j \cos\phi_j,\sin\theta_j \sin\phi_j,\cos\theta_j)^T$.
%equal exchange fields in two ferromagnetic layers, $|\boldsymbol{h}_L|=|\boldsymbol{h}_R|=h$, with arbitrary oriented magnetization directions defined by spherical angles $\theta_j$ and $\phi_j$ (see Fig. \ref{shema}), then $\boldsymbol{h}_j=h(\sin\theta_j \cos\phi_j,\sin\theta_j \sin\phi_j,\cos\theta_j)^T$.
The third term in the single particle Hamiltonian  models the interfaces between superconducting and ferromagnetic materials using Dirac delta-like potentials, $H_i=[W_i+R_i(\mathbf{k}\times\mathbf{e}_x)\boldsymbol{\sigma}]\delta(x-x_i)$, where $i\in\{L,R\}$ denotes the left and right S/F interfaces. Here, the first term, $W_i$ is the insulating potential, while the second term represents the   Rashba SOC potential \cite{Rashba1960,Bychkov1984}, which is modelled assuming that the scalar potential gradient is along the $x$-direction, as the contact boundary lines between different materials are parallel to the $y$-axis. The strength of the Rashba SOC at the interfaces is described by the phenomenological parameter $R_i$, and we assume $R_L=-R_R=R$. The hole part Hamiltonian, $\tilde{H}$ from Eq. (\ref{HBdG}), is obtained by substituting $\boldsymbol{\sigma}$ in $H$ with $\boldsymbol{\tilde{\sigma}}=(-\sigma_x,\sigma_y,\sigma_z)^T$. The pairing potential in the superconductors is defined as $\hat{\Delta}=\Delta_L \sigma_x \Theta(-x)+\Delta_R \sigma_x \Theta(x-d)$. In contrast to $s$-wave superconductors, where the pairing potential is isotropic, the $d$-wave pairing potential depends on the direction of quasiparticle motion and, in the left and right superconducting electrodes, is given by \cite{Kashivaya2000,Nas2025}
\begin{equation}
	\begin{aligned}
		\Delta_{L[R]}(\gamma)&=\Delta(T)\cos(2(\gamma - \beta_{L[R]}))e^{i\varphi_{L[R]}},\\
		\tilde{\Delta}_{L[R]}(\gamma)&=\Delta(T)\cos{(2(\gamma +\beta_{L[R]}))}e^{i\varphi_{L[R]}}.\label{op}
	\end{aligned}
\end{equation}
For quasiparticles propagating at angles $\gamma$ and $\pi-\gamma$, measured with respect to the $x$-axis, we use $\Delta_{L[R]}$ and $\tilde{\Delta}_{L[R]}$, respectively. The temperature dependence of the pair potential amplitude is given by $\Delta(T)=\Delta_0\tanh{(1.74\sqrt{T_c/T-1})}$, where $T_c$ is the critical temperature of the superconductor \cite{Muhlschlegel1959}. $\varphi_L$ and $\varphi_R$ are the macroscopic phases of the left and right superconducting condensates, respectively.

To determine elementary propagating states of quasiparticles at excitation energy $E$ we solve the stationary BdG equation, $H_{BdG}\Psi(\mathbf{r})=E\Psi(\mathbf{r})$, where  $\Psi(\mathbf{r})=[u_{\uparrow}(\mathbf{r}),u_{\downarrow}(\mathbf{r}),v_{\uparrow}(\mathbf{r}),v_{\downarrow}(\mathbf{r})]^T$ is a four component wave vector. As translational invariance holds along the $y$-direction, the wave function can be written in the form $\Psi(\mathbf{r})=e^{ik_y y}\psi(x,\gamma)$, where transverse momentum parallel to the interface, $k_y=k_F\sin \gamma$, is conserved.

After constructing wave functions in each region of the junction and solving the boundary conditions, we obtained the coefficients of Andreev reflection required for the calculation of CPR using the generalised Furusaki-Tsukada technique \cite{Furusaki1991} based on the McMillan Green's function approach \cite{McMillan1968}. Additional details of the mathematical formalism are provided in Appendix \ref{apendix1}. The dc Josephson current in a 2D junction is given by \cite{Pajovic2006, Tanaka97,Costa2017,Nas2025}
\begin{widetext}
	\begin{equation} I(\varphi)=\frac{ek_{B}T}{2\hbar}\sum_{\omega_{n}}\sum_{k_{y}}\left(\frac{a_{1n}^{\uparrow\downarrow}+a_{1n}^{\downarrow\uparrow}}{2\Omega_{Ln}}|\Delta_{L}(\gamma)|\frac{k_{Ln}^{-}+k_{Ln}^{+}}{k_{Fx}}
		-\frac{a_{2n}^{\uparrow\downarrow} +a_{2n}^{\downarrow\uparrow}}{2\tilde{\Omega}_{Ln}}|\tilde{\Delta}_{L}(\gamma)|\frac{\tilde{k}_{Ln}^{-}+\tilde{k}_{Ln}^{+}}{k_{Fx}}\right).\label{struja}
	\end{equation}
\end{widetext}
Here, $\varphi=\varphi_R-\varphi_L$ is the macroscopic phase difference between superconductors, while $k_{Ln}^{\pm}$, $\tilde{k}_{Ln}^{\pm}$, $a_{1n}^{\uparrow\downarrow}$, $a_{1n}^{\downarrow\uparrow}$, $a_{2n}^{\uparrow\downarrow}$, $a_{2n}^{\downarrow\uparrow}$, $\Omega_{Ln}$, $\tilde{\Omega}_{Ln}$ are obtained from $k_{L}^{\pm}$, $\tilde{k}_{L}^{\pm}$, $a_{1}^{\uparrow\downarrow}$, $a_{1}^{\downarrow\uparrow}$, $a_{2}^{\uparrow\downarrow}$, $a_{2}^{\downarrow\uparrow}$, $\Omega_{L}$, $\tilde{\Omega}_{L}$ (given in Appendix \ref{apendix1}), respectively, by the analytical continuation $E\rightarrow\pm i\omega_n$ where $\omega_n=\pi k_B T (2n+1)$ are Matsubara frequencies. Note that $a_1^{\uparrow\downarrow} (a_1^{\downarrow\uparrow})$ is the probability amplitude of Andreev reflection for an incoming spin-up (spin-down) electron-like quasiparticle at the left interface with spin flipped orientation, while $a_2^{\uparrow\downarrow} (a_2^{\downarrow\uparrow})$ is the corresponding probability amplitude of Andreev reflection for an incoming hole-like quasiparticle.

Another formalism for obtaining the dc Josephson current is by using the phase dependence of ABS spectra. Since we consider Josephson junctions whose barrier thickness is much smaller than the superconducting coherence length, the dominant contribution to the transport properties comes from ABS \cite{Beenakker1991,Bagwell1992}. The boundary conditions (see Eqs. (\ref{granicni}) in Appendix \ref{apendix1}) can be written in matrix form
\begin{equation}\label{matricni oblik}
	AX=B,
\end{equation}
where $A$ is $24\times24$ matrix whose elements are functions of $E$, $\varphi$ and $\gamma$, $X$ is a column vector of 24 unknown scattering coefficients, and $B$ is a column vector of 24 terms originating from the propagating states of the incoming quasiparticles. The energy spectra of ABSs can be obtained from the condition for the existence of nontrivial solutions of Eq. (\ref{matricni oblik}),
\begin{equation}\label{det}
	\det(A)=0.
\end{equation}
For the given $k_y$, the relation between $E$ and $\varphi$ can be obtained as a numerical solution of previous equation. The ABS energy spectra have energies below the gap,
\begin{equation}
	|E|\leq \min\left\{|\Delta_L|,|\tilde{\Delta}_L,|\Delta_R|,|\tilde{\Delta}_R|\right\}.
\end{equation}
From $E(\varphi)$, the free energy can be found as \cite{Beenakker1991,Beenaker1992,Beenaker2023,Alipourzadeh2025}
\begin{equation}\label{slobodna}
	F(\varphi)=-\frac{1}{2}\sum_{k_y=-k_F}^{k_F}\sum_{i}E_i (\varphi) \tanh\left(\frac{1}{2} \frac{E_i}{k_B T}\right),
\end{equation}
where $E_i(\varphi)$ are the positive subgap energies of the ABS spectra, enumerated by $i$. The Josephson current originating from quantized subgap states is given by
\begin{equation}\label{sl}
	I(\varphi)=\frac{2e}{\hbar}\frac{dF(\varphi)}{d\varphi}.
\end{equation}

\section{Symmetry analysis} \label{sec3}
In this section, we consider the conditions for the appearance of the anomalous and diode effects by using symmetry analysis of the junction Hamiltonian since  any two Hamiltonian connected by a unitary or antiunitary transformations have identical spectra. Usually, the CPR is an odd function of the phase difference between two superconducting condensates, $I(\varphi)=-I(-\varphi)$. That implies that there exists a transformation $U$ such that the Hamiltonian satisfies $UH(\varphi)U^\dagger=H(-\varphi)$. In this case, $I(\varphi=0)=0$, and the AJE is absent. This is, for example, the case in a Josephson junction with a normal metal barrier and superconducting electrodes with isotropic $s$-wave, or anisotropic $d$-wave symmetry of the order parameter with $(\beta_L, \beta_R)=(0,0)$ orientation and  without spin-orbit coupling at the interface. Here, there is a set of 16 unitary or antiunitary transformations that map $H(\varphi)$ to $H(-\varphi)$: $\sigma_i T$, $P_y \sigma_i T$, $P_x \sigma_i$ and $P_x P_y \sigma_i$, where $\sigma_i$ ($i=0,x,y,z$) are the identity and Pauli matrices in spin space, $T$ is the time reversal operation, $P_x$ and $P_y$ are parity operations along the $x$- and $y$-axes, respectively, resulting in the absence of AJE \cite{Flensberg2016,Nas2025}. The necessary condition for the AJE is that all these symmetry transformations are broken.

Our model of the Josephson junction introduces additional elements, two monodomain ferromagnetic layers with arbitrary orientations in the barrier, interfacial Rashba SOC between superconducting and ferromagnetic materials, and the possibility of arbitrary orientation of the order parameter in $d$-wave superconductors. This requires further analysis to determine the conditions under which the AJE can occur.

Appendix \ref{appendix2}, Table \ref{table2}, presents the transformations of the Hamiltonian used in this analysis and describes how the ingredients on which the Hamiltonian depends are altered. Some transformations that change $H(\varphi)$ to $H(-\varphi)$ are broken by the presence of SOC. To break the remaining symmetries, an appropriate mutual orientation of the superconducting electrodes and the spin splitting directions in the two ferromagnetic layers is required. These transformations and the corresponding conditions that violate them are listed in Table \ref{table1}.  For simplicity, as mentioned, we assume that the exchange fields in the two ferromagnets are equal, $h_L=h_R=h$.
\begin{table}[h]
	\centering
	\caption{Table of transformations that yield $UH(\varphi)U^\dagger=H(-\varphi)$ and ingredients that break that symmetry. This table does not include transformations that are already broken by the presence of interfacial Rashba SOC.}
	\centering
	\begin{tabular}{|l|l|l|}
		\hline
		%\multicolumn{3}{|c|}{\centering $UH(\varphi)U^\dagger=H(-\varphi)$}\\ \hline\hline
		%\multicolumn{1}{|c|}{} & \multicolumn{2}{|c|}{What breaks symmetry}\\ \hline\hline
		U& Superconductors & Ferromagnets \\ \hline\hline
		$T$ &  & At least one $F$ is present.\\ \hline
		$\sigma_z T$ & & $h_{zL}\not=0$ or $h_{zR}\not=0$  \\ \hline
		$P_x \sigma_x$ & $\beta_L\not=-\beta_R$ & $(\theta_L,\phi_L)\not=(\pi-\theta_R,-\phi_R)$\\ \hline
		$P_x \sigma_y$ & $\beta_L\not=-\beta_R$ & $(\theta_L,\phi_L)\not=(\pi-\theta_R,\pi-\phi_R)$\\ \hline
		$P_x P_y$ &  $\beta_L\not=\beta_R$ &
		$(\theta_L,\phi_L)\not=(\theta_R,\phi_R)$\\ \hline
		$P_y \sigma_x T$ & $\beta_L\not=0$ or $\beta_R\not=0$ &
		$h_{xL}\not=0$ or $h_{xR}\not=0$\\ \hline
		$P_y \sigma_y T$ & $\beta_L\not=0$ or $\beta_R\not=0$ & $h_{yL}\not=0$ or $h_{yR}\not=0$\\ \hline
		$P_y P_x \sigma_z$ & $\beta_L\not=\beta_R$ & $(\theta_L,\phi_L)\not=(\theta_R,\phi_R+\pi)$\\ \hline
	\end{tabular}
	\label{table1}
\end{table}

In a Josephson junction with $s$-wave or $d$-wave superconducting electrodes in the  $(\beta_L, \beta_R)=(0,0)$ orientation, the required breaking of symmetries from Table \ref{table1} is achieved solely by appropriate orientations of the spin splitting directions in the ferromagnets. The symmetry transformations that are trivially broken are: $T$ (if at least one ferromagnet exists), $\sigma_z T$ (at least one ferromagnet has a nonzero component of the exchange field in the $z$-direction), $P_y \sigma_x T$ (if at least one ferromagnet has a nonzero component of the exchange field in the $x$-direction), and $P_y \sigma_y T$ (if at least one ferromagnet has a nonzero component of the exchange field in the $y$-direction). Therefore, a necessary condition for  breaking these symmetries is the presence of all three components of the exchange field in the barrier.

From Table \ref{table1} it appears that the transformations $P_x \sigma_x$ and $P_x \sigma_y$ are broken when $\theta_L\not=\pi-\theta_R$ or $\phi_L\not=-\phi_R$ and $\phi_L\not=\pi-\phi_R$. Note that these transformations map $\phi_L-\phi_R$ to $\phi_L-\phi_R$. However, since the Josephson current depends only on the relative azimuthal angle between the ferromagnetic exchange fields due to preferred axis of spin quantization  (the $z$-axis defined by Rashba SOC at the interfaces),  these symmetries are preserved for all values of $\phi_L$ and $\phi_R$ when $\theta_L=\pi-\theta_R$.

Additionally, it appears that the transformations $P_x P_y$ and $P_x P_y \sigma_z$ are broken for $\theta_L\not=\theta_R$ or $\phi_L\not=\phi_R$ and $\phi_L\not=\pi+\phi_R$. In the case where $\theta_L=\theta_R$, breaking symmetry requires that $\phi_L\not=\phi_R$ and $\phi_L\not=\pi+\phi_R$, i.e. the vectors $\mathbf{e}_z$, $\mathbf{h}_L$ and $\mathbf{h}_R$ must be noncoplanar.  In other cases, for $\theta_L\not=\theta_R$, noncoplanarity is not required to break these symmetries.

However, to clarify requirements for the existence of the AJE in these types of junctions, we need to consider the transformations under which the Hamiltonian is mapped from $H(\varphi,\beta_L,\beta_R)$ to $H(-\varphi,-\beta_L,-\beta_R)$, and whether these are broken. Besides the transformations already broken by Rashba SOC, $P_y \sigma_x T$ and $P_y \sigma_y T$ remain preserved for all $\theta_L$ and $\theta_R$ when $\phi_L=\phi_R$ and $\phi_L=\pi+\phi_R$, i.e. in the case of coplanarity of the vectors $\mathbf{e}_z$, $\mathbf{h}_L$ and $\mathbf{h}_R$. This means that the Josephson current is antisymmetric with respect the point $(\beta_L,\beta_R)=(0,0)$, i.e., $I(\varphi,\beta_L,\beta_R)=-I(-\varphi,-\beta_L,-\beta_R)$ so we have $I(0,0,0)=-I(0,0,0)$ leading to a zero anomalous Josephson current.

Therefore, considering all previous conclusions, we can establish the conditions for the occurrence of AJE in Josephson junctions with $s$-wave or $(\beta_L,\beta_R)=(0,0)$ oriented $d$-wave superconductors, in addition to the presence of interfacial Rashba SOC: the orientations of the magnetizations in the two ferromagnets and the $z$-axis must be noncoplanar, and additionally $\theta_L\not=\pi-\theta_R$ i.e. the $z$-components of the exchange fields in the ferromagnetic layers cannot be opposite.

For the achievement of conditions for the existence AJE in Josephson junctions of the type $\beta_L=-\beta_R$ additional considerations are required. Specifically, the transformations $P_x \sigma_x T$ and $P_x \sigma_y T$ which exchange the variables $\theta_L$ and $\theta_R$ in junction Hamiltonian (see Table \ref{table3} in Appendix \ref{appendix2}) and yield $I(\varphi,\theta_L,\theta_R)=I(\varphi,\theta_R,\theta_L)$ are preserved for $\beta_L=-\beta_R$. Additionally, the transformation $P_x P_y \sigma_z$ which maps $H(\varphi,\theta_L,\theta_R,\beta_L,\beta_R)$ to $H(-\varphi,\theta_R,\theta_L,-\beta_R,-\beta_L)$ and yields $I(\varphi,\theta_L,\theta_R,\beta_L,\beta_R)=-I(-\varphi,\theta_R,\theta_L,-\beta_R,-\beta_L)$ is also preserved for $\beta_L=-\beta_R$. Since for $\beta_L=-\beta_R$ current is an even function under the exchange of $\theta_L$ and $\theta_R$ we have $I(\varphi,\theta_L,\theta_R,\beta_L,\beta_R)=-I(-\varphi,\theta_L,\theta_R,-\beta_R,-\beta_L)$, and consequently $I(\varphi=0,\beta_L,-\beta_L)=-I(\varphi=0,\beta_L,-\beta_L)$. Furthermore, $P_x P_y \sigma_z$ maps $\phi_L-\phi_R$ into $\phi_R-\phi_L$ and it is preserved when $\phi_L=\phi_R$ or $\phi_L=\pi+\phi_R$. This implies that for coplanar orientation of the spin splitting directions with $\beta_L=-\beta_R$, the AJE is absent.

Additionally, transformations $P_x \sigma_x$ and $P_x \sigma_y$ which map $H(\varphi,\beta_L,-\beta_L)$ to $H(-\varphi,\beta_L,-\beta_L)$ are preserved for $\theta_L=\pi-\theta_R$. In the case of coplanar orientation of the vectors $\mathbf{e}_z$, $\mathbf{h}_L$ and $\mathbf{h}_R$ the symmetry is also preserved, whereas for noncoplanar orientation, the symmetry is broken when $\theta_L\not=\pi-\theta_R$.

Based on previous considerations, we can establish general conditions for the occurrence of  AJE in $s$-wave and $(\beta_L,-\beta_L)$ including $(0,0)$ type of junction: noncoplanarity of $\boldsymbol{e}_z$, $\boldsymbol{h}_L$ and $\boldsymbol{h}_R$ ($\phi_L\not=\phi_R, \phi_L\not=\pi+\phi_R$) together with the condition $\theta_L\not=\pi-\theta_R$.

\begin{figure*}[!t]
	\centering
	\includegraphics[width=18cm]{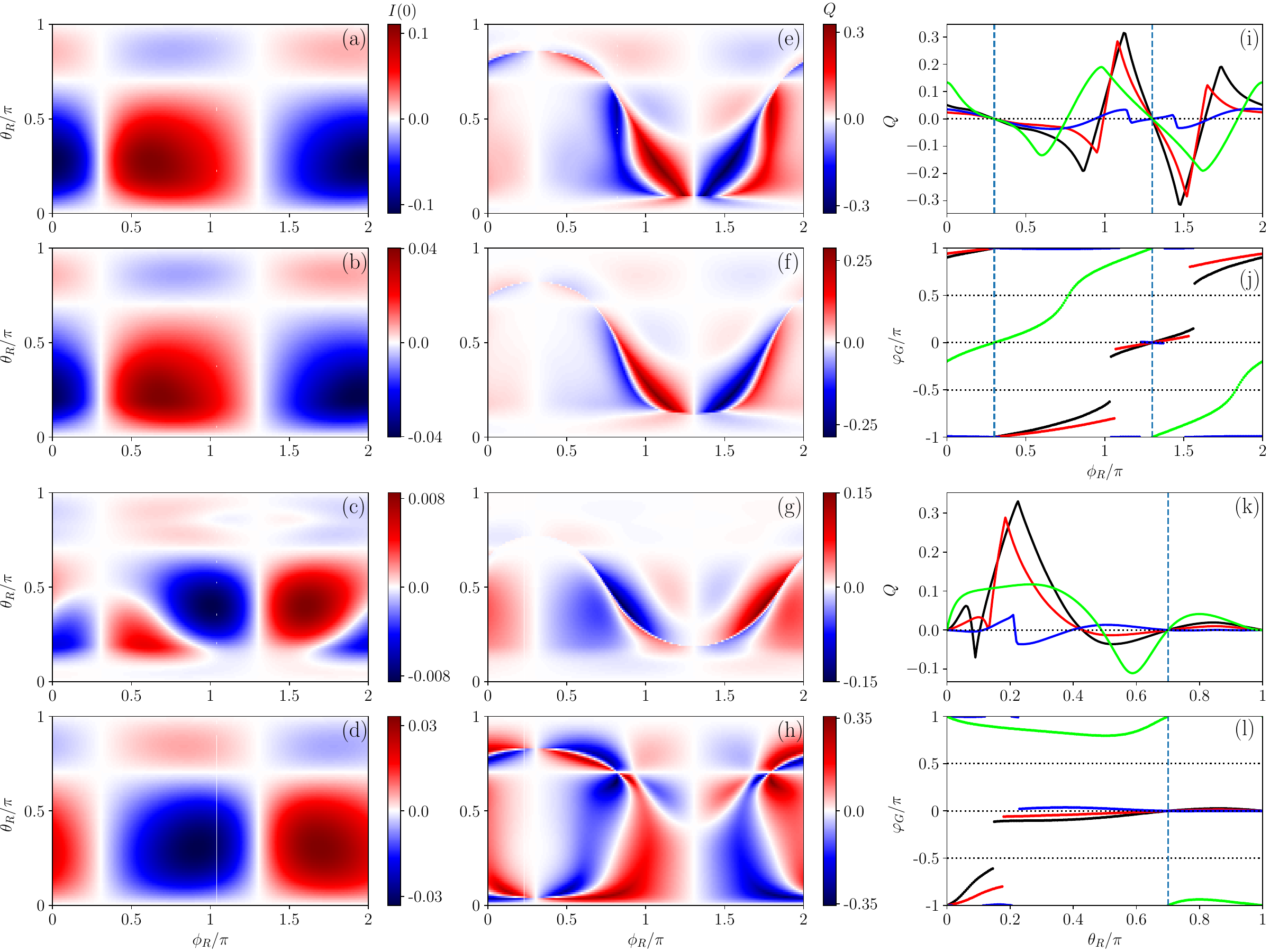}
	\caption{(a)-(d) Anomalous Josephson current $I(0)$ and (e)-(h) diode efficiency $Q$  as functions of the angles $\phi_R$ and $\theta_R$ for a fixed orientation of the exchange field in the left ferromagnet ($\phi_L=0.3\pi$ and $\theta_L=0.3\pi$), for s-wave ((a) and (e)), $(\beta_L,\beta_R)=(0,0)$  d-wave ((b) and (f)), $(\beta_L,\beta_R)=(0.1\pi,-0.1\pi)$  d-wave ((c) and (g)), $(\beta_L,\beta_R)=(0.2\pi,-0.2\pi)$  d-wave ((d) and (h)), types of Josephson junction, respectively. Diode efficiency $Q$ and ground state phase difference $\varphi_G$ as a function of $\phi_R$ for a fixed angle $\theta_R=0.2\pi$ (panels (i) and (j)), and as  functions of $\theta_R$ for fixed angle $\phi_R=1.1\pi$ (panels (k) and (l)), for s-wave (black), $(0,0)$ d-wave (red), $(0.1\pi,-0.1\pi)$ d-wave (blue) and $(0.2\pi,-0.2\pi)$ d-wave (green) types of junction. The vertical dashed blue lines indicate coplanar orientation of the spin splitting field in the two ferromagnetic layers. Parameters: $h/\mu=0.8$, $dk_F=10$, $T/T_c=0.1$, $\chi=1.25$, $Z=0.5$.}\label{Fig2}
	%\caption{(a)$s$-wave AJE, (b)$(0,0)$ $d$-wave AJE, (c)$(0.1\pi,-0.1\pi)$ $d$-wave, (d)$(0.2\pi,-0.2\pi)$ $d$-wave AJE, (e)$s$-wave JDE, (f)$(0,0)$ $d$-wave JDE, (g)$(0.1\pi,-0.1\pi)$ $d$-wave JDE, (h)$(0.2\pi,-0.2\pi)$ $d$-wave JDE. For $h/\mu=0.8$, $dk_F=10$, $T/T_c=0.1$, $\chi=1.25$, $Z=0.5$, $\theta_L=0.3\pi$, $\phi_L=0.3\pi$. (i), (j), (k) CPR for $\theta_R=0.2$ and $\phi_R=1.1$, diode efficiency in dependence of $\phi_R$, ground state phase difference in dependence of $\phi_r$, respectively. Black for $s$-wave, red for $(0,0)$ $d$-wave, blue for $(0.1\pi,-0.1\pi)$ $d$-wave and green for $(0.2\pi,-0.2\pi)$ $d$-wave. (l) Diode efficiency dependence of $\beta_L=-\beta_R=\beta$ for $\theta_R=0.2\pi$ and $\phi_R=1.1\pi$ (black) and $\theta_R=0.25\pi$ and $\phi_R=1.1\pi$ (red).}\label{Fig2}		
\end{figure*}

For junctions of the type $\beta_L=\beta_R\not=0$, breaking of symmetry transformations $P_x P_y$ and $P_x P_y \sigma_z$ which map $H(\varphi)$ into $H(-\varphi)$ is achieved by $\theta_L\not=\theta_R$ or $\phi_L\not=\phi_R$ and $\phi_R\not=\pi+\phi_L$.

For the $\beta_L\not=\pm\beta_R$ type of the junction, all transformations from Table \ref{table1} except $T$ and $\sigma_z T$ are broken by the orientation of the superconducting electrodes. To break these symmetries, at least one magnet with a nonzero $z$-component of the exchange field is required.

To find the conditions for JDE, we need to consider the transformations $H(\varphi,h)\rightarrow H(-\varphi,h)$. If this symmetry is preserved, then $I_c^+(h)=-I_c^-(h)$, indicating that JDE is absent. These transformations are the same as those considered for AJE, and one can see that the conditions for the appearance of JDE are the same as those for the appearance of AJE.

Therefore, from the point of view of the conditions for the appearance of the anomalous and diode effects in Josephson junctions with two monodomain ferromagnetic layers and interfacial Rashba SOC between different materials, there are three classes: 1) junctions with $s$-wave or $\beta_L=-\beta_R$ oriented $d$-wave superconductors; 2) junctions with $\beta_L=\beta_R\not=0$; and 3) junctions with $\beta_L\not=\pm\beta_R$.

The conclusions obtained from the symmetry analysis are fully confirmed by our numerical results presented below.

\section{Numerical results} \label{sec4}

\begin{figure*}[t]
	\centering
	\includegraphics[width=18cm]{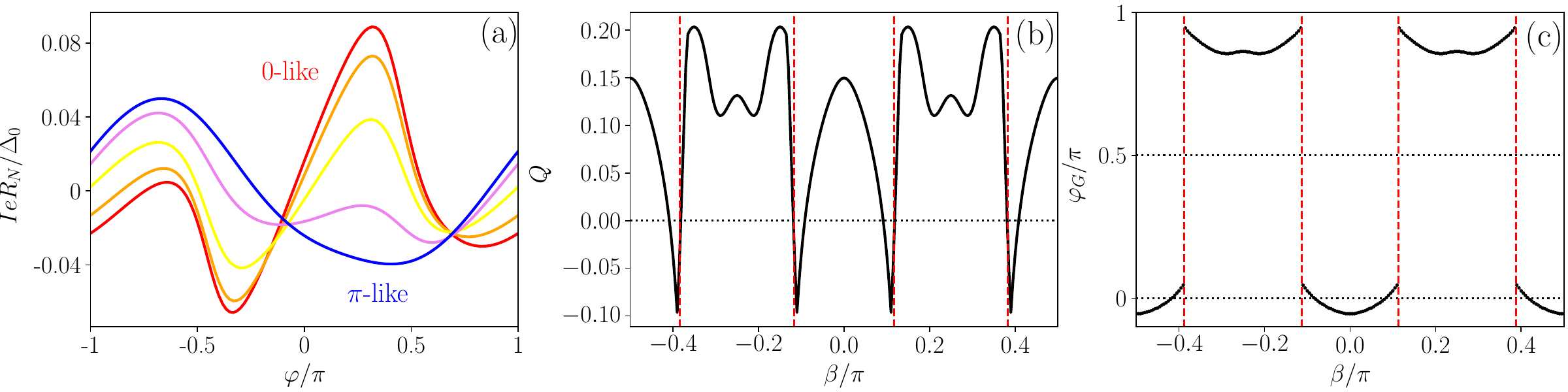}
	\caption{(a) CPR for $\beta_L=-\beta_R=\beta$ type of junction for $\beta=0$ (red), $\beta=0.05\pi$ (orange), $\beta=0.1\pi$ (yellow), $\beta=0.15\pi$ (purple) and $\beta=0.2\pi$ (blue). The junction undergoes a $0-\pi$ like transition when the superconducting electrodes orientation is changed. (b) Diode efficiency $Q$  and (c) ground state phase difference $\varphi_G$  as functions of $\beta$. The vertical red dashed lines indicate the values of $\beta$ at which $0-\pi$ phase transitions occur. Parameters: $\phi_L=0.3\pi$, $\theta_L=0.3\pi$, $\phi_R=1.1\pi$, $\theta_R=0.2\pi$, and the other parameters are the same as in Fig. \ref{Fig2}.} \label{Fig3}		
\end{figure*}
In this section, we analyze how the orientations of spin-splitting fields in two ferromagnetic layers affect nonreciprocal transport features in three previously established classes of Josephson junctions. In our calculations, we assume that the Fermi energy is 1000 times higher than the superconducting gap. The exchange field strengths and lengths are normalised by the inverse Fermi energy, $\mu$, and the Fermi wave vector, $k_F$, respectively. The presented results are obtained for a short Josephson junction with $dk_F=10$, and the exchange field strength in both ferromagnets is $h/\mu=0.8$. For numerical purposes, we introduce dimensionless parameters to quantify the strength of the insulating, $Z_{L(R)}=2mW_{L(R)}/(\hbar^2 k_F)$, and Rashba SOC, $\chi_{L(R)}=2mR_{L(R)}/\hbar^2k_F$, potentials at the left and right S/F interfaces. For simplicity, we illustrate results for symmetric case with identical transparencies at both interfaces ($Z_L=Z_R=Z$). To model highly transparent interfaces, but a realistic ballistic junction due to modern fabrication techniques, we choose $Z=0.5$. We assume equal magnitudes of Rashba SOC with opposite directions at the two interfaces, $\chi_L=-\chi_R=\chi=1.25$. The results shown are calculated for low temperature, $T/T_c=0.1$.

%Properties of diode efficiency and anomalous current are determined by the characteristics of the CPR. In systems with nonreciprocal Josephson current, in addition to the standard $\sin\varphi$ term, a $\cos\varphi$ component emerges in the CPR which is essential for the AJE. Furthermore, the coexistence of $\sin{\varphi}$, $\sin{2\varphi}$ and $\cos{\varphi}$ terms is required to enable the JDE \cite{Tanaka2022,Nas2025}.

We begin by analyzing the first class of  Josephson junctions defined in the previous section. The anomalous Josephson current and diode efficiency, as functions of the orientation of the magnetization vector in the right ferrmagnetic layer (i.e. as  functions of the angles $\theta_R$ and $\phi_R$), are shown in Fig. \ref{Fig2}: (a) and (e) for junctions with $s$-wave superconductors; (b) and (f) for $(\beta_L,\beta_R)=(0,0)$ oriented $d$-wave superconductors;  (c) and (g) for $(\beta_L,\beta_R)=(0.1\pi,-0.1\pi)$ oriented $d$-wave superconductors; and (d) and (h) for $(\beta_L,\beta_R)=(0.2\pi,-0.2\pi)$ $d$-wave oriented superconductors, with fixed orientation of the exchange field in the left ferromagnet defined by $\theta_L=0.3\pi$ and $\phi_L=0.3\pi$. The value and sign of nonreciprocal transport properties strongly depend on both the mutual orientations of the ferromagnetic layers and the orientation of the order parameter in the two superconducting electrodes. All panels on the left side of Fig. \ref{Fig2} show that both $I(0)$ and $Q$ vanish for $\phi_R=\phi_L=0.3\pi$ and $\phi_R=\phi_L+\pi=1.3\pi$, independent of the angle $\theta_R$ (see two vertical white lines), and for $\theta_R=0$, $\theta_R=\pi-\theta_L=0.7\pi$ and $\theta_R=\pi$ independent of the angle $\phi_R$ (see three horizontal white lines). The appearance of  vertical white lines with $I(\varphi=0)=0$ and $Q=0$, confirms that for coplanar orientation of the $z$-axis and spin-splitting fields in two ferromagnetic layers both AJE and JDE are forbidden  in this class of Josephson junction. On the basis of coplanarity the appearance of two horizontal white lines for $\theta_R=0$ and $\theta_R=\pi$ independent of the angle $\phi_R$ is also explained, while the third horizontal white line for $\theta_R=0.7\pi$ corresponds to the ferromagnetic orientation when the $z$-components of the exchange fields in the two ferromagnetic layers are equal but have opposite signs. The absence of AJE and JDE is in accordance with the conclusions of the symmetry analysis given in the previous section.

The right column of Fig. \ref{Fig2} shows the dependence of $Q$ and ground state phase difference $\varphi_G$ on the azimuthal angle $\phi_R$ for fixed polar angle $\theta_R=0.2\pi$ (Figs. \ref{Fig2} (i) and (j)), as well as on the polar angle $\theta_R$ for a fixed azimuthal angle $\phi_R=1.1\pi$ (Figs. \ref{Fig2} (k) and (l)), respectively. Due to the presence of Rashba SOC at the S/F interfaces, the total CPR arises from all transverse $k_y$ channels that contribute to transport. Consequently, individual phase shifts, different for different $k_y$, lead to a cumulative ground state phase difference that is not purely $0$ or $\pi$, but can take any value within the interval $\varphi_G\in[0,\pi]$ \cite{Costa2023,Nas2025}. Note that for coplanar orientation of the spin-splitting fields in the two ferromagnetic layers in the barrier, i.e. for $\phi_R=0.3\pi$ and $\phi_R=1.3\pi$, the ground state phase difference is purely $0$ or $\pi$ while $Q$ vanish for all considered junction (as indicated by the dashed lines in Figs. \ref{Fig2} (j) and (i), respectively). A similar behavior is found when the $z$-components of the exchange field in the two ferromagnets are equal in magnitudes but have opposite signs, i.e. $\theta_R=\pi-\theta_L=0.7\pi$, as indicated by the dash line in Figs. \ref{Fig2} (k) and (l).

The $0-\pi$ like phase transitions can be induced by tunning the mutual orientation of the two ferromagnets in the barrier, i. e. by varying azimuthal angle $\phi_R$ or the polar angle $\theta_R$ (see Figs. \ref{Fig2} (j) and (l), respectively). The phase transition occurs near the sharp sign reversal of  diode efficiency, leading to an enhanced value of $Q$ near the phase transition point. Besides the phase transition induced by $\phi_R$, which occurs via a jump as in the case of s-wave, $(0,0)$ and $(0.1\pi,-0.1\pi)$ types of junctions, the phase transition can also be achieved continuously. This is the case for the $(0.2\pi,-0.2\pi)$ type of junction, where a continuous phase transition induced by $\phi_R$ occurs at $\varphi_G=\pi/2$ or $-\pi/2$. For  $|\varphi_G|<\pi/2$ junction is in $0$-like state, while for $|\varphi_G|>\pi/2$ junction is in $\pi$-like state. For this type of junction, a phase transition induced by the polar angle $\theta_R$ does not occur for $\phi_R=1.1\pi$, and the junction remains in the $\pi$-like state for all values of  $\theta_R$ (see Fig. \ref{Fig2} (l)). This implies that $0-\pi$ like phase transition can also be induced by varying the mutual orientation of the $d$-wave superconducting electrodes ($\beta_L=-\beta_R=\beta$) \cite{Nas2025}. In Fig. \ref{Fig3} (a) CPRs are shown for arbitrarily oriented magnetizations in  ferromagnetic layers characterized by $\phi_L=0.3\pi$, $\theta_L=0.3\pi$, $\phi_R=1.1\pi$ and $\theta_R=0.2\pi$,  for several Josephson junctions belonging to the first class, indicating the appearance of $0$-$\pi$ like phase transitions with changing $\beta$. The corresponding ground state phase difference dependence on $\beta$ given in Fig. \ref{Fig3} (c) demonstrates that $0-\pi$ like phase transitions occurs at values of $\beta$ at which the diode efficiency sharply passes through zero, as indicated by dashed lines in Fig. \ref{Fig3} (b).

\begin{figure}[t]
	\centering
	\includegraphics[width=9cm]{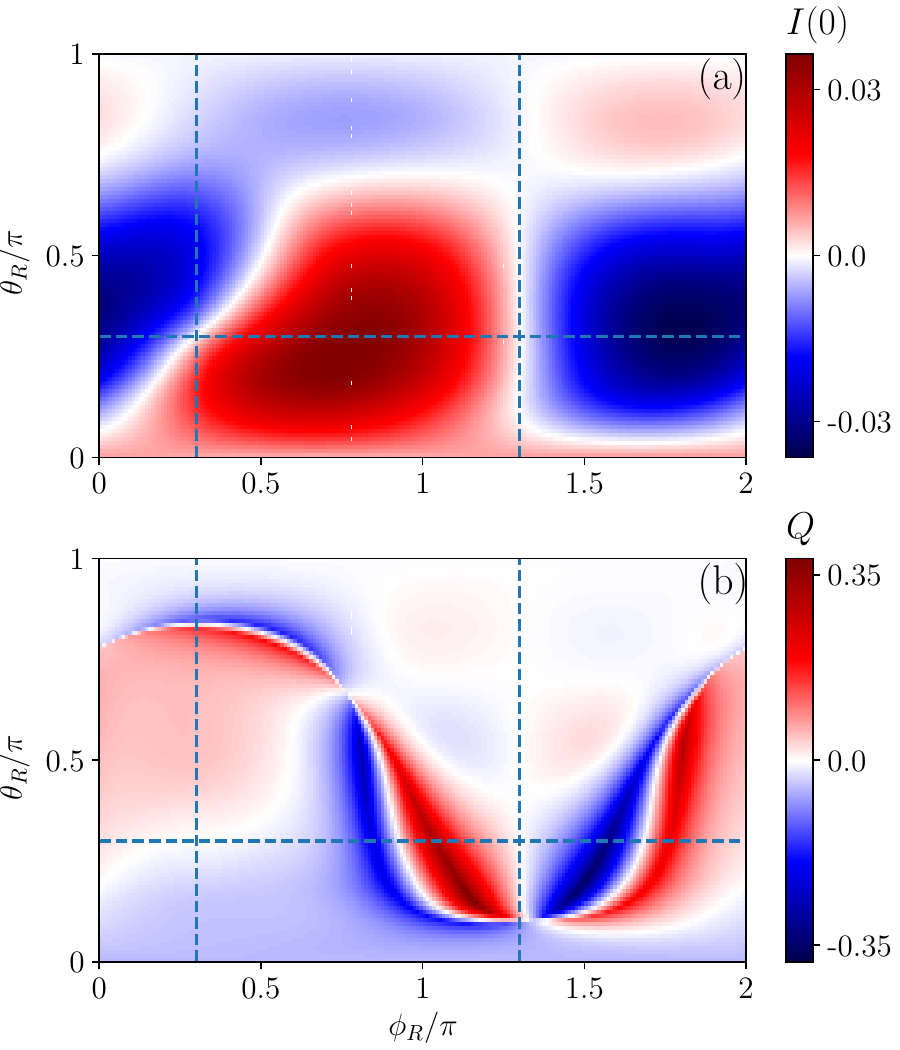}
	\caption{(a) Anomalous Josephson current $I(0)$  and (b) diode efficiency $Q$   as  functions of the angles $\phi_R$ and $\theta_R$, with a  fixed orientation of the exchange field in the left ferromagnet ($\phi_L=0.3\pi$ and $\theta_L=0.3\pi$), for $(\beta_L,\beta_R)=(0.15\pi,0.15\pi)$ type of junction. The intersection of the dashed lines indicates the orientation of the right ferromagnet for which AJE and JDE are forbidden. The other parameters are the same as in Fig. \ref{Fig2}.} \label{Fig4}	
	%\caption{(a)$(0.1\pi,0.1\pi)$ $d$-wave AJE, (b)$(0.1\pi,0.1\pi)$ $d$-wave JDE, (c) diode efficiency in dependence of $\phi_R$, for $\theta_R=0.05\pi$ (black), $\theta_R=0.3\pi$ (red), $\theta_R=0.8\pi$ (blue).}\label{Fig4}		
\end{figure}
\begin{figure}[t]
	\centering
	\includegraphics[width=9cm]{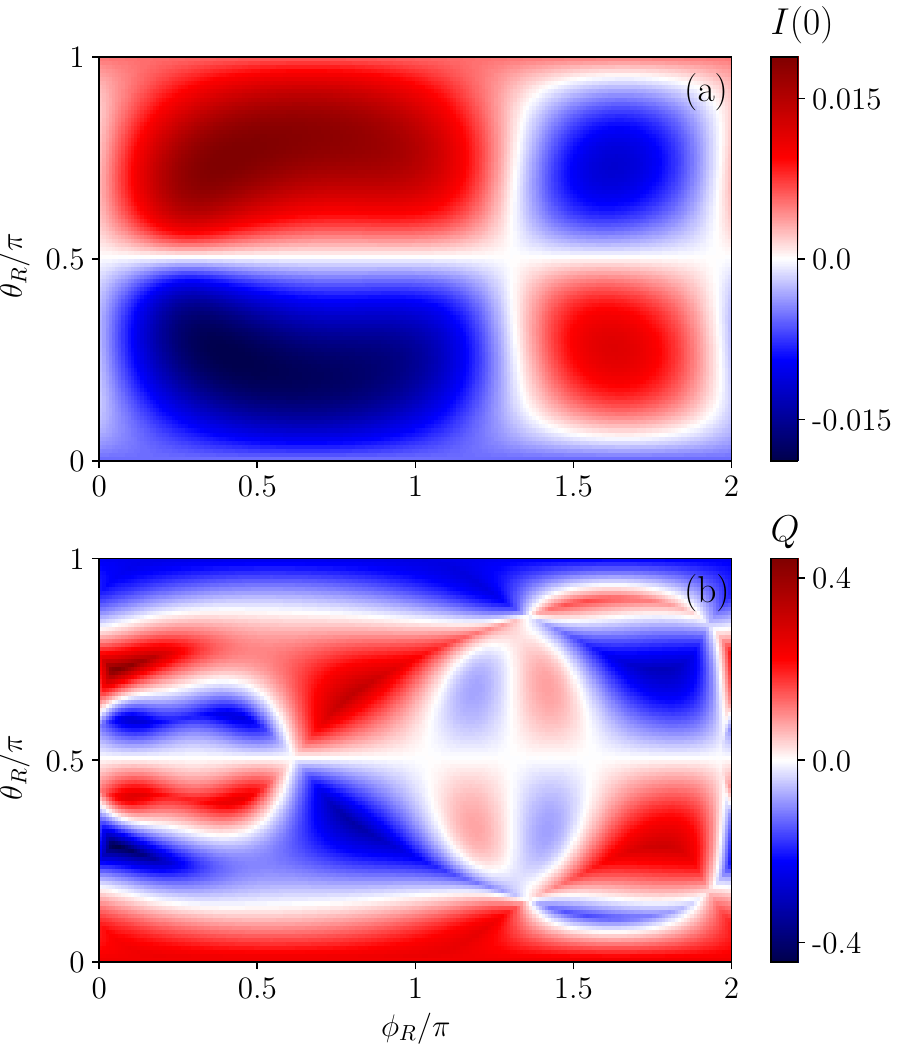}
	\caption{(a) Anomalous Josephson current $I(0)$  and (b) diode efficiency $Q$   as  functions of the angles $\phi_R$ and $\theta_R$ for a fixed orientation of the exchange field in the left ferromagnet ($\phi_L=0.3\pi$ and $\theta_L=0.3\pi$), for $(\beta_L,\beta_R)=(0.09\pi,-0.21\pi)$ type of junction. The other parameters are the same as in Fig. \ref{Fig2}.} \label{Fig5}
	%\caption{(a)$(0.09\pi,-0.21\pi)$ $d$-wave AJE, (b)$(0.09\pi,-0.21\pi)$ $d$-wave JDE, (c) curent-phase relations for $\theta_R=0.3\pi$ $\phi_R=0.3\pi$ (black), $\theta_R=0.3\pi$ $\phi_R=1.3\pi$ (red), $\theta_R=0.7\pi$ $\phi_R=0.3\pi$ (blue), $\theta_R=0.7\pi$ $\phi_R=1.3\pi$ (green).}\label{Fig5}		
\end{figure}
\begin{figure*}[!t]
	\centering
	\includegraphics[width=18cm]{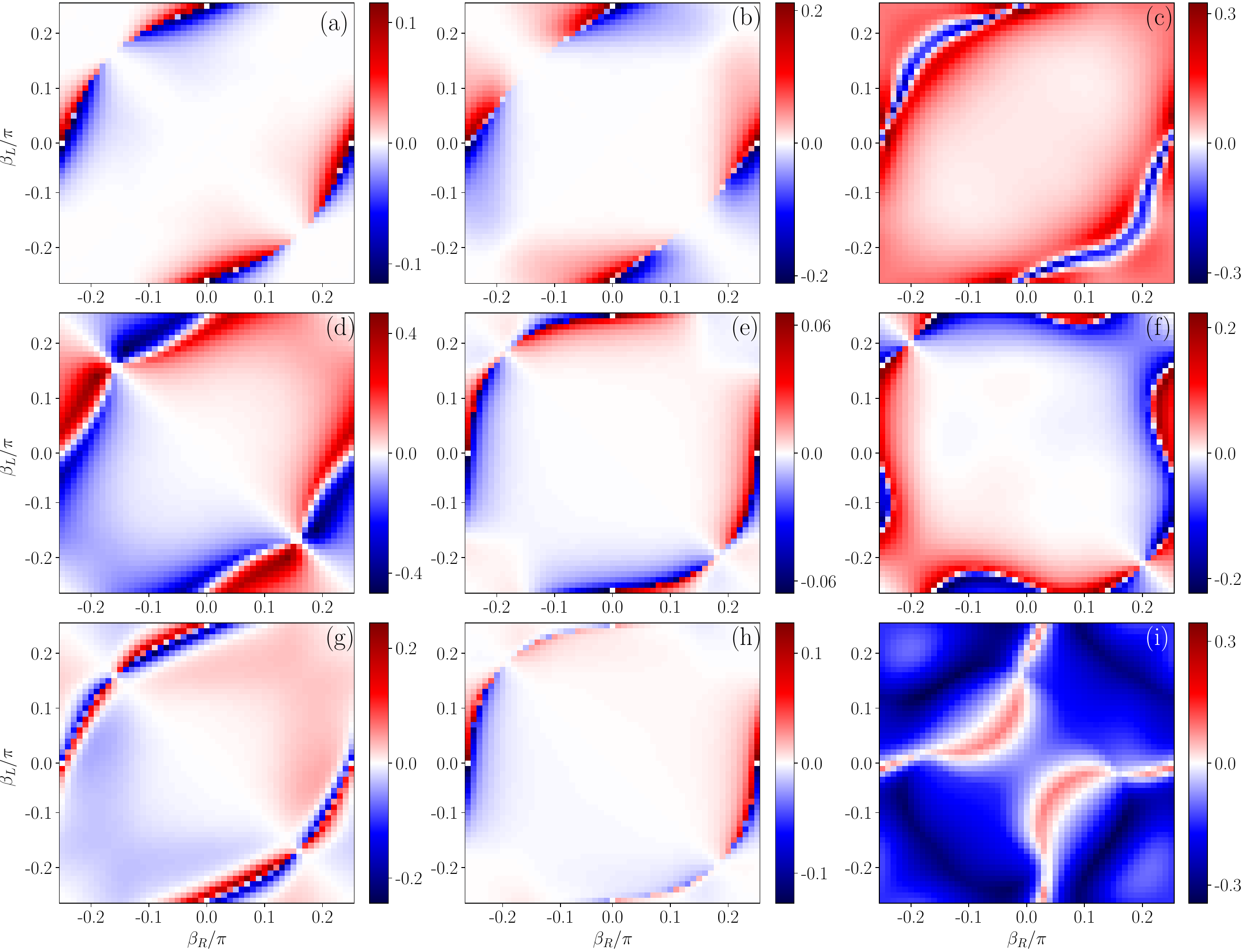}
	\caption{Diode efficiency $Q$ versus left and right d-wave superconductor crystal orientation $\beta_L$ and $\beta_R$ for $\theta_L=0.3\pi$ and $\phi_L=0.3\pi$. (a) $\theta_R=0.3\pi$ and $\phi_R=0.3\pi$, (b) $\theta_R=0.3\pi$ and $\phi_R=1.3\pi$, (c) $\theta_R=0.3\pi$ and $\phi_R=0$, (d) $\theta_R=0.7\pi$ and $\phi_R=0.3\pi$, (e) $\theta_R=0.7\pi$ and $\phi_R=1.3\pi$, (f) $\theta_R=0.7\pi$ and $\phi_R=0.8\pi$, (g) $\theta_R=0.4\pi$ and $\phi_R=0.3\pi$, (h) $\theta_R=0.6\pi$ and $\phi_R=1.3\pi$, (i) $\theta_R=0.5\pi$ and $\phi_R=0.8\pi$. The remaining parameters are the same as  in Fig. \ref{Fig2}.}\label{Fig6}
	%\caption{JDE in Josephson junctions with $d$-wave superconductors for $\theta_L=0.3\pi$ and $\phi_L=0.3\pi$. (a) $\theta_R=0.3\pi$ and $\phi_L=0.3\pi$, (b) $\theta_R=0.3\pi$ and $\phi_L=1.3\pi$, (c) $\theta_R=0.3\pi$ and $\phi_L=0$, (d) $\theta_R=0.7\pi$ and $\phi_L=0.3\pi$, (e) $\theta_R=0.7\pi$ and $\phi_L=1.3\pi$, (f) $\theta_R=0.7\pi$ and $\phi_L=0.8\pi$, (g) $\theta_R=0.4\pi$ and $\phi_L=0.3\pi$, (h) $\theta_R=0.6\pi$ and $\phi_L=1.3\pi$, (i) $\theta_R=0.5\pi$ and $\phi_L=0.8\pi$.}\label{Fig6}		
\end{figure*}

To analyze the nonreciprocal characteristics in the second class of Josephson junctions, we present results for the $(\beta_L,\beta_R)=(0.15\pi , 0.15\pi)$ type of Josephson junction. The orientation of the left ferromagnet layer is fixed at $\theta_L=0.3\pi$ and $\phi_L=0.3\pi$. The anomalous Josephson current and diode efficiency, as function of the orientation of the right ferromagnet layer (i.e., the angles $\theta_R$ and $\phi_R$), are shown in Figs. \ref{Fig4} (a) and (b). According to the symmetry analysis, there are two characteristic orientations of the right ferromagnet for which both AJE and JDE are prohibited. For the considered Josephson junction, the first orientation is given by $\theta_R=\theta_L=0.3\pi$ and $\phi_R=\phi_L=0.3\pi$, while the second is given by $\theta_R=\theta_L=0.3\pi$ and $\phi_R=\phi_L+\pi=1.3\pi$, as indicated by the intersection of the dashed lines in Figs. \ref{Fig4} (a) and (b). It can be observed that there are other points in Fig. \ref{Fig4} where $I(0)$ and $Q$ vanish, which depend on the values of other junction parameters. However, only for the orientations of the ferromagnetic layers at which the dashed lines intersect both AJE and JDE are absent regardless of changes in other parameters of the considered junction.

The third class of Josephson junctions refers to junctions with arbitrary oriented $d$-wave superconducting electrodes satysfying $\beta_R\neq \pm\beta_L$. The anomalous Josephson current and diode efficiency, as functions of the orientation of the  exchange field  in the right ferromagnet, for the $(0.09\pi,-0.21\pi)$ type of junction, are shown in Figs. \ref{Fig5} (a) and (b), respectively. Since the established condition from symmetry analysis for the appearance of AJE and JDE is a nonzero $z$-component of the exchange field in at least one ferromagnetic layer, we fixed the orientation of the left magnetization to be in the $xy$-plane ($\theta_L=0.5\pi$ and $\phi_R=0.3\pi$), achieving absence of nonzero $z$-component. As a result, $I(0)$ and $Q$ vanish when the right ferromagnet has $\theta_R=0.5\pi$ orientation, which is confirmed by the presence of horizontal white line at $\theta_R=0.5\pi$ independent of the angle $\phi_R$, as shown in Figs. \ref{Fig5} (a) and (b).

The mutual influence of the orientation of magnetization in two ferromagnetic layers and the orientation of $d$-wave superconducting electrodes significantly affects nonreciprocal features and deserves further investigation, as we have previously focused mainly on conditions under which both AJE and JDE are forbidden for symmetry reasons.  Panels in Fig. \ref{Fig6} show the dependence of diode efficiency on the superconducting orientation angles $\beta_L$ and $\beta_R$ for different mutual configurations of spin splitting directions in the ferromagnetic barrier. The first two columns of panels correspond to coplanar orientations, $\phi_R=\phi_L$ and $\phi_R=\pi+\phi_L$, while the third column represents the noncoplanar case. The first row of panels is for $\theta_R=\theta_L$, the second for $\theta_R=\pi-\theta_L$, and the third for the case with different $z$-components of the exchange fields in the ferromagnetic layers.

In the following, we analyze the symmetry properties of the diode efficiency distributions in the $(\beta_L, \beta_R)$ panels shown in Fig. \ref{Fig6}.

The transformations $P_y \sigma_x T$ and $P_y \sigma_y T$ (see Table \ref{table2} in Appendix \ref{appendix2}) map $H(\varphi,\beta_L,\beta_R)$ to $H(-\varphi,-\beta_L,-\beta_R)$. These transformations are preserved for all $\theta_L$ and $\theta_R$ whenever $\phi_L=\phi_R$ or $\phi_L=\phi_R+\pi$ leading $I_c^+(\beta_L,\beta_R)=-I_c^-(-\beta_L,-\beta_R)$. Therefore, for coplanar orientation of $\boldsymbol{e}_z$, $\boldsymbol{h}_L$ and $\boldsymbol{h}_R$ vectors, the diode efficiency is an odd function of $\beta_L$ and $\beta_R$ with respect to the point $(\beta_L,\beta_R)=(0,0)$ satisfying
\begin{equation}
	Q(\beta_L,\beta_R)=-Q(-\beta_L,-\beta_R).
\end{equation}
This is shown in panels (a), (b), (d), (e), (g) and (h) of Fig. \ref{Fig6} confirming the symmetry requirements.

On the other hand, $P_x P_y$ transforms $H(\varphi,\beta_L,\beta_R)$ to $H(-\varphi,-\beta_L,-\beta_R)$ and is preserved for $\theta_L=\theta_R$ and $\phi_L=\phi_R$ or $\phi_L=\phi_R+\pi$. This leads to the odd character of the diode efficiency with respect to the antidiagonal, as can be confirmed from panels (a) and (b) in Fig. \ref{Fig6}.

Further, the transformation $P_x \sigma_x T$ (see Table \ref{table3} in Appendix \ref{appendix2}) maps $H(\varphi,\beta_L,\beta_R)$ to $H(\varphi,-\beta_R,-\beta_L)$ yielding $I_c^\pm(\beta_L,\beta_R)=I_c^\pm(-\beta_R,-\beta_L)$, what is preserved for $\theta_L=\theta_R$ for all azimuthal orientations $\phi_L$ and $\phi_R$ of the exchange fields in the ferromagnets. Consequently, the diode efficiency exhibit even parity with respect to the main diagonal, as can be seen from panels (a), (b) and (c) in Fig. \ref{Fig6}.

If there are transformations that map the Hamiltonian  $H(\varphi,\beta_L,\beta_R)$ to $H(-\varphi,-\beta_L,-\beta_R)$, then the diode efficiency is an odd function around the main diagonal in the $(\beta_L, \beta_R)$ plane. These transformations are $P_x \sigma_x$ and $P_x \sigma_y$, and they are preserved for $\theta_L=\pi-\theta_R$, as shown in panels (d), (e) and (f) of Fig. \ref{Fig6}.

For the case of noncoplanar orientation and different magnitudes of the $z$-component of the exchange fields in the two ferromagnets, there is no parity symmetry with respect to the point $(0,0)$ and both diagonals, as can be seen from panel (i) of Fig. \ref{Fig6} .

%By carefully setting orientation of superconducting electrodes it can be achieved significant values of diode efficiency of Josephson junction. Ja se nadam da ima tehnika za ovo. For example more than 30\% on panel (f) or more than 40\% on panel (d).

\begin{figure}[!t]
	\centering
	\includegraphics[width=9cm]{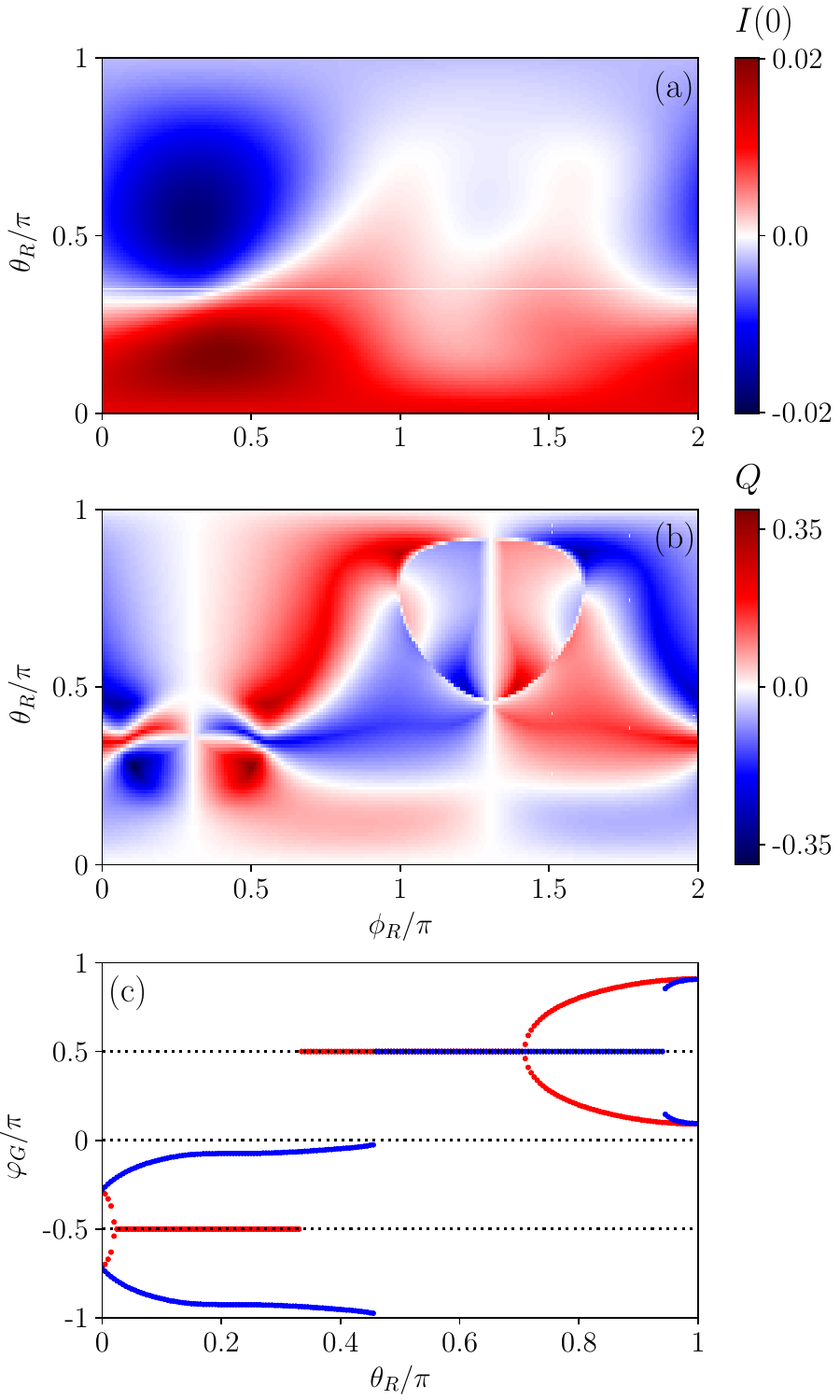}
	\caption{(a) Anomalous Josephson current $I(0)$ and (b) diode efficiency $Q$  as  functions of the angles $\phi_R$ and $\theta_R$, together with (c) the ground state phase difference $\varphi_G$ as a function of $\theta_R$ for $\phi_R=0.3\pi$ (red) and $\phi_R=1.3\pi$ (blue), calculated for $(\beta_L,\beta_R)=(0,0.25\pi)$ d-wave type junction. The  orientation of the exchange field in the left ferromagnet is fixed at $\phi_L=0.3\pi$ and $\theta_L=0.3\pi$. The remaining parameters are the same as in Fig. \ref{Fig2}.}\label{Fig7}	
	%\includegraphics[width=9cm]{numerika_slika7}
	%\caption{(a)$(0,0.25\pi)$ $d$-wave AJE, (b)$(0,0.25\pi)$ $d$-wave JDE, (c) ground state phase difference vs $\theta_R$ for $\phi_R=0.3\pi$-red and $\phi_R=1.3\pi$-blue. other parameters like in Fig. \ref{Fig2}.}\label{Fig7}		
\end{figure}

\begin{figure}[!t]
	\centering
	\includegraphics[width=8cm]{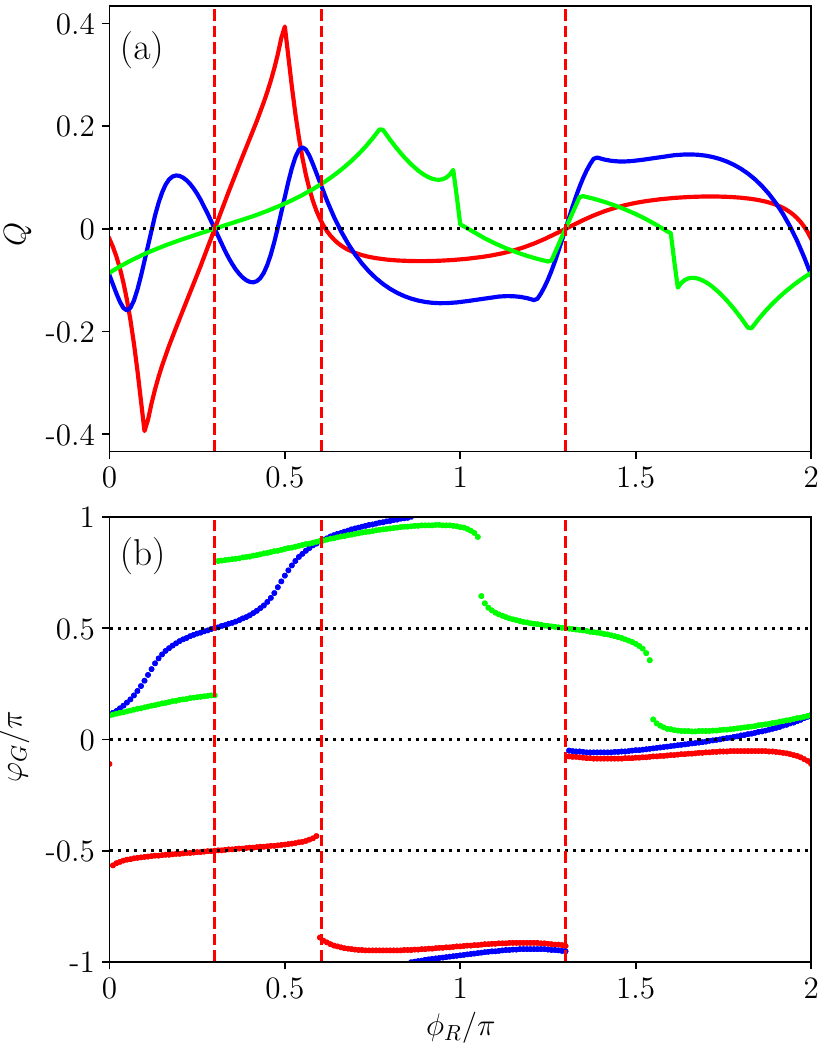}
	\caption{(a)Diode efficiency and (b) ground state phase difference $\varphi_G$ as functions of $\phi_R$ for  $\theta_R=0.27\pi$ (red), $\theta_R=0.4\pi$ (blue), and  $\theta_R=0.8\pi$ (green) for $(0,0.25\pi)$ d-wave junction. The  orientation of the exchange field in the left ferromagnet is fixed at $\phi_L=0.3\pi$ and $\theta_L=0.3\pi$. The remaining parameters are the same as  in Fig. \ref{Fig2}.} \label{Fig8}
	%\caption{(a)diode efficiency vs $\phi_R$ for $(0,0.25\pi)$ type of junction. $\theta_R=0.27\pi$-red, $\theta_R=0.4\pi$-blue, $\theta_R=0.8\pi$-lightgreen. (b) $\varphi_G$ za iste parametre.}
	%\caption{slika0pi4}
	\label{Fig8}		
\end{figure}

For coplanar orientation of spin splitting directions, the corresponding panels from Fig. \ref{Fig6} show that the diode efficiency for $(0,\pm0.25\pi)$ and $(\pm0.25\pi,0)$ oriented superconducting electrodes is zero, due to the odd character of $Q$ with respect to the $(0,0)$ point in the $(\beta_L,\beta_R)$ plane and the periodic behavior in dependence on superconducting orientations \cite{Nas2025}
\begin{equation}
	Q(\beta_L,\beta_R)=Q(\beta_L\pm\pi/2,\beta_R\pm\pi/2).
\end{equation}
Diode efficiency for a $(0,0.25\pi)$ type junction as a function of the magnetization orientation in right ferromagnet for $\theta_L=0.3\pi$ and $\phi_L=0.3\pi$, shown in Fig. \ref{Fig7} (b), displays two vertical white lines at $\phi_R=\phi_L=0.3\pi$ and $\phi_R=\phi_L+\pi=1.3\pi$, and two horizontal white lines at $\theta_R=0$ and $\pi$, since the diode effect is forbidden for coplanar orientation of spin splitting directions. In contrast, the corresponding anomalous Josephson current has a nonzero value for coplanar orientation, as seen in Fig. \ref{Fig7} (a), which is consistent with its classification in the third class of junctions defined in the previous section. For coplanar orientation, the spin splitting direction transformations $P_x\sigma_xT$ and $P_y\sigma_yT$ are Hamiltonian symmetries that map $H(\varphi,0,\pi/4)$ to $H(-\varphi\pm\pi,0,\pi/4)$, leading to $2\pi$ periodicity of the CPR ($I(\varphi)=-I(-\varphi+\pi)$), and the current at phase difference $\varphi=\pm\pi/2$ becomes zero. This means that the  CPR contains $\sin{2n\varphi}$ and $\cos{(2n-1)\varphi}$  Fourier components \cite{Lu2015}, and the free energy minima indicate that the junction ground state can be: $\pi/2$, coexisting $0$-like and $\pi$-like states near $\pi/2$, $-\pi/2$, or coexisting $0$-like and $\pi$-like states near $-\pi/2$ (see Fig. \ref{Fig7} (c)). This differs from the case of a $(0,\pi/4)$ type junction without SOC, where the phase difference can be $\pi/2$ or a coexistence of $0$ and $\pi$ states, which is a consequence of $\pi$ periodicity with only $\sin 2n\varphi$ Fourier components in the CPR \cite{Nas21,Zikic2007}.

%\textcolor{red}{Napomena: Ovdje je mogice govoriti o razlikovanju dva stanja, a ne cetiri kako smo govorili. 1. stanje $0$-like i $\pi$-like koegzistencije u okolini $-\pi/2$ sto ukljucuje i stanje $-\pi/2$, i 2.stanje $0$-like i $\pi$-like koegzistencije u okolini $+\pi/2$ sto ukljucuje i stanje $+\pi/2$. Naime, ako se pogleda crvena linija i tacka faznog prelaza skokom sa $-\pi/2$ na $+\pi/2$, i plava sa prelazom, naizgled sa $0-\pi$ koegzistencije na $+\pi/2$ stanje, onda su to ustvari promjene fazne razlike osnovnog stanja za $\pi$. U tim tackama postoji $\pi$ periodicnost CPR i vazi $I(\varphi)=I(\varphi+\pi)$. U slucaju kad postoji AJE ali ne i JDE CPR se moze dekomponavati kao
%$$I(\varphi)=\sum_{n\geq1}I_{2n}\sin(2n\varphi)+J_{2n-1}\cos((2n-1)\varphi)$$
%Sa druge strane je:
%$$I(\varphi\pm\pi)=\sum_{n\geq1}I_{2n}\sin(2n(\varphi\pm\pi))+J_{2n-1}\cos((2n-1)(\varphi\pm\pi))$$,
%$$I(\varphi\pm\pi)=\sum_{n\geq1}I_{2n}\sin(2n\varphi)-J_{2n-1}\cos((2n-1)\varphi)$$
%U tacki faznog prelaza postoji koegzistencija pa je $I(\varphi)=I(\varphi+\pi)$ odakle slijedi da je
%$$J_{2n-1}=0, \forall n\in\mathbb{N}$$
%pa je $I(\varphi=0)=0$. Dakle, u tackama faznog prelaza gdje je promjena fazne razlike osnovnog stanja za $\pm\pi$ anomalna struja mijenja znak. Ovo sto smo pokazali vazi kad nema JDE.}

The phase transitions depicted in Fig. \ref{Fig7} (c) correspond to effective $\pi$ jump in the ground state phase difference. The CPR in this type of junction can be decomposed into Fourier harmonics
\begin{equation}
I(\varphi)=\sum_{n\geq1}I_{2n}\sin(2n\varphi)+J_{2n-1}\cos((2n-1)\varphi).
\end{equation}
Since, shifting the phase by $\pi$ yields
\begin{equation}
I(\varphi\pm\pi)=\sum_{n\geq1}I_{2n}\sin(2n\varphi)-J_{2n-1}\cos((2n-1)\varphi),
\end{equation}
and at the crossover point, where the condition $I(\varphi)=I(\varphi\pm\pi)$ holds, it follows that all odd cosine harmonics must vanish
\begin{equation}
	J_{2n-1}=0, \forall n\in\mathbb{N}.
\end{equation}
This implies that at the phase transition point, where JDE is absent, the anomalous Josephson current is zero and reverse its sign across the transition. This behavior can be confirmed numerically by comparing panels (a), (b) and (c) in Fig. \ref{Fig7}.

In contrast, for a $(0,\pi/4)$ type of junction with a noncoplanar orientation of spin splitting directions, JDE emerges, resulting to presence of both sine and cosine Fourier harmonics of all order, i.e. terms of the form $\sin n\varphi$ and $\cos n\varphi$. As a result, the free energy minimum is unique (nondegenerate) and defines the so-called $0$-like and $\pi$-like states. In Figs. \ref{Fig8} (a) and (b), the diode efficiency and the corresponding ground state phase difference as a function of $\phi_R$ for several values of $\theta_R=0.27\pi$ (red), $\theta_R=0.4\pi$ (blue), and $\theta_R=0.8\pi$ (green), with a fixed orientation of the left ferromagnet ($\theta_L=0.3\pi$ and $\phi_L=0.3\pi$), are shown, respectively. A phase transition between the $0$-like and $\pi$-like states can be induced  by tuning $\phi_R$. In contrast to the first class of junctions shown in Fig. \ref{Fig2} (j), the crossover in this case occurs at $\phi_R=0.3\pi$ and $\phi_R=1.3\pi$, corresponding to a coplanar configuration.

\begin{figure*}[!t]
	\centering
	\includegraphics[width=18cm]{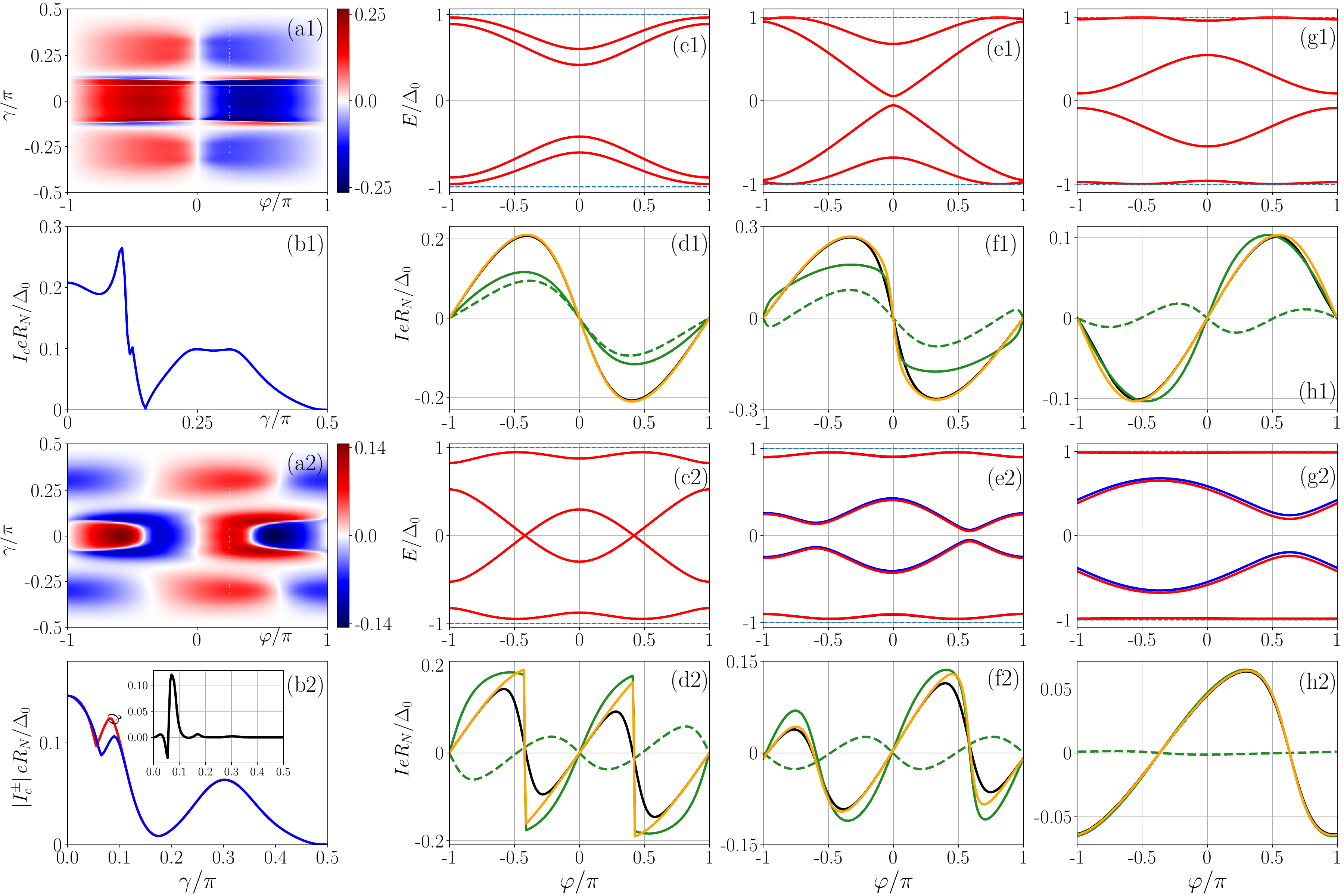}
	\caption{Josephson junction with s-wave superconducting electrodes. Panels (a1)-(h1) correspond to a homogeneous ferromagnetic  barrier with $\theta_L=\theta_R=0.3\pi$ and $\phi_L=\phi_R=0.3\pi$. Panels (a2)-(h2) correspond to a junction containing two ferromagnets with noncolinear magnetizations, with $\theta_L=0.3\pi$, $\phi_L=0.3\pi$, $\theta_R=0.26\pi$ and $\phi_R=1.06\pi$.  (a1), (a2) Angle-resolved CPRs $I(\varphi,\gamma)$ for transverse channels with incident angle $\gamma\in\left[-\pi/2,\pi/2\right]$. (b1), (b2) Josephson critical current as a function of incident angle $\gamma$, where current in opposite directions  $I_c^{+}$ (red) and $|I_c^{-}|$ (blue) are shown. Inset in (b2) shows diode efficiency $Q$ as a function of $\gamma$.  ABS  and the corresponding CPR are shown for: (c1), (c2) and (d1), (d2) $\gamma=0$; (e1) and (f1) $\gamma=\pm0.105\pi$; (g1) and (h1) $\gamma=\pm0.125\pi$; (e2) and (f2) $\gamma=\pm0.07\pi$; (g2) and (h2) $\gamma=\pm0.3\pi$, respectively. The red and blue curves denote the ABS bands for $\gamma$ and $-\gamma$, respectively. Note that, for homogeneous ferromagnet barrier, curves for $\pm\gamma$ nearly overlap. The solid and dashed green curves represent the contribution from the lower and upper ABS bands, respectively. The orange and black curves represent the total CPRs for the given channel obtained using ABS and F-T approaches, respectively. The remaining parameters are the same as in Fig. \ref{Fig2}. } \label{Fig9}		
\end{figure*}

\section{Andreev bound states} \label{sec5}

In this section, we examine how different orientations of exchange fields in two ferromagnetic layers affect the low energy ABS spectra and their signatures in nonreciprocal transport. We present results for short junctions, as in this regime only a few in-gap states appear within the superconducting gap\cite{Beenakker1991,Meng2022}, as shown below. Since we are dealing with a 2D junction, there are many  transverse channels, defined by $k_y=k_F\sin\gamma$, that contribute to charge transport. Channels contribute differently to the  CPR, not only in current intensity, but also, due to Rashba SOC, in the ground state phase differences of individual channels \cite{Costa2023}. The total CPR is obtained by summing over all $k_y$ resolved channel contributions. Consequently, the combined impact of different transverse channels on the ABS spectra must be considered. Because we present results at low temperatures, the free energy of a single channel (based on Eq. (\ref{slobodna})) can be approximated as
\begin{equation}\label{sl2}
	F(\gamma,\varphi)=-\frac{1}{2}\sum_i E_i(\gamma,\varphi),
\end{equation}
where $E_i(\gamma,\varphi)$ are positive ABS energies, satisfying $0\leq E_i\leq \Delta$. The CPR for each channel is then obtained from corresponding ABS spectrum using Eq. (\ref{sl}). The channels defined with opposite $\gamma$ are needed for the calculation of free energy from phase-dependent ABS spectra, in accordance to particle-hole symmetry conservation. The conditions under which Andreev bound states predominantly determine charge transport will be established through comparisons of the Furusaki-Tsukada technique and the ABS approach.

\subsection{Junctions with $s$-wave pairing}

We first consider a Josephson junction with a homogeneous ferromagnet in the barrier,  with $\theta_L=\theta_R=0.3\pi$ and $\phi_L=\phi_R=0.3\pi$. In Fig. \ref{Fig9} (a1), the incident angle-resolved CPR shows that the dominant harmonic in all channels is $\pm \sin \varphi$, and consequently, both $I(0)$ and $Q$. this is consequence of the necessity of the cosine term in CPR for the occurance of AJE, while coexistence of $\sin{\varphi}$, $\sin{2\varphi}$ and $\cos{\varphi}$ terms is required for JDE \cite{Tanaka2022,Nas2025}.  The dependence of the maximal Josephson current on $\gamma$ depicted in Fig. \ref{Fig9} (b1), exhibits a nonmonotonic behavior, where observed dips indicate a sign reversals of the current across channels, with $\gamma\in\left[0,\pi/2\right)$, since the angle-resolved CPR show even symmetry with respect  to the incident angle $\gamma$.

To consider the contribution of ABSs to the Josephson current for different channels, we show phase-dependent energy spectra of ABS in Figs. \ref{Fig9} (c1), (e1), and (g1) for incident angles $\pm\gamma=0$, $0.105\pi$ and $0.125\pi$, respectively. The ABS spectra are spin nondegenerate and display two positive energy ABS bands, together with their negative energy counterparts in accordance with electron-hole symmetry, with almost matching curves for $\pm \gamma$ incident angles. Both the lower and upper  ABS phase-dependent energies are symmetric with respect to $\varphi=0$. In the case of a quasiparticle  incident normal to the boundary ($\gamma=0$), there is no influence of Rashba SOC, and the nondegeneracy arises solely from the exchange field.  The contributions of the energy bands to the total supercurrent are comparable, as can be seen by comparing green curves, solid for the lower and dashed for the upper band, in Fig. \ref{Fig9} (d1), where the corresponding CPRs are shown. The CPRs from both bands have the same ground state phase difference ($\pi$ state) for the chosen parameter values.

With increasing quasiparticle incident angles ($\gamma\not=0$), the phase-dependent spectrum remains symmetrical with respect to $\varphi=0$. The ABS spectrum for the $\pm\gamma=0.105\pi$ channel, corresponding to the maximal critical current, show that lower band is shifted towards the centre of the superconducting gap near $\varphi=0$, while the  upper band moves towards the gap edge (see Fig. \ref{Fig9} (e1)). For the $\pm\gamma=0.125\pi$ channel, which corresponds to the angle at which maximal critical current with reversed sign occurs, the lower band is pushed towards low energies near $\varphi=\pm\pi$, while the upper band is pinned to the gap edge, becoming almost flat with respect to $\varphi$ (see Fig. \ref{Fig9} (g1)). The corresponding CPRs are depicted in Figs. \ref{Fig9} (f1) and (h1), and show that CPRs originating from different bands posses different ground state phase differences.

The good agreement between total CPRs derived from ABS phase-dependent bands for a single channel (orange curves) and those obtained by the F-T technique (black curves) shown in Figs. \ref{Fig9} (d1), (f1), and (h1) indicates that coherent charge transport is predominantly governed by subgap ABSs. The absence of both the AJE and JDE shown in CPR curves is consistent with our symmetry analysis, and is confirmed within both F-T and ABS approaches.

Next, we discuss the case of two ferromagnets in the barrier with mutually noncollinear magnetizations, where our symmetry analysis predict that both AJE and JDE are present. We chose the orientation of the left and right ferromagnet magnetizations ($\theta_L=0.3\pi$, $\phi_L=0.3\pi$, $\theta_R=0.26\pi$ and $\phi_R=1.06\pi$) corresponding to the configuration where diode efficiency reaches its maximum value, in accordance with Fig. \ref{Fig2} (e). In this case, the angle-resolved CPR shown in Fig. \ref{Fig9} (a2) becomes more complex  because single-channel contributions are diverse and show the presence of $\sin \varphi$, $\sin 2\varphi$ and $\cos \varphi$ harmonics. The difference in the dependence of maximal critical currents in opposite directions, $I_c^{+}$ (red curve) and $|I_c^{-}|$ (blue curve), on the incident angle $\gamma$ shown in Fig. \ref{Fig9} (b2) becomes more pronounced at smaller quasiparticles incidence angles (see also the inset displaying diode efficiency as a function of $\gamma$). ABS spectrum is nondegenerate and symmetric with respect to $\varphi=0$ (see Fig. \ref{Fig9} (c2)). The lower band undergoes a zero energy crossing around $\varphi=0$, which means that the junction is in the vicinity of $0-\pi$ like crossover \cite{Yokoyama2014}. This results in a discontinuity in the current-phase relation with a sawtooth-like profile and the presence of significant second harmonics (solid green curve in Fig. \ref{Fig9} (d2)). The upper band has a less prominent contribution to the CPR with second harmonic of opposite sign compared to the lower band (dashed green curve in Fig. \ref{Fig9} (d2)). Consequently, the total CPR obtained from the ABS retains sawtooth-like profile (orange curve in Fig. \ref{Fig9} (d2)), with noticeable difference from the CPR obtained by the F-T technique (black curve in Fig. \ref{Fig9} (d2)) particularly near the phase difference where the zero energy crossing appears.

The ABS spectra display an asymmetric nature around $\varphi=0$ for channels where both the diode and anomalous effects are finite. This behavior is illustrated in Figs. \ref{Fig9} (e2) and (g2), which show the ABS spectrum for incidence angles $\pm\gamma=0.07\pi$, where diode efficiency is significant (see Fig. \ref{Fig9} (b2)), and for $\pm\gamma=0.3\pi$, where the anomalous effect is maximal (see Fig. \ref{Fig9} (a2)), respectively. In this case,  the ABS bands for $-\gamma$ and $\gamma$ differ slightly, as seen by comparing the blue (for $-\gamma$) and red (for $\gamma$) curves (see \ref{Fig9} (e2) and (g2)). The symmetry relation $E(-\varphi)=E(\varphi)$ is therefore broken, and the skewed phase-dependent ABS spectrum reflects the presence of both JDE and AJE, since both the difference between the absolute values of maximal and minimal currents, as well as the anomalous supercurrent at $\varphi=0$, are observed, as shown in Figs. \ref{Fig9} (f2) and (h2) \cite{Wang2025a,Cayao2024}. The dominant contribution to the CPR comes from the lower bands, since the upper band is almost flat. For incident angles $\pm\gamma=0.07\pi$, these  low-energy states are close to zero energy around $\varphi=0$, resulting in  significant second harmonic contributions to the total CPR (Figs. \ref{Fig9} (e2) and (f2)). Since zero energy crossing is absent, the profile of CPR remains smooth here. As before, the discrepancy between the CPRs obtained from F-T technique and that obtained by the ABS approach is enhanced near the phase difference where the low energy ABS emerge.

For larger incident angle $\pm\gamma=0.3\pi$, where the anomalous effect is pronounced, CPR exhibits a clear anomalous characteristic, due to asymmetric nature and significant slope at $\varphi=0$ in lower band, resulting in a dominant $\cos \varphi$ term  (see Fig. \ref{Fig9} (h2)). In this case, the agreement between the CPRs obtained by the F-T technique and ABS is almost complete.

\begin{figure}[!t]
	\centering
	\includegraphics[width=8.7cm]{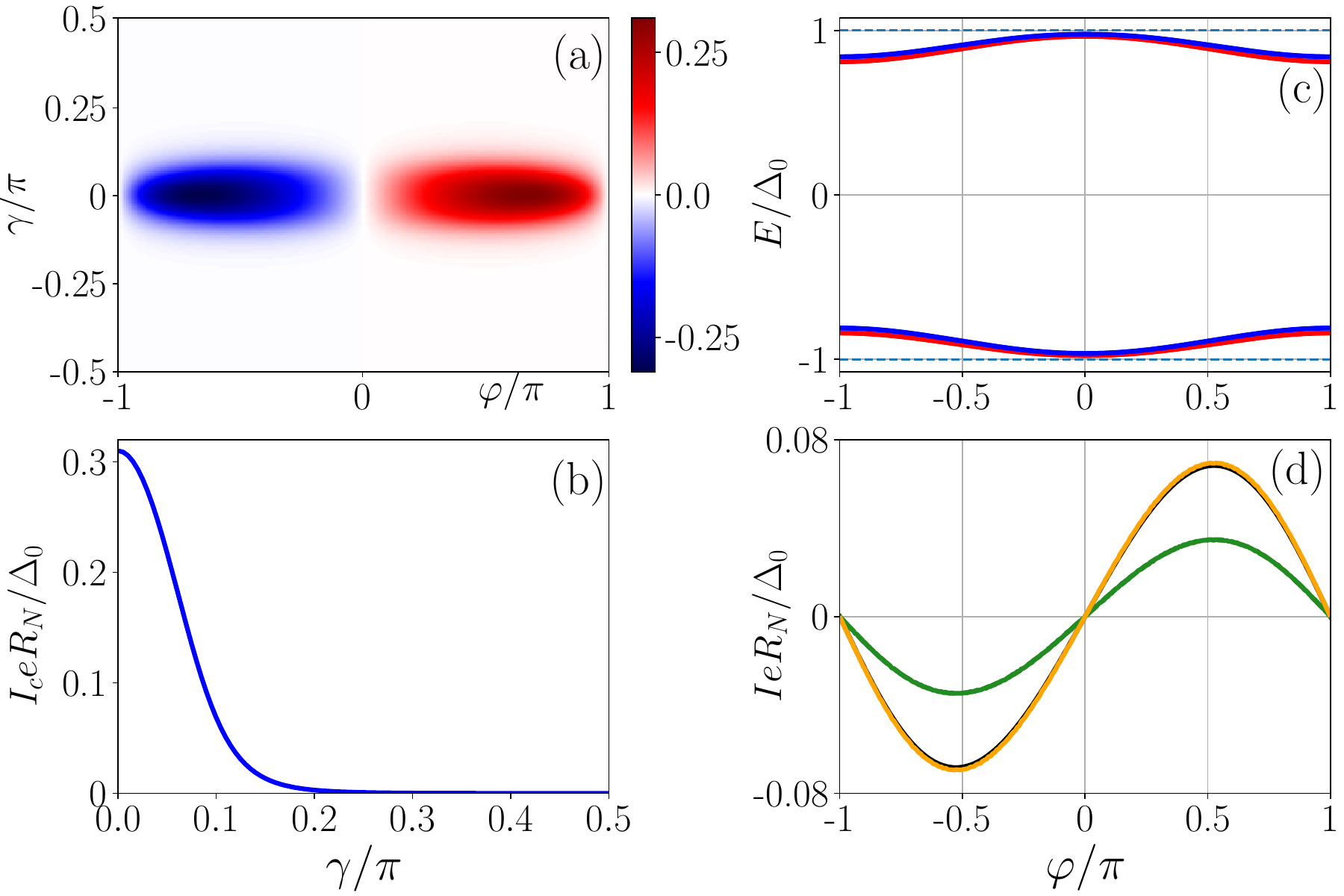}
	\caption{All panels correspond to a Josephson junction with s-wave superconducting electrodes and an inhomogeneous ferromagnetic barrier where the magnetization vectors in the two ferromagnetic layers are oppositely oriented  ($\theta_L=\phi_L=0.3\pi$, $\theta_R=\pi-\theta_L=0.7\pi$ and $\phi_R=\pi+\phi_L=1.3\pi$). (a) Angle-resolved CPRs $I(\varphi,\gamma)$ for transverse channels with incident angle $\gamma\in\left[-\pi/2,\pi/2\right]$. (b) Josephson critical current as a function of incident angle $\gamma$. (c) ABS energies as a function of the superconducting phase difference $\varphi$. The red  and blue curves represent ABS bands for $\gamma$ and $-\gamma$, respectively. (d) the corresponding CPR for $\gamma=0.1\pi$.  The green curve correspond to currents from ABS bands, while the orange and black curves represent the total CPRs for a given channel from the ABS and F-T approaches, respectively. The remaining parameters are the same as in Fig. \ref{Fig2}.}\label{Fig10}		
\end{figure}
Further, we consider the junction with an inhomogeneous ferromagnet in which the magnetization vectors in the two ferromagnetic layers are oriented oppositely, namely $\theta_L=\phi_L=0.3\pi$, $\theta_R=\pi-\theta_L=0.7\pi$ and $\phi_R=\pi+\phi_L=1.3\pi$. In this configuration both AJE and JDE are absent, as confirmed by our symmetry analysis. The angle-resolved CPR (Fig. \ref{Fig10} (a)) shows that the dominant contribution to the current comes from channels with a small quasiparticle incident angles. All channels exhibit a $0$-like phase with dominant $\sin\varphi$ harmonic, while the maximal current through the channels is monotonically decreasing function of $\gamma$ (see Fig. \ref{Fig10} (b)). The spectrum of ABS remains symmetric with respect to $\varphi=0$  (Fig. \ref{Fig10} (c)), in accordance with the absence of both JDE and AJE. The ABS spectrum contains only one positive energy band indicating a spin degenerate state.  This degeneracy originated from to the specific orientations of the spin splitting directions in the junction. Namely, the two exchange fields act oppositely on the quasiparticle spin, while the Rashba SOC at the two interfaces  also have opposite orientations (in the $\mathbf{e}_z$ direction on the left S/F interface, and $-\mathbf{e}_z$ on the right S/F interface).  The agreement between the F-T and ABS approaches is significant (overlapped black and orange curves in Fig. \ref{Fig10} (d)), indicating that the current is predominantly carried by subgap ABSs, whereas the contribution from the continuum part of the spectrum remains negligible.

\subsection{Junctions with $d$-wave pairing}
\begin{figure*}[!t]
	\centering
	\includegraphics[width=18cm]{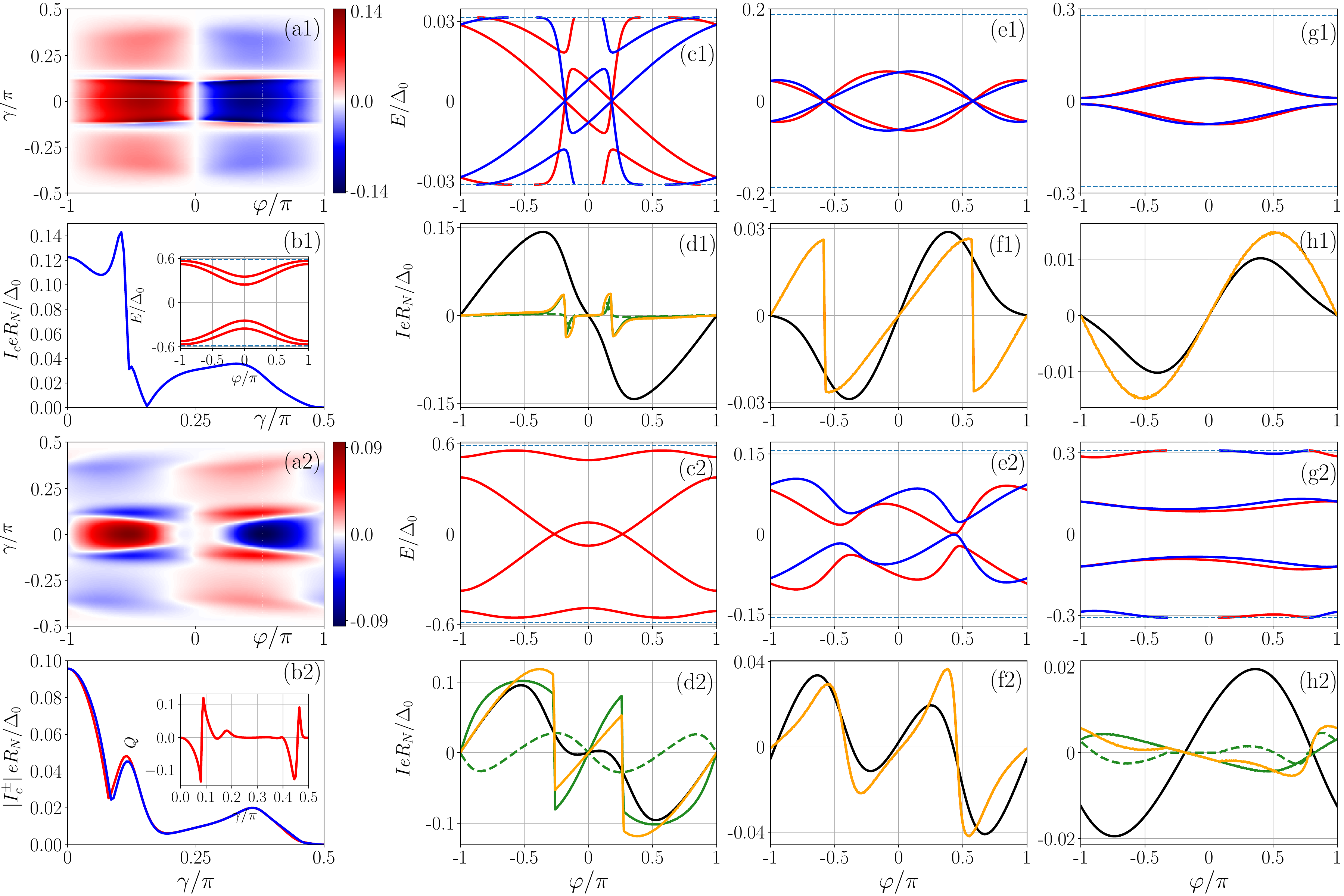}
	\caption{All panels correspond to ($0.15\pi, 0.15\pi$) type $d$-wave Josephson junction. Panels (a1)-(h1) correspond to homogeneous ferromagnetic barrier, with $\theta_L=\theta_R=0.3\pi$ and $\phi_L=\phi_R=0.3\pi$. Panels (a2)-(h2) correspond to a junction containing two ferromagnets with noncolinear magnetizations, with $\theta_L=0.3\pi$, $\phi_L=0.3\pi$, $\theta_R=0.13\pi$ and $\phi_R=1.16\pi$. The remaining parameters are the same as in Fig. \ref{Fig2}. (a1), (a2) Angle-resolved CPRs $I(\varphi,\gamma)$ for transverse channels with incident angle $\gamma\in\left[-\pi/2,\pi/2\right]$. (b1), (b2) Josephson critical current as a function of incident angle $\gamma$, where current in opposite directions  $I_c^{+}$ (red) and $|I_c^{-}|$ (blue) are shown. ABS  and the corresponding CPR are shown for: (c1) and (d1) $\pm\gamma=0.105\pi$; (e1) and (f1) $\pm\gamma=0.13\pi$; (g1) and (h1) $\pm\gamma=0.145\pi$; (c2) and (d2) $\gamma=0$; (e2) and (f2) $\pm\gamma=0.075\pi$; (g2) and (h2) $\pm\gamma=0.35\pi$, respectively. The red and blue curves denote the ABS bands for $\gamma$ and $-\gamma$, respectively.  The solid and dashed green curves represent the contribution from the lower and upper ABS bands, respectively. The orange and black curves represent the total CPRs for the given channel obtained using ABS and F-T approaches, respectively. ABS for $\gamma=0$, and diode efficiency as a function of $\gamma$, are shown in panels (b1) and (b2), respectively, as insets. The remaining parameters are the same as in Fig. \ref{Fig2}.}\label{Fig11}		
\end{figure*}
%Further, we discuss the junction with inhomogeneous ferromagnet when the magnetization vectors in two ferromagnetic layers are mutually directed opposite, with $\theta_L=\phi_L=0.3\pi$, $\theta_R=\pi-\theta_L=0.7\pi$ and $\phi_R=\pi+\phi_L=1.3\pi$, where AJE and JDE are absent, too. Angle resolved CPR (see Fig. \ref{Fig10} (a)) shows that the dominant contribution to the current comes from the channels with a small angle of incidence, while the maximal current through the channels is monothonicaly decreasing function of $\gamma$ with $0$-like phase in all channels (see Fig. \ref{Fig10} (b)). The spectrum of ABS is symmetric with respect to $\varphi=0$, Fig. \ref{Fig10} (c). In the present case only one positive energy band appears meaning spin degenerated state. The degeneracy appears due to the specific orientations of spin splitting directions in the junction. Namely, two exchange fields acts opposite on the quasiparticle spin, and the Rashba SOC at the interfaces are also opposite (in direction $\mathbf{e}_z$ on the left S/F interface, and $-\mathbf{e}_z$ on the right S/F interface). With increasing $\gamma$ the ABS is pushed to the gap edges, and the agreement between Furusaki-Tsukada and ABS approach is significant (Fig. \ref{Fig10} (d)).

To analyze the ABS spectra in the case of a junction with $d$-wave superconducting electrodes, we first consider the Josephson junction with $\beta_L=\beta_R=0.15\pi$. For homogeneous ferromagnet in the barrier, with $\theta_L=\theta_R=\phi_L=\phi_R=0.3\pi$, the angle resolved CPR  and the incident angle dependence of the maximal Josephson current are shown in  Figs. \ref{Fig11} (a1) and (b1), respectively. The global behavior is very similar to that found in the corresponding junction with $s$-wave superconducting electrodes, since the dominant harmonic in all channels is $\pm\sin \varphi$, with equal magnitudes of positive $I_c^+$ and negative $I_c^-$ critical current, resulting in the absence of both AJE and JDE. Similarly, the ABS spectrum for normal incidence (inset in Fig. \ref{Fig11} (b1)) and the corresponding CPRs are qualitatively similar to the $s$-wave case, even though  current values is reduced due to the narrower superconducting gap arising from the angle dependent superconducting order parameter \cite{Lofwander2001}.

The ABS spectra for channels with maximal critical current,  $\pm\gamma=0.105\pi$, are shown in Fig. \ref{Fig11} (c1), where the red curve corresponds to $\gamma=0.105\pi$ and the blue curve to $\gamma=-0.105\pi$.  In these regime, the  superconducting gap becomes very narrow, and the phase dependent energy bands of the ABS appears complex. Unlike the Josephson junction with $s$-wave electrodes, in the case of $d$-wave superconductors, the difference between the ABS spectra for $\gamma$ and $-\gamma$ is more pronounced due to the  anisotropy of the order parameter. For the $s$-wave junction symmetric phase dependent energy around $\varphi=0$, for both $\gamma$ and $-\gamma$ channels, results in the absence of both an anomalous phase shift  and a difference between the critical current in opposite directions. In the junction with $d$-wave electrodes, although the ABS spectra themselves becomes asymmetric with respect to $\varphi=0$, the total free energy obtained from Eq. (\ref{sl2}), by including contributions from $\gamma$ and $-\gamma$, satisfies the symmetry relation $F(\varphi)=F(-\varphi)$). Consequently, nether AJE nor JDE emerges. For certain values of the phase difference, the upper band is pushed out of the superconducting gap and does not contribute  to the currents carried by the ABS. Despite the rich phase dependent ABS spectra, because  the superconducting gap becomes nearly closed ($\sim0.03\Delta_0$), the contribution of continuum states to the Josephson current is the most significant (see Appendix \ref{appendix3}), as seen in Fig. \ref{Fig11} (d1), where the CPR obtained from the F-T approach (black curve) significantly exceeds the CPR obtained from the ABS (orange curve).

In the case of the $\pm\gamma=0.13\pi$ channel, where the maximal current with reversed sign appears for chosen values of other parameters, the upper band is complitely expelled from the superconducting gap while the lower band is pushed toward low energies, and exhibits a zero energy crossing at the same $\varphi$ for $\pm\gamma$ (see Fig. \ref{Fig11} (e1)), leading to a sawtooth like CPR with a pronounced second harmonic term (orange curve in Fig. \ref{Fig11} (f1)). The appearance of this term can be attributed to the vicinity to the angle $\gamma$ at which the current changes sign. This behavior is in contrasts with the CPR obtained by the F-T technique, which does not show a significant second harmonic term (compare the orange and black curves in Fig. \ref{Fig11} (f1)).  The discrepancy between these two methods originate from the combined influence of the small superconducting gap ($\approx0.2\Delta_0$) and the presence  of a zero energy crossing in the ABS spectrum. Moving away from the angle where the current changes sign, for example at $\pm\gamma=0.145\pi$, the ABS spectrum has only a low energy band, which is slightly different for $\gamma$ and $-\gamma$ channels and no longer exhibits a zero energy crossing (Fig. \ref{Fig11} (g1)). In this regime the first harmonic $\sin\varphi$ is again dominantly promoted (Fig. \ref{Fig11} (h1)), and  better agreement between the  F-T and ABS approaches is observed.

%\begin{figure}[!t]
%	\centering
%	\includegraphics[width=8.5cm]{ABS_d_slika3}
%	\caption{ABS D3}\label{slikaD3}		
%\end{figure}

When the homogeneous ferromagnet in the barrier is replaced by two ferromagnets with inhomogeneous magnetization,  with $\theta_L=\phi_L=0.3\pi$, $\theta_R=0.13\pi$ and $\phi_R=1.16\pi$, the diode efficiency reaches its maximal value, as shown in Fig. \ref{Fig4} (b). The angle resolved CPR for this junction is presented in Fig. \ref{Fig11} (a2), illustrating the diverse contributions of  individual channels to the cumulative CPR. In Fig. \ref{Fig11} (b2) and its inset, the incident angle dependence of the critical current and diode efficiency are shown, respectively. The ABS spectra for three representative incident angles $\pm\gamma=0$ (normal incidence), $0.075\pi$ (corresponding to the maximal diode efficiency), and $0.35\pi$ (corresponding to the strongest anomalous current), are given in Figs. \ref{Fig11} (c2), (e2), and (g2), while the corresponding CPRs are shown in Figs. \ref{Fig11} (d2), (f2) and (h2). These results exhibit qualitatively similar features to the case with $s$-wave superconducting electrodes (Fig. \ref{Fig9}), but with a more significant difference in energies for  $\pm\gamma$ directions of quasiparticles, contributing to asymmetric free energy with respect to $\varphi=0$ and thus to presence of both AJE and JDE. The most prominent discrepancy between the currents obtained from the F-T and ABS approaches can be seen in Fig. \ref{Fig11} (h2), due to the combined influence of a small gap ($\sim0.3\Delta_0$) and  ABS bands with a small slope (see Fig. \ref{Fig11} (g2)), leading to a small current carried by the ABS and a dominant contribution from states in the continuum part of spectrum.

\begin{figure}[!t]
	\centering
	\includegraphics[width=8.5cm]{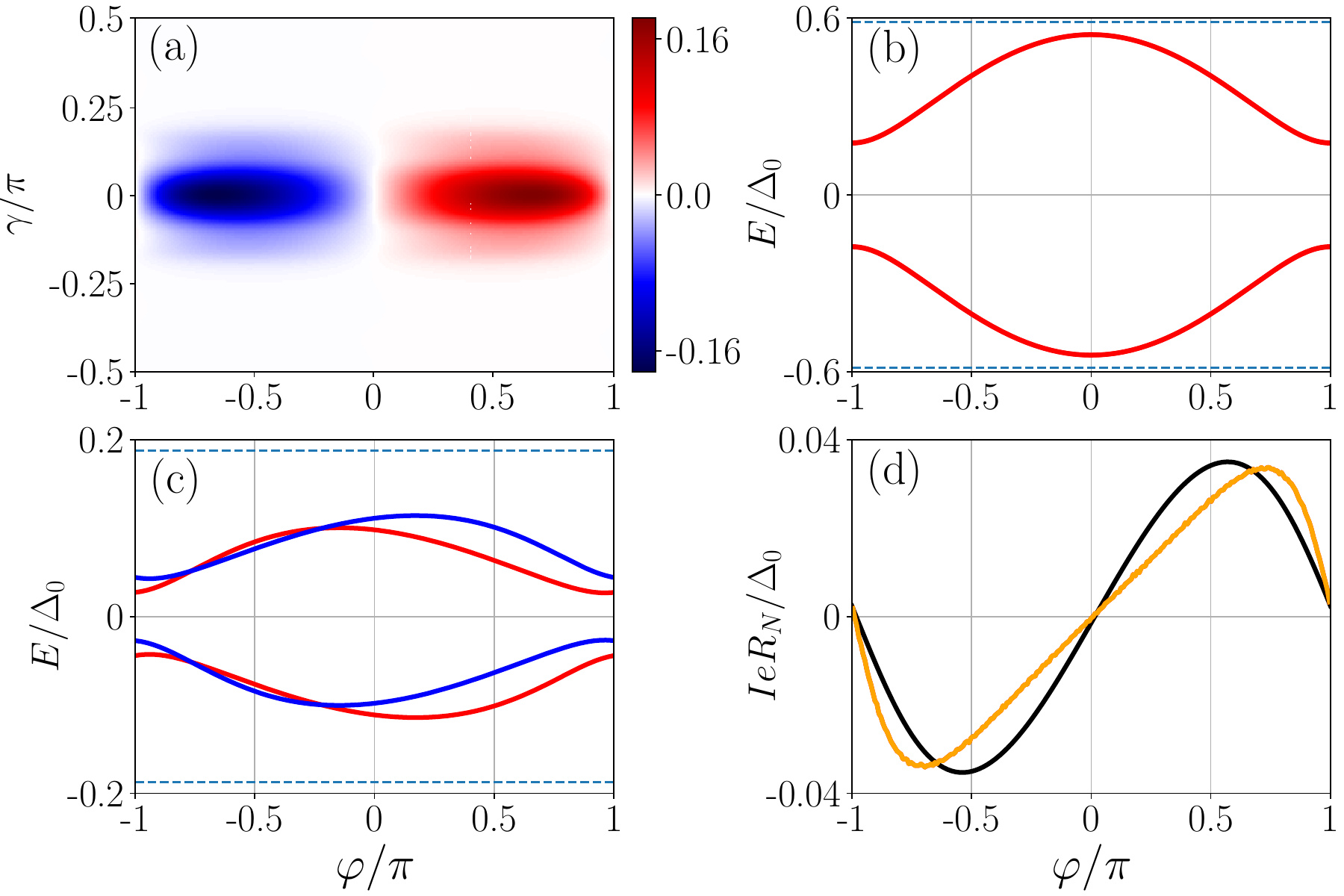}
	\caption{All panels correspond to  ($0.15\pi, 0.15\pi$) type  $d$-wave Josephson junction with an inhomogeneous ferromagnet in the barrier where the magnetization vectors in the two ferromagnetic layers are oriented oppositely ($\theta_L=\phi_L=0.3\pi$, $\theta_R=\pi-\theta_L=0.7\pi$ and $\phi_R=\pi+\phi_L=1.3\pi$). (a) Angle-resolved CPRs $I(\varphi,\gamma)$ for transverse channels with incident angle $\gamma\in\left[-\pi/2,\pi/2\right]$.  (b) and (c) ABS for $\gamma=0$ and $\gamma=\pm0.1\pi$, respectively.  The red and blue curves represent ABS bands for $\gamma$ and $-\gamma$, respectively. (d) CPR for $\gamma=0.1\pi$, where  the orange and black curves denotes the results obtained by using the ABS and F-T approaches, respectively. The remaining parameters are the same as in Fig. \ref{Fig2}. }\label{Fig12}		
\end{figure}

When the two ferromagnets in the barrier of a Josephson junction with $d$-wave electrodes are oriented oppositely, in contrast to corresponding $s$-wave junction, both AJE and JDE are present, in agreement with our symmetry analysis.  The angle-resolved CPR, shown in Fig. \ref{Fig12} (a), is similar to the $s$-wave junction case, with a monotonically decreasing critical current as a function of the incident angle $\gamma$ of quasiparticles. In all channels, the CPRs are in a $0$-like state. There is now a slight difference between $I_c^+$ and $|I_c^-|$, resulting in JDE with weak diode efficiency.   The ABS are degenerate in all channels due to the opposite effective action of spin splitting fields through the junction. In contrast to the $\gamma=0$ channel, where the ABS spectrum is symmetric with respect to $\varphi=0$ (Fig. \ref{Fig12} (b)), for $\gamma\not=0$ the phase dependent energy band becomes asymmetric, thereby enabling the appearance of both AJE and JDE (Fig. \ref{Fig12} (c)). As before, for a narrowed superconducting gap, there is noticable difference between the F-T and ABS methods (see Fig\ref{Fig11}(d)). The CPR obtained from the ABS (orange curve in Fig. \ref{Fig12} (d)) shows a diode efficiency $Q=0.0054$, while the Furusaki-Tsukada CPR (black curve in Fig. \ref{Fig12} (d)) shows $Q=0.0042$.

%\begin{figure*}[!t]
%	\centering
%	\begin{subfigure}{0.45\textwidth}
%		\centering
%		\includegraphics[width=8cm]{D4_slika1}
%		%\caption{\label{fig:image1}}
%	\end{subfigure}
%	\quad
%	\begin{subfigure}{0.45\textwidth}
%		\centering
%		\includegraphics[width=8cm]{D4_slika2}
%		%\caption{\label{fig:image2}}
%	\end{subfigure}
%\caption{Fig13}\label{Fig13}
%\end{figure*}

\begin{figure}[!t]
	\centering
	\includegraphics[width=8.5cm]{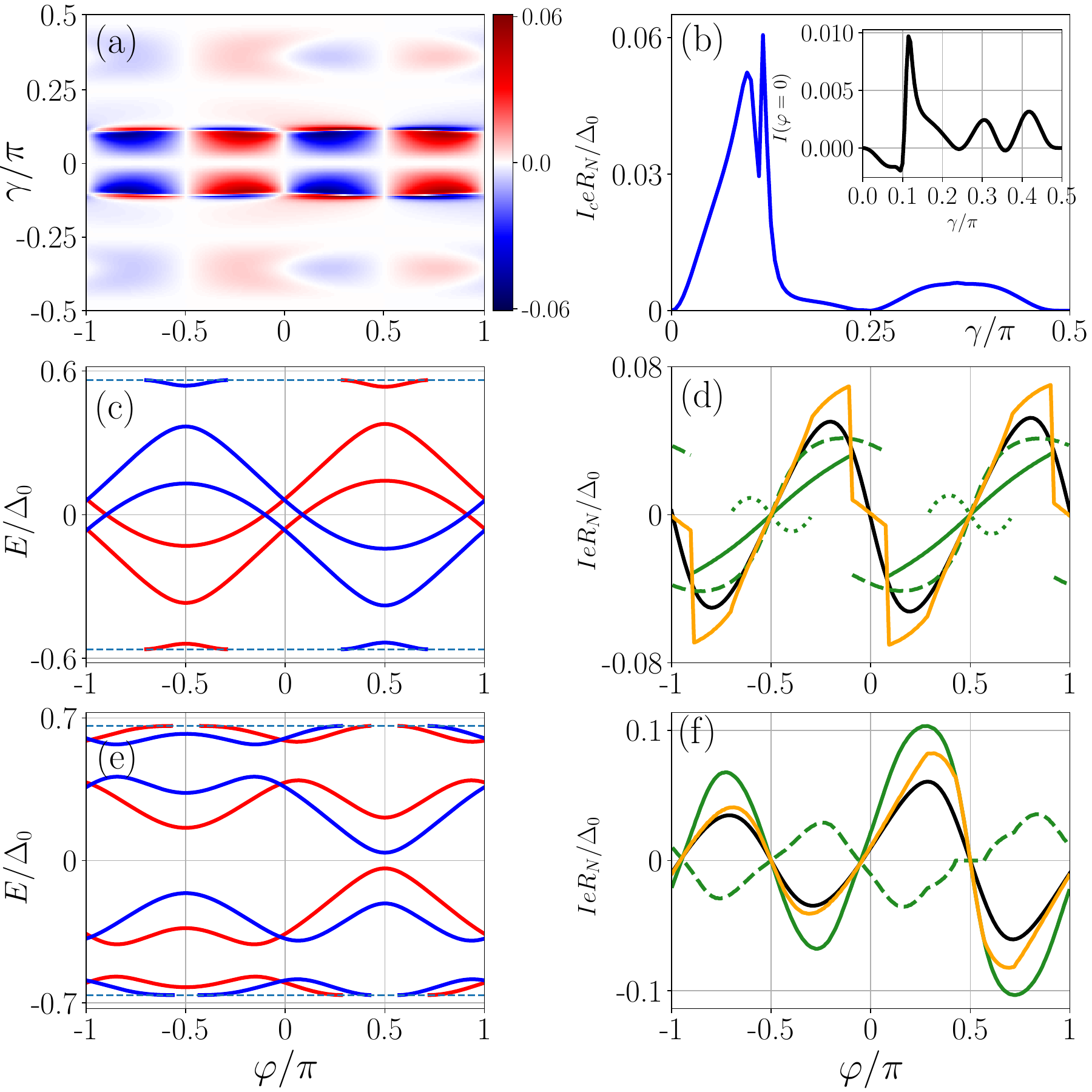}
	\caption{All panels correspond to ($0, 0.25\pi$) type  $d$-wave Josephson junction with homogeneous ferromagnet in the barrier, characterized by $\theta_L=\theta_R=0.3\pi$ and $\phi_L=\phi_R=0.3\pi$. (a) Angle-resolved CPRs $I(\varphi,\gamma)$ for transverse channels with incident angle $\gamma\in\left[-\pi/2,\pi/2\right]$. (b) Josephson critical current as a function of incident angle $\gamma$. In the inset anomalous current as a function of $\gamma$ is shown.   ABS  and the corresponding CPR are shown for: (c) and (d) $\gamma=\pm0.095\pi$; (e) and (f) $\gamma=\pm0.115\pi$, respectively. The red and blue curves represent ABS bands for $\gamma$ and $-\gamma$, respectively. The green solid, dashed and dotted curves correspond to currents from the first, second and third bands, respectively. The orange and black curves represent the total CPRs for the given channels, obtained from ABS and F-T approaches, respectively. The remaining parameters are the same as in Fig. \ref{Fig2}.}\label{Fig13}		
\end{figure}

We further consider a special case of a Josephson junction between $d_{x^2-y^2}$ and $d_{xy}$ superconducting electrodes, i.e., a $(\beta_L,\beta_R)=(0,\pi/4)$ type of junction, where new features emerge. The incident angle-resolved CPR for homogeneous magnetization in the barrier, characterized by $\theta_L=\theta_R=0.3\pi$ and $\phi_L=\phi_R=0.3\pi$, shown in Fig. \ref{Fig13} (a), displays a nonzero anomalous Josephson current in single channels, while the current vanishes at phase differences $\varphi=\pm\pi/2$. This behavior originate from symmetry that leads to $I(\varphi)=-I(-\varphi+\pi)$, as  discussed previously. The CPR in channels $\gamma=0$ and $\pm\gamma=0.25\pi$ is zero because, these directions coincide with nodal lines of the superconducting order parameters in the right and left superconductors, respectively. The critical currents in opposite directions, $I_c^+$ and $|I_c^-|$, are equal (see Fig. \ref{Fig13} (b)), resulting in the absence of JDE. The ABS spectrum for channels, $\pm\gamma=0.095\pi$ (corresponding to the first peak in critical current) and $\pm\gamma=0.115\pi$ (corresponding to the second peak in critical current on  one side and the strongest anomalous Josephson current on the other, as seen in Fig. \ref{Fig13} (b) and inset), is shown in Figs. \ref{Fig13} (c) and (e). The phase-dependent energy exhibit vanishing derivatives, for all bands, with respect to $\varphi$ at $\varphi=\pm\pi/2$, corresponding to zero current at those phase differences (see Fig. \ref{Fig13} (d) and (f)), indicating the presence of only even harmonics in the sine part and odd harmonics in the cosine part in the Fourier spectrum of the CPR. Consequently, a small but finite anomalous current is present, whereas JDE is absent. In Fig. \ref{Fig13} (c), for $\pm\gamma=0.095\pi$,  three bands at positive subgap energies appear for phase differences near $\varphi=\pm\pi/2$. The lower bands exhibit zero energy crossings positioned symmetrically around $\varphi=\pm\pi/2$, leading to a sawtooth like corresponding CPR shown by the orange curve in Fig. \ref{Fig13} (d). For the $\pm\gamma=0.115\pi$ channel, there are two spin bands in positive energies. The upper band is pushed toward the superconducting gap edge, contributing weakly to the CPR, while the lower band does not show zero energy crossing (see Fig. \ref{Fig13} (e)). In this regime, it is evident that the main contribution to the Josephson current comes from ABS, whereas the contribution from continuum  states becomes comparatively small (see Fig. \ref{Fig13} (f)). This behavior can be attributed to  the relatively wide superconducting gap for the presented channel and the absence of zero energy crossings in the bands. It is also obvious  that the CPR for $\pm\gamma=0.115\pi$ has the opposite sign in the sinusoidal part of the Fourier spectrum compared to that for $\pm\gamma=0.095\pi$. This behavior may be interpreted as Josephson ground state phase difference crossover with respect to the transverse momentum, manifested by the dip separating the corresponding peaks in Fig. \ref{Fig13} (b).

\begin{figure}[!t]
	\centering
	\includegraphics[width=8.5cm]{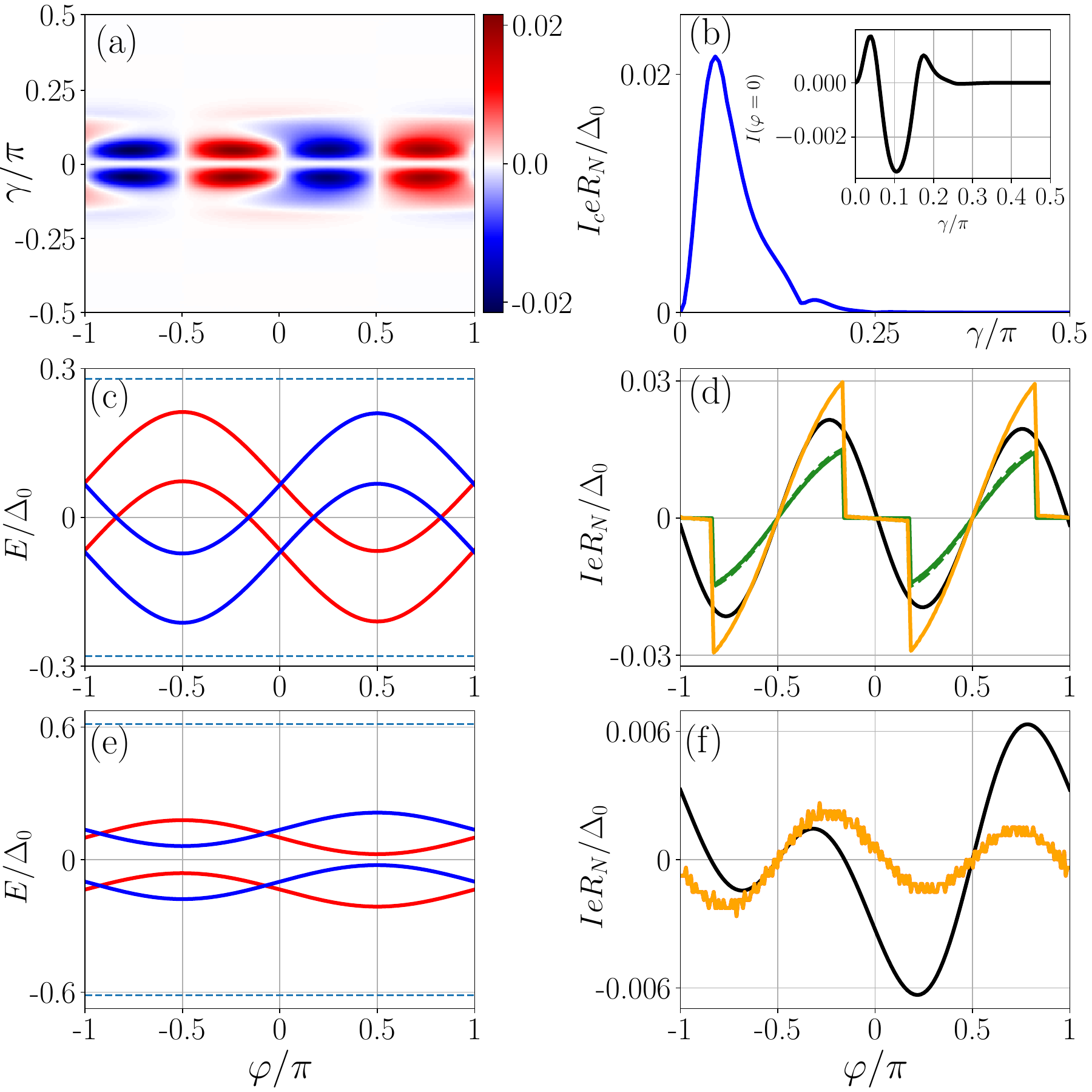}
	\caption{All panels correspond to ($0, 0.25\pi$) type  $d$-wave Josephson junction with an inhomogeneous ferromagnet in the barrier, where the magnetization vectors in the two ferromagnetic layers are oriented oppositely ($\theta_L=\phi_L=0.3\pi$, $\theta_R=\pi-\theta_L=0.7\pi$ and $\phi_R=\pi+\phi_L=1.3\pi$).  (a) Angle-resolved CPRs $I(\varphi,\gamma)$ for transverse channels with incident angle $\gamma\in\left[-\pi/2,\pi/2\right]$. (b) Josephson critical current as a function of incident angle $\gamma$. In the inset anomalous current as a function of $\gamma$ is shown.   ABS  and the corresponding CPR are shown for: (c) and (d) $\gamma=\pm0.045\pi$; (e) and (f) $\gamma=\pm0.105\pi$, respectively. The red and blue curves represent ABS bands for $\gamma$ and $-\gamma$, respectively. The solid and dashed green curves correspond to currents from lower and upper bands, respectively. The orange and black curves represent the total CPRs for the given channel obtained from ABS and F-T approaches, respectively. The remaining parameters are the same as in Fig. \ref{Fig2}.}\label{Fig14}		
\end{figure}

When the magnetizations of the two ferromagnets are oppositely oriented in the $(0, \pi/4)$ type of Josephson junction, the anomalous Josephson current remains finite but relatively weak, while the critical currents in the opposite direction are equal in magnitude. This is shown in Figs. \ref{Fig14} (a) and (b), where the angle-resolved CPR and the critical current as a function of the incident angle $\gamma$ are presented. The phase dependent ABS spectra and corresponding CPR for the channels $\pm\gamma=0.045\pi$ (corresponding to maximal critical current) and $\pm\gamma=0.105\pi$ (corresponding to maximal anomalous current), are shown in Figs. \ref{Fig14} (c)-(f).  The ABS energies remains spin degenerate and exhibit properties similar to those previously discussed: vanishing slope with respect to the phase difference for $\varphi=\pm\pi/2$. The CPR for the $\pm\gamma=0.045\pi$ channels exhibits a sawtooth profile due to the zero energy crossing in the ABS spectra around $\varphi=\pm \pi/2$, leading to disagreement between the CPR obtained by the F-T and ABS approaches. For the $\pm\gamma=0.105\pi$ channels, the superconducting gap becomes wider ($\sim 0.6\Delta_0$), the ABS energies do not exhibit zero energy crossing. In this regime, the contributions of bands for opposite transverse channels are almost opposite, resulting in a small Josephson current carried by the ABS. The contribution of the continuum part of the spectra is significant, leading to disagreement between the F-T and ABS approaches and an enhanced anomalous current in this channel.

\begin{figure}[!t]
	\centering
	\includegraphics[width=8.5cm]{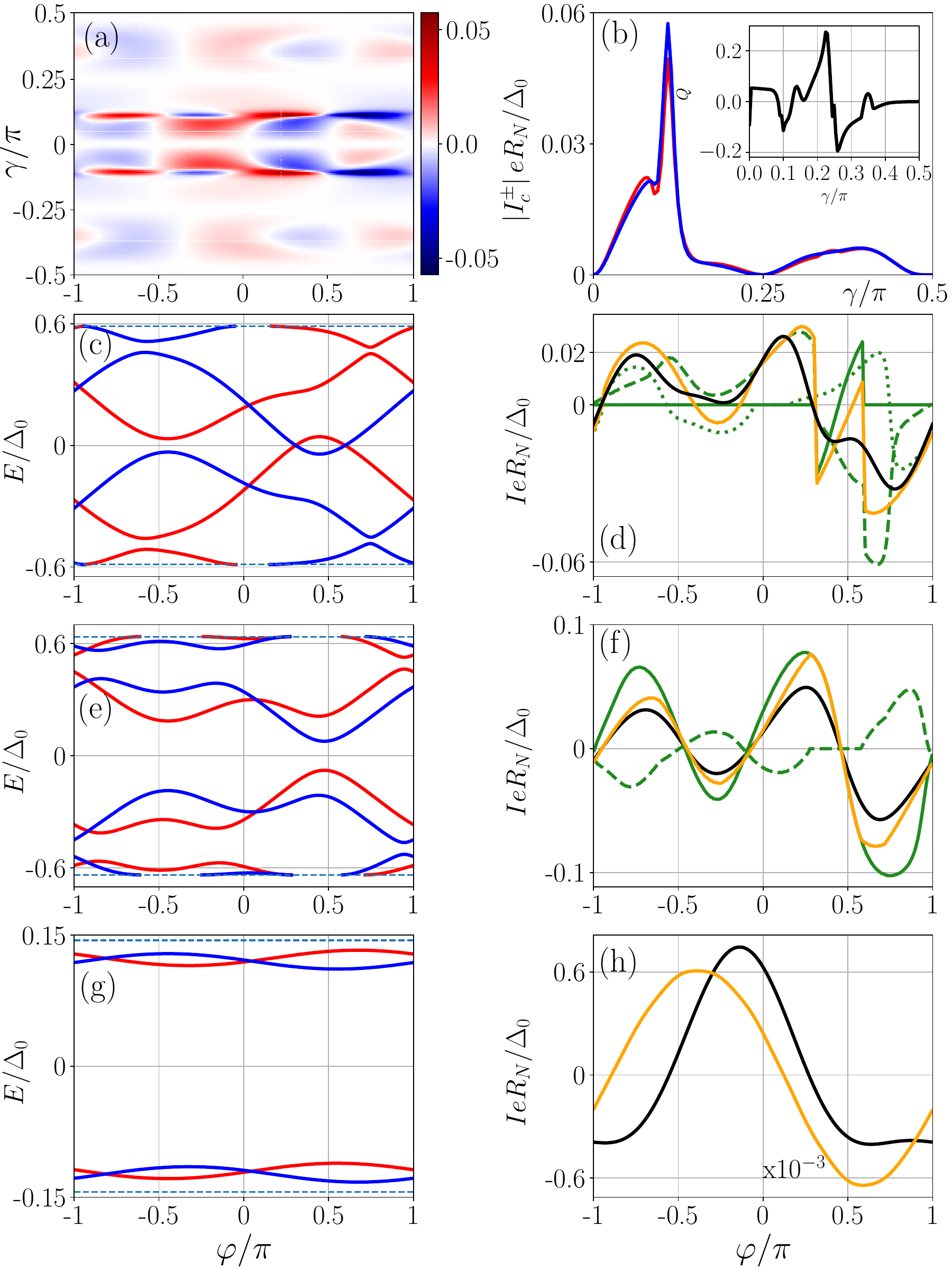}
	\caption{All panels correspond to ($0, 0.25\pi$) type  $d$-wave Josephson junction containing two ferromagnets with noncolinear magnetizations, whit $\theta_L=0.3\pi$, $\phi_L=0.3\pi$, $\theta_R=0.27\pi$ and $\phi_R=0.51\pi$.  (a) Angle-resolved CPRs $I(\varphi,\gamma)$ for transverse channels with incident angle $\gamma\in\left[-\pi/2,\pi/2\right]$. (b) Josephson critical current as a function of incident angle $\gamma$. In the inset anomalous current as a function of $\gamma$ is shown.   ABS  and the corresponding CPR are shown for: (c) and (d) $\gamma=\pm0.1\pi$; (e) and (f) $\gamma=\pm0.11\pi$; (g) and (h) $\gamma=\pm0.227\pi$ respectively. The red and blue curves represent ABS bands for $\gamma$ and $-\gamma$, respectively. The solid, dashed and dotted green curves correspond to currents from lower and upper bands, respectively.
 The orange and black curves represent the total CPRs for the given channel obtained from ABS and F-T approaches, respectively. The remaining parameters are the same as in Fig. \ref{Fig2}.}\label{Fig15}		
\end{figure}

The diode efficiency shown in Fig. \ref{Fig7} (b) reaching its maximum for magnetization configuration $\theta_L=\phi_L=0.3\pi$, $\theta_R=0.27\pi$ and $\phi_R=0.51\pi$. The angle-resolved CPR for this case is presented in Fig. \ref{Fig15} (a), showing that the current is finite at $\varphi=\pm\pi/2$ due to the  symmetry breaking mentioned above. The critical currents in both directions, $I_c^+$ and $|I_c^-|$, as functions of $\gamma$ are shown in Fig. \ref{Fig15} (b), along with the diode efficiency in the inset. The largest anomalous current appears in the $\pm\gamma=0.1\pi$ channel and the ABS spectra for this channel are shown in Fig. \ref{Fig15} (c). The phase-dependent energy bands have finite slope at $\varphi=\pm\pi/2$. For certain phase differences, three positive energy bands for $\gamma=0.1\pi$ appear between the zero energy crossings  (red curves in Fig. \ref{Fig15} (c)). At the same time, no subgap ABS bands exist for $\gamma=-0.1\pi$  at the same phase (blue curves in Fig. \ref{Fig15} (c)). The pronounced asymmetry between the spectra for opposite transverse momenta explains the emergence of both AJE and JDE. Individual band contributions to the CPR, as well as the CPR obtained from ABS and F-T approaches, are shown in Fig. \ref{Fig15} (d). The disagreement between them arises from  the zero energy crossing and from the fact that the upper band close to the gap edge is expelled for certain values of $\varphi$ without contributing to the CPR. For the channel $\pm\gamma=0.11\pi$ (corresponding to maximal current), the ABS spectra are shown in Fig. \ref{Fig15} (e). Bands are pushed toward higher energies, and there are no zero energy crossings, leading to better agreement in CPR obtained by F-T and ABS approaches, indicating that transport through subgap ABSs is significant (see Fig. \ref{Fig15} (f)).

The diode efficiency is maximal for $\gamma=0.227\pi$. The corresponding ABS spectra and CPR are shown in Figs. \ref{Fig15} (g) and (h). The superconducting gap is very small, and only one band remains below the gap edge strongly pushed toward the continuum, which explains the  small contribution of ABS to the total Josephson current. The CPR obtained from F-T technique is three order of magnitude larger than the one obtained from the  ABS approach. A large discrepancy appears in the diode efficiency, F-T approach yields $Q=0.297$, while the ABS approuch gives $Q=0.029$ (10 times smaller).

\section{Conclusion} \label{sec6}

We studied the emergence of the anomalous and diode Josephson effect in planar 2D Josephson junctions containing two ferromagnetic layers with arbitrarily oriented magnetization vectors and Rashba SOC induced at the interfaces between superconducting and ferromagnetic materials. The superconductors were considered to have either $s$-wave or arbitrarily oriented $d$-wave lobes of the order parameter. Through a systematic symmetry analysis of the junction Hamiltonian we identified the conditions  for appearance of AJE and JDE  and classified the junctions into three classes. In junction with $s$-wave or $\beta_L=-\beta_R$ oriented $d$-wave superconductors, appearance of both AJE and JDE require a noncoplanar orientation of spin-splitting directions in the ferromagnets and SOC interfaces with spin quantization along the $z$-axis, together with the requirement of  unequal and opposite $z$-components of the exchange fields. For $\beta_L=\beta_R\neq 0$  $d$-wave superconductors, both AJE and JDE are forbidden when the  polar orientations are equal and  the azimuthal orientations of the magnetizations in the ferromagnets are equal or opposite. In all remaining orientations of $d$-wave electrodes ($\beta_L\neq \pm\beta_R$), both AJE and JDE emerge whenever at least one ferromagnetic layer has a nonzero component of magnetization perpendicular to the junction plane. We confirmed all symmetry derived conditions  numerically, using the Furusaki-Tsukada technique,  generalized to spin-active materials and anisotropic superconductors, for calculating the Josephson current.

For particular case of a Josephson junction between $d_{x^2-y^2}$ and $d_{xy}$ oriented superconductors, calculations of diode efficiency revealed that, for coplanar orientation of spin splitting directions in the junction, JDE is prohibited due to additional symmetries specific to this type of junction. In the absence of  JDE, the CPR can be decomposed into a Fourier series containing $\sin 2n\varphi$ and $\cos(2n-1)\varphi$ terms, explaining both the degeneracy of the ground state phase differences and the  coexistence of $0$-like and $\pi$-like states around $\varphi=\pm\pi/2$. We found that AJE vanishes at the phase transition between these states, while the change in  the sign of the anomalous Jospehson current can be used as an indicator of the junction phase transition.

%(We also discussed a specific case of a Josephson junction between $d_{x^2-y^2}$ and $d_{xy}$ oriented superconductors through a ferromagnetic bilayer in the presence of SOC. By numerically calculating the anomalous Josephson current, we confirmed that it belongs to the third class. However, calculations of diode efficiency show that, in the case of coplanar orientation between spin splitting directions in the junction, JDE is forbidden due to additional symmetries in this type of junction. Here, in the absence of  JDE, the CPR can be decomposed into a Fourier series with $\sin 2n\varphi$ and $\cos(2n-1)\varphi$ terms, which explains the degenerate ground state phase differences with coexistence of $0$-like and $\pi$-like states around $\varphi=\pm\pi/2$. We found that at phase crossovers, the AJE is absent; at the points of phase transition, the anomalous Josephson current changes sign. On the other hand, when JDE is present, all sine and cosine terms appear in the Fourier decomposition of the CPR.)

In addition, we analyzed how the orientations of spin splitting fields in two ferromagnets affect the phase-dependent energy spectrum of the ABSs and their signatures in nonreciprocal transport, together with the contributions of individual transverse channels to the total CPR. The ABS spectrum becomes degenerate for oppositely oriented magnetizations in the two ferromagnetic layers, since Rashba SOC spin splitting at two S/F interfaces acts in opposite directions. We demonstrated that the difference between the bands in the ABS spectrum for conjugate transverse channels is more pronounced in a junction with $d$-wave superconductors due to the anisotropy of the superconducting order parameter and possible intersections between phase-dependent energy bands for the two directions ($-\gamma$ and $\gamma$). By contrast, ABS phase dependent bands in a junction with $s$-wave superconductors, for conjugate quasiparticle directions follow each other and differ only by small energy shift. The asymmetric phase-dependent ABS spectra  around $\varphi=0$ indicates presence of anomalous phase shift and unequal critical current in opposite directions. By analyzing angle-resolved transverse CPRs, we found that whenever JDE emerges, different transverse channels can be dominated by $\sin\varphi$, $\sin2\varphi$ or $\cos\varphi$ harmonics. The coexistence of these components constitutes a necessary condition for the emergence of the JDE \cite{Tanaka2022,Nas2025}.
  %while in junctions with $d$-wave electrodes, the asymmetry around $\varphi=0$ of the free energy, which accounts for the contribution of opposite incidence angles, indicates the presence of AJE and JDE.

The conditions under which ABSs dominate charge transport were identified by comparing the Furusaki-Tsukada technique and ABS approaches. In most cases, a minor contribution from continuum states leads to good agreement between the CPRs obtained using these two methods. Noticeable differences occur primarily in the presence of zero energy crossings in the ABS spectrum, where the ABS approach produces sawtooth like CPRs, whereas the corresponding F-T profiles remain smooth. In junctions with $d$-wave superconductors, differences mostly arise due to the narrower superconducting gap, which may become nearly closed, enhancing the role of the continuum part of the spectrum in charge transport.

Our results establish the symmetry based conditions and microscopic mechanism for realizing nonreciprocal charge transport in spin-active Josephson junctions, and show that significant diode efficiency, higher than 40\%, can be obtained by tuning the orientation of the superconducting electrodes and the directions of the exchange fields in the ferromagnetic layers.

\begin{acknowledgments}
Numerical calculations were performed using the computational resources of The Laboratory for Theoretical Condensed Matter Physics at University of Montenegro. The work of S. Dj. and Z. P. was supported by Ministry of Education, Science and Innovation of Montenegro through the Project SUPERMAG2D No. 0402-082/23-1121/6. The work of Z. P. was supported by Serbian Ministry of Science, Technological Development and Innovation through Project No. 451-03-34/2026-03/200162. Z. P. also acknowledges support from the COST Action SUPERQUMAP, CA21144, as well as the hospitality of the University of Montenegro during her visit.
\end{acknowledgments}

\appendix
\onecolumngrid
\section{Formalism}\label{apendix1}

In this appendix, we present some details for calculating the CPR based on the Furusaki-Tsukada approach. The calculation begins by solving the eigenproblem of the BdG Hamiltonian from Eq. (\ref{HBdG}) in each part of the considered Josephson junction, using the translational invariance of the junction under translations along the $y$-axis, so that the wave function can be written as $\Psi(\mathbf{r})=e^{ik_yy}\psi(x,\gamma)$. The longitudinal ($x$) components for the elementary propagating states in the superconducting regions are given, in the same way as in Ref. \cite{Nas2025}, by

 %By solving eigenproblem of the BdG Hamiltonian we obtain elementary propagating states in each part of the Josephson junction. Since the junction is invariant under translations along $y$-axis, transversal momentum $k_y=k_F\sin \gamma$ is conserved, and  the solutions of BdG equation are given by $\Psi(\mathbf{r})=e^{ik_yy}\psi(x,\gamma)$. Longitudinal components in superconducting regions are given by
\begin{equation}
	\begin{aligned}
		\psi_{\alpha j}^{\uparrow}(x,\gamma)&=\left(u_{j}e^{i\varphi_j/2}, 0, 0, \frac{\Delta_j^{*}}{|\Delta_{j}|}v_je^{-i\varphi_j/2}\right)^Te^{ik_j^{+}x},\hspace{1cm}\psi_{\delta j}^{\uparrow}(x,\gamma)=\left(\tilde{u}_{j}e^{i\varphi_j/2}, 0, 0, \frac{\tilde{\Delta}_j^{*}}{|\tilde{\Delta}_{j}|}\tilde{v}_je^{-i\varphi_j/2}\right)^Te^{-i\tilde{k}_j^{+}x},\\
		\psi_{\alpha j}^{\downarrow}(x,\gamma)&=\left(0, u_{j}e^{i\varphi_j/2}, \frac{\Delta_j^{*}}{|\Delta_{j}|}v_je^{-i\varphi_j/2}, 0\right)^Te^{ik_j^{+}x},\hspace{1cm} \psi_{\delta j}^{\downarrow}(x,\gamma)=\left(0, \tilde{u}_{j}e^{i\varphi_j/2}, \frac{\tilde{\Delta}_j^{*}}{|\tilde{\Delta}_{j}|}\tilde{v}_je^{-i\varphi_j/2}, 0\right)^Te^{-i\tilde{k}_j^{+}x},\\
		\psi_{\bar{\alpha} j}^{\uparrow}(x,\gamma)&=\left(0, v_{j}e^{i\varphi_j/2}, \frac{\Delta_j^{*}}{|\Delta_{j}|}u_je^{-i\varphi_j/2},0\right)^Te^{ik_j^{-}x},\hspace{1cm} \psi_{\bar{\delta} j}^{\uparrow}(x,\gamma)=\left(0, \tilde{v}_{j}e^{i\varphi_j/2}, \frac{\tilde{\Delta}_j^{*}}{|\tilde{\Delta}_{j}|}\tilde{u}_je^{-i\varphi_j/2}, 0\right)^Te^{-i\tilde{k}_j^{-}x}, \\
		\psi_{\bar{\alpha} j}^{\downarrow}(x,\gamma)&=\left(v_{j}e^{i\varphi_j/2}, 0, 0, \frac{\Delta_j^{*}}{|\Delta_{j}|}u_je^{-i\varphi_j/2}\right)^Te^{ik_j^{-}x},\hspace{1cm} \psi_{\bar{\delta} j}^{\downarrow}(x,\gamma)=\left(\tilde{v}_{j}e^{i\varphi_j/2}, 0, 0, \frac{\tilde{\Delta}_j^{*}}{|\tilde{\Delta}_{j}|}\tilde{u}_je^{-i\varphi_j/2}\right)^Te^{-i\tilde{k}_j^{-}x}.
	\end{aligned}
\end{equation}
Here, the subscript $j\in\{L,R\}$ is used to denote the left and right superconductors, $\alpha(\bar\alpha)$ and $\delta(\bar\delta)$ denote quasiparticles propagating in directions defined by the angles $\gamma(-\gamma)$ and $\pi-\gamma (\gamma-\pi)$, respectively, which are related to the dependence of the d-wave pairing potential on the direction of quasiparticle motion, as indicated by Eq. (\ref{op}). The BCS coherence factors are
% to differ them by the order parameter (\ref{op}). BCS coherence factors are
\begin{equation}
	u_j[\tilde{u}_j]=\sqrt{\frac{1}{2}\left(1+\frac{\Omega_j[\tilde{\Omega}_j]}{E}\right)},\hspace{1cm}
	v_j[\tilde{v}_j]=\sqrt{\frac{1}{2}\left(1-\frac{\Omega_{j}[\tilde{\Omega}_j]}{E}\right)},
\end{equation}
and $\Omega_j=\sqrt{E^{2}-\left|\Delta_j(\gamma)\right|^{2}}$, while $\tilde{\Omega}_j=\sqrt{E^{2}-\left|\tilde{\Delta}_j(\gamma)\right|^{2}}$.
Longitudinal ($x$) components of the wave vectors in superconductors are
\begin{equation}
	k_{j}^{\pm}=\sqrt{\frac{2m}{\hbar^2}\left(\mu\pm\sqrt{E^2-|\Delta_j|^2}\right)-k_y^2},\hspace{1cm}
	\tilde{k}_{j}^{\pm}=\sqrt{\frac{2m}{\hbar^2}\left(\mu\pm\sqrt{E^2-|\tilde{\Delta}_j|^2}\right)-k_y^2}.
\end{equation}
Elementary propagating states in the ferromagnetic region are \cite{Nas2025}
\begin{equation}
	\begin{aligned}
		\psi_{\alpha[\delta]j}^{\uparrow}(x,\gamma)&=\left(\cos{\frac{\theta_j}{2}}, e^{i\phi_j}\sin{\frac{\theta_j}{2}}, 0, 0\right)^Te^{+[-]iq_{\uparrow}^{+}x},\\
		\psi_{\alpha[\delta]j}^{\downarrow}(x,\gamma)&=\left(-e^{-i\phi_j}\sin{\frac{\theta_j}{2}}, \cos{\frac{\theta_j}{2}}, 0, 0\right)^Te^{+[-]iq_{\downarrow}^{+}x},\\
		\psi_{\bar{\alpha}[\bar{\delta}]j}^{\downarrow}(x,\gamma)&=\left(0, 0, e^{i\phi_j}\sin{\frac{\theta_j}{2}}, \cos{\frac{\theta_j}{2}}\right)^Te^{+[-]iq_{\downarrow}^{-}x},\\
		\psi_{\bar{\alpha}[\bar{\delta}]j}^{\uparrow}(x,\gamma)&=\left(0, 0, \cos{\frac{\theta_j}{2}}, -e^{-i\phi_j}\sin{\frac{\theta_j}{2}}\right)^Te^{+[-]iq_{\uparrow}^{-}x}.
	\end{aligned}
\end{equation}
Here, $j\in\{F_L,F_R\}$ denotes the left and right ferromagnetic layers. The directions of the exchange fields are determined by the spherical angles $\theta_j$ and $\phi_j$. The longitudinal ($x$) components of the wave vectors are
\begin{equation}
	q_{\sigma}^{\pm}=\sqrt{\frac{2m}{\hbar^2}(\mu+\rho_{\sigma}h\pm E)-k_y^2},
\end{equation}
with $\rho_{\sigma}=+1(-1)$ for spin projection $\sigma=\uparrow(\downarrow)$.

For the derivation and numerical calculations of the Josephson current, all possible scattering processes of the incident quasiparticle need to be considered. For a given $E>0$ and $k_y$ four such processes are present: 1) an electron-like quasiparticle (ELQ) injected from the left superconductor; 2) a hole-like quasiparticle (HLQ) injected from the left superconductor; 3) ELQ injected from the right superconductor; and 4) HLQ injected from the right superconductor. Since the incident quasiparticle has two possible spin projections (spin-up or spin-down), each of four processes must be considered twice. For the first process when spin-up ELQ is incident from the left superconductor the scattering state $\psi_1^\uparrow(x,\gamma)$ is
% Each of these cases have two variants, for two spin projections of incident quasiparticle (spin-up or spin-down). The scattering states $\psi_1^\uparrow(x,\gamma)$ for incident spin-up ELQ from the left superconductor is
\begin{equation}
	\psi_1^\uparrow(x,\gamma)=\left\{
	\begin{aligned}
		\psi_{\alpha L}^{\uparrow}+a_1^{\uparrow\uparrow}\psi_{\bar{\alpha} L}^{\uparrow}+a_1^{\uparrow\downarrow}\psi_{\bar{\alpha} L}^{\downarrow}+b_1^{\uparrow\uparrow}\psi_{\beta L}^{\uparrow}+b_1^{\uparrow\downarrow}\psi_{\beta L}^{\downarrow},&\hspace{0.5cm} x<0\\
		f_1\psi_{\alpha F_L}^{\uparrow}+f_2\psi_{\alpha F_L}^{\downarrow}+f_3\psi_{\beta F_L}^{\uparrow}+f_4\psi_{\beta F_L}^{\downarrow}+f_5\psi_{\bar{\alpha} F_L}^{\downarrow}+f_6\psi_{\bar{\alpha} F_L}^{\uparrow}+f_7\psi_{\bar{\beta} F_L}^{\downarrow}+f_8\psi_{\bar{\alpha} F_L}^{\uparrow},&\hspace{0.5cm} 0<x<d/2\\
		g_1\psi_{\alpha F_R}^{\uparrow}+g_2\psi_{\alpha F_R}^{\downarrow}+g_3\psi_{\beta F_R}^{\uparrow}+g_4\psi_{\beta F_R}^{\downarrow}+g_5\psi_{\bar{\alpha} F_R}^{\downarrow}+g_6\psi_{\bar{\alpha} F_R}^{\uparrow}+g_7\psi_{\bar{\beta} F_R}^{\downarrow}+g_8\psi_{\bar{\alpha} F_R}^{\uparrow},&\hspace{0.5cm} d/2<x<d\\
		c_1^{\uparrow\uparrow}\psi_{\alpha R}^{\uparrow}+c_1^{\uparrow\downarrow}\psi_{\alpha R}^{\downarrow}+d_1^{\uparrow\uparrow}\psi_{\bar{\beta} R}^{\uparrow}+d_1^{\uparrow\downarrow}\psi_{\bar{\beta} R}^{\downarrow},&\hspace{0.5cm} x>d
	\end{aligned}\right.
\end{equation}
while for incident spin-down ELQ from the left superconductor the scattering state  $\psi_1^\downarrow(x,\gamma)$ is
\begin{equation}
	\psi_1^\downarrow(x,\gamma)=\left\{
	\begin{aligned}
		\psi_{\alpha L}^{\downarrow}+a_1^{\downarrow\uparrow}\psi_{\bar{\alpha} L}^{\uparrow}+a_1^{\downarrow\downarrow}\psi_{\bar{\alpha} L}^{\downarrow}+b_1^{\downarrow\uparrow}\psi_{\beta L}^{\uparrow}+b_1^{\downarrow\downarrow}\psi_{\beta L}^{\downarrow},&\hspace{0.5cm} x<0\\
		f'_1\psi_{\alpha F_L}^{\uparrow}+f'_2\psi_{\alpha F_L}^{\downarrow}+f'_3\psi_{\beta F_L}^{\uparrow}+f'_4\psi_{\beta F_L}^{\downarrow}+f'_5\psi_{\bar{\alpha} F_L}^{\downarrow}+f'_6\psi_{\bar{\alpha} F_L}^{\uparrow}+f'_7\psi_{\bar{\beta} F_L}^{\downarrow}+f'_8\psi_{\bar{\alpha} F_L}^{\uparrow},&\hspace{0.5cm} 0<x<d/2\\
		g'_1\psi_{\alpha F_R}^{\uparrow}+g'_2\psi_{\alpha F_R}^{\downarrow}+'\psi_{\beta F_R}^{\uparrow}+g'_4\psi_{\beta F_R}^{\downarrow}+g'_5\psi_{\bar{\alpha} F_R}^{\downarrow}+g'_6\psi_{\bar{\alpha} F_R}^{\uparrow}+g'_7\psi_{\bar{\beta} F_R}^{\downarrow}+g'_8\psi_{\bar{\alpha} F_R}^{\uparrow},&\hspace{0.5cm} d/2<x<d\\
		c_1^{\downarrow\uparrow}\psi_{\alpha R}^{\uparrow}+c_1^{\downarrow\downarrow}\psi_{\alpha R}^{\downarrow}+d_1^{\downarrow\uparrow}\psi_{\bar{\beta} R}^{\uparrow}+d_1^{\downarrow\downarrow}\psi_{\bar{\beta} R}^{\downarrow},&\hspace{0.5cm} x>d
	\end{aligned}\right.
\end{equation}
The other scattering states can be obtained analogously.

The scattering coefficients $a_1^{\uparrow\uparrow}$ and $a_1^{\uparrow\downarrow}$ ($a_1^{\downarrow\downarrow}$ and $a_1^{\downarrow\uparrow}$) represent the probability amplitudes for Andreev reflection of an incoming spin-up (spin-down) ELQ at the left interface without and with spin flipping, respectively. Similarly, $b_1^{\uparrow\uparrow}$ and $b_1^{\uparrow\downarrow}$ ($b_1^{\downarrow\downarrow}$ and $b_1^{\downarrow\uparrow}$), denote the corresponding probability amplitudes for normal reflection without and with spin flip, respectively. The corresponding probability amplitudes for normal transmission are $c_1^{\uparrow\uparrow}$ and $c_1^{\uparrow\downarrow}$ ($c_1^{\downarrow\downarrow}$ and $c_1^{\downarrow\uparrow}$), respectively, and the probability amplitudes for Andreev transmission without and with spin flip are $d_1^{\uparrow\uparrow}$ and $d_1^{\uparrow\downarrow}$ ($d_1^{\downarrow\downarrow}$ and $d_1^{\downarrow\uparrow}$) \cite{Nas2025}. The coefficients $f_i$ ($f'_i$) and $g_i$ ($g'_i$) ($i=1,...,8$) represent possible propagating states in the left and right ferromagnetic layers, respectively.

For each process there are 24 unknown scattering coefficients, which are determined by the interfacial boundary conditions at $x=0$, $d/2$ and $d$, and can be written as
\begin{equation}
	\begin{aligned}
		\psi_{S_L}& (0) =\psi_{F_L} (0), \\
		\psi_{F_L}^\prime& (0)-\psi_{S_L}^\prime(0)=\frac{2m}{\hbar^2}(W_L\sigma_0 \tau_0-\hbar v_FR_L k_y\sigma_z \tau_z)\psi (0),  \\
		\psi_{F_L}&(d/2)=\psi_{F_R}(d/2),  \\
		\psi_{F_R}^\prime& (d/2)-\psi_{F_L}^\prime
		(d/2)=0,  \\
		\psi_{F_R}&(d)=\psi_{S_R} (d),  \\
		\psi_{S_R}^\prime&(d)-\psi_{F_R}^\prime
		(d)=\frac{2m}{\hbar^2}(W_R\sigma_0 \tau_0-\hbar v_FR_R k_y\sigma_z \tau_z)\psi (d),
	\end{aligned}\label{granicni}
\end{equation}
where $\tau_0$ is the identity matrix in particle-hole space.

The Josephson CPRs are then computed from the coefficients of Andreev reflection only, according to the Green's function based Furusaki-Tsukada formula, Eq. (\ref{struja}), which is extended to the case of an anisotropic superconductor and the presence of Rashba SOC.

\section{Symmetry transformations}\label{appendix2}

\setcounter{table}{0}
\renewcommand{\thetable}{B\arabic{table}}
For Josephson junctions with $s$-wave or $(\beta_L,\beta_R)=(0,0)$ oriented $d$-wave superconducting electrodes, and in the absence of exchange and SOC fields, the single particle Hamiltonian is given by
\begin{equation}
	H=\left(\frac{\hbar^2 k^2}{2m}-\mu\right)\tau_z+\hat{\Delta}\Theta (-x)\Theta (x-d),
\end{equation}
where there are 16 symmetry transformations $UH(\varphi)U^\dagger=H(-\varphi)$. These transformations are listed in Table \ref{table2}.

When additional ingredients are introduced into the junctions, such as a ferromagnetic exchange field with arbitrary orientation in two layers, interfacial Rashba SOC at the S/F interfaces, and arbitrarily oriented $d$-wave superconducting electrodes, symmetry can be broken. This leads to the appearance of a nonzero Josephson current at zero phase difference between the superconducting condensates. Table \ref{table2} shows how these transformations affect the ingredients of the junction Hamiltonian. The analysis of these transformations and their consequences is provided in the main text. Additional transformations used in Sections \ref{sec3} and \ref{sec4} are listed in Table \ref{table3}.
\begin{table}[!h]
	\centering
	\caption{Table of transformations $UHU^\dagger$ and how the variables on which the Hamiltonian depends change under those transformations.}
	\centering
	\begin{tabular}{|l||l|l|l|l|l|l|l|l|}
		\hline
		$U$ & $\varphi$ & $\beta_L$ & $\beta_R$ & $h\cos\theta_L$ & $h\sin\theta_L e^{i\phi_L}$ & $h\cos\theta_R$ & $h\sin\theta_R e^{i\phi_R}$ & $R$ \\ \hline\hline
		$T$ & $-\varphi$ & $\beta_L$ & $\beta_R$ & $h\cos(\pi-\theta_L)$ & $h\sin\theta_L e^{i(\pi+\phi_L)}$ & $h\cos(\pi-\theta_R)$ & $h\sin\theta_R e^{i(\pi+\phi_R)}$ & $R$ \\ \hline
		$P_x$ & $-\varphi$ & $-\beta_R$ & $-\beta_L$ & $h\cos\theta_R$ & $h\sin\theta_R e^{i\phi_R}$ & $h\cos\theta_L$ & $h\sin\theta_L e^{i\phi_L}$ & $-R$ \\ \hline
		
		$\sigma_x T$ & $-\varphi$ & $\beta_L$ & $\beta_R$ & $h\cos\theta_L$ & $h\sin\theta_L e^{i(\pi-\phi_L)}$ & $h\cos\theta_R$ & $h\sin\theta_R e^{i(\pi-\phi_R)}$ & $-R$ \\ \hline
		$\sigma_y T$ & $-\varphi$ & $\beta_L$ & $\beta_R$ & $h\cos\theta_L$ & $h\sin\theta_L e^{-i\phi_L}$ & $h\cos\theta_R$ & $h\sin\theta_R e^{-i\phi_R}$ & $-R$ \\ \hline
		$\sigma_z T$ & $-\varphi$ & $\beta_L$ & $\beta_R$ & $h\cos(\pi-\theta_L)$ & $h\sin\theta_L e^{i\phi_L}$ & $h\cos(\pi-\theta_R)$ & $h\sin\theta_R e^{i\phi_R}$ & $R$ \\ \hline
		
		$P_y T$ & $-\varphi$ & $-\beta_L$ & $-\beta_R$ & $h\cos(\pi-\theta_L)$ & $h\sin\theta_L e^{i(\pi+\phi_L)}$ & $h\cos(\pi-\theta_R)$ & $h\sin\theta_R e^{i(\pi+\phi_R)}$ & $-R$ \\ \hline
		$P_x \sigma_x$ & $-\varphi$ & $-\beta_R$ & $-\beta_L$ & $h\cos(\pi-\theta_R)$ & $h\sin\theta_R e^{-i\phi_R}$ & $h\cos(\pi-\theta_L)$ & $h\sin\theta_L e^{-i\phi_L}$ & $R$ \\ \hline
		
		$P_x \sigma_y$ & $-\varphi$ & $-\beta_R$ & $-\beta_L$ & $h\cos(\pi-\theta_R)$ & $h\sin\theta_R e^{i(\pi-\phi_R)}$ & $h\cos(\pi-\theta_L)$ & $h\sin\theta_L e^{i(\pi-\phi_L)}$ & $R$ \\ \hline
		
		$P_x \sigma_z$ & $-\varphi$ & $-\beta_R$ & $-\beta_L$ & $h\cos\theta_R$ & $h\sin\theta_R e^{i(\pi+\phi_R)}$ & $h\cos\theta_L$ & $h\sin\theta_L e^{i(\pi+\phi_L)}$ & $-R$ \\ \hline
		
		$P_x P_y$ & $-\varphi$ & $\beta_R$ & $\beta_L$ & $h\cos\theta_R$ & $h\sin\theta_R e^{i\phi_R}$ & $h\cos\theta_L$ & $h\sin\theta_L e^{i\phi_L}$ & $R$ \\ \hline
		
		$P_y \sigma_x T$ & $-\varphi$ & $-\beta_L$ & $-\beta_R$ & $h\cos\theta_L$ & $h\sin\theta_L e^{i(\pi-\phi_L)}$ & $h\cos\theta_R$ & $h\sin\theta_R e^{i(\pi-\phi_R)}$ & $R$ \\ \hline
		
		$P_y \sigma_y T$ & $-\varphi$ & $-\beta_L$ & $-\beta_R$ & $h\cos\theta_L$ & $h\sin\theta_L e^{-i\phi_L}$ & $h\cos\theta_R$ & $h\sin\theta_R e^{-i\phi_R}$ & $R$ \\ \hline
		$P_x P_y \sigma_z$ & $-\varphi$ & $-\beta_R$ & $-\beta_L$ & $h\cos\theta_R$ & $h\sin\theta_R e^{i(\pi+\phi_R)}$ & $h\cos\theta_L$ & $h\sin\theta_L e^{i(\pi+\phi_L)}$ & $R$ \\ \hline
		$P_y \sigma_z T$ & $-\varphi$ & $-\beta_L$ & $-\beta_R$ & $h\cos(\pi-\theta_L)$ & $h\sin\theta_L e^{i\phi_L}$ & $h\cos(\pi-\theta_R)$ & $h\sin\theta_R e^{i\phi_R}$ & $-R$ \\ \hline
		$P_x P_y \sigma_x$ & $-\varphi$ & $\beta_R$ & $\beta_L$ & $h\cos(\pi-\theta_R)$ & $h\sin\theta_R e^{-i\phi_R}$ & $h\cos(\pi-\theta_L)$ & $h\sin\theta_L e^{-i\phi_L}$ & $-R$ \\ \hline
		$P_x P_y \sigma_y$ & $-\varphi$ & $\beta_R$ & $\beta_L$ & $h\cos(\pi-\theta_R)$ & $h\sin\theta_R e^{i(\pi-\phi_R)}$ & $h\cos(\pi-\theta_L)$ & $h\sin\theta_L e^{i(\pi-\phi_L)}$ & $-R$ \\ \hline
	\end{tabular}
	\label{table2}
\end{table}

\begin{table}[!h]
	\centering
	\caption{Table of additional transformations $UHU^\dagger$ and how the variables on which the Hamiltonian depends change under these transformations used in the analysis.}
	\centering
	\begin{tabular}{|l||l|l|l|l|l|l|l|l|}
		\hline
		$U$ & $\varphi$ & $\beta_L$ & $\beta_R$ & $h\cos\theta_L$ & $h\sin\theta_L e^{i\phi_L}$ & $h\cos\theta_R$ & $h\sin\theta_R e^{i\phi_R}$ & $R$ \\ \hline\hline
		$\sigma_x$ & $\varphi$ & $\beta_L$ & $\beta_R$ & $h\cos(\pi-\theta_L)$ & $h\sin\theta_L e^{-i\phi_L}$ & $h\cos(\pi-\theta_R)$ & $h\sin\theta_R e^{-i\phi_R}$ & $-R$ \\ \hline
		$\sigma_y$ & $\varphi$ & $\beta_L$ & $\beta_R$ & $h\cos(\pi-\theta_L)$ & $h\sin\theta_L e^{i(\pi-\phi_L)}$ & $h\cos(\pi-\theta_R)$ & $h\sin\theta_R e^{i(\pi-\phi_R)}$ & $-R$ \\ \hline
		$\sigma_z$ & $\varphi$ & $\beta_L$ & $\beta_R$ & $h\cos\theta_L$ & $h\sin\theta_L e^{i(\pi+\phi_L)}$ & $h\cos\theta_R$ & $h\sin\theta_R e^{i(\pi+\phi_R)}$ & $R$ \\ \hline
		$P_y$ & $\varphi$ & $-\beta_L$ & $-\beta_R$ & $h\cos\theta_L$ & $h\sin\theta_L e^{i\phi_L}$ & $h\cos\theta_R$ & $h\sin\theta_R e^{i\phi_R}$ & $-R$ \\ \hline
		$P_x T$ & $\varphi$ & $-\beta_R$ & $-\beta_L$ & $h\cos(\pi-\theta_R)$ & $h\sin\theta_R e^{i(\pi+\phi_R)}$ & $h\cos(\pi-\theta_L)$ & $h\sin\theta_L e^{i(\pi+\phi_L)}$ & $-R$ \\ \hline
		$P_y \sigma_x$ & $\varphi$ & $-\beta_L$ & $-\beta_R$ & $h\cos(\pi-\theta_L)$ & $h\sin\theta_L e^{-i\phi_L}$ & $h\cos(\pi-\theta_R)$ & $h\sin\theta_R e^{-i\phi_R}$ & $R$ \\ \hline
		$P_y \sigma_y$ & $\varphi$ & $-\beta_L$ & $-\beta_R$ & $h\cos(\pi-\theta_L)$ & $h\sin\theta_L e^{i(\pi-\phi_L)}$ & $h\cos(\pi-\theta_R)$ & $h\sin\theta_R e^{i(\pi-\phi_R)}$ & $R$ \\ \hline
		$P_y \sigma_z$ & $\varphi$ & $-\beta_L$ & $-\beta_R$ & $h\cos\theta_L$ & $h\sin\theta_L e^{i(\pi+\phi_L)}$ & $h\cos\theta_R$ & $h\sin\theta_R e^{i(\pi+\phi_R)}$ & $-R$ \\ \hline
		$P_x \sigma_x T$ & $\varphi$ & $-\beta_R$ & $-\beta_L$ & $h\cos\theta_R$ & $h\sin\theta_R e^{i(\pi-\phi_R)}$ & $h\cos\theta_L$ & $h\sin\theta_L e^{i(\pi-\phi_L)}$ & $R$ \\ \hline
		$P_x \sigma_y T$ & $\varphi$ & $-\beta_R$ & $-\beta_L$ & $h\cos\theta_R$ & $h\sin\theta_R e^{-i\phi_R}$ & $h\cos\theta_L$ & $h\sin\theta_L e^{-i\phi_L}$ & $R$ \\ \hline
	\end{tabular}
	\label{table3}
\end{table}
%\tableofcontents
\newpage
\section{Additional numerical results} \label{appendix3}
\begin{figure*}[!h]
	\centering
	\includegraphics[width=17cm]{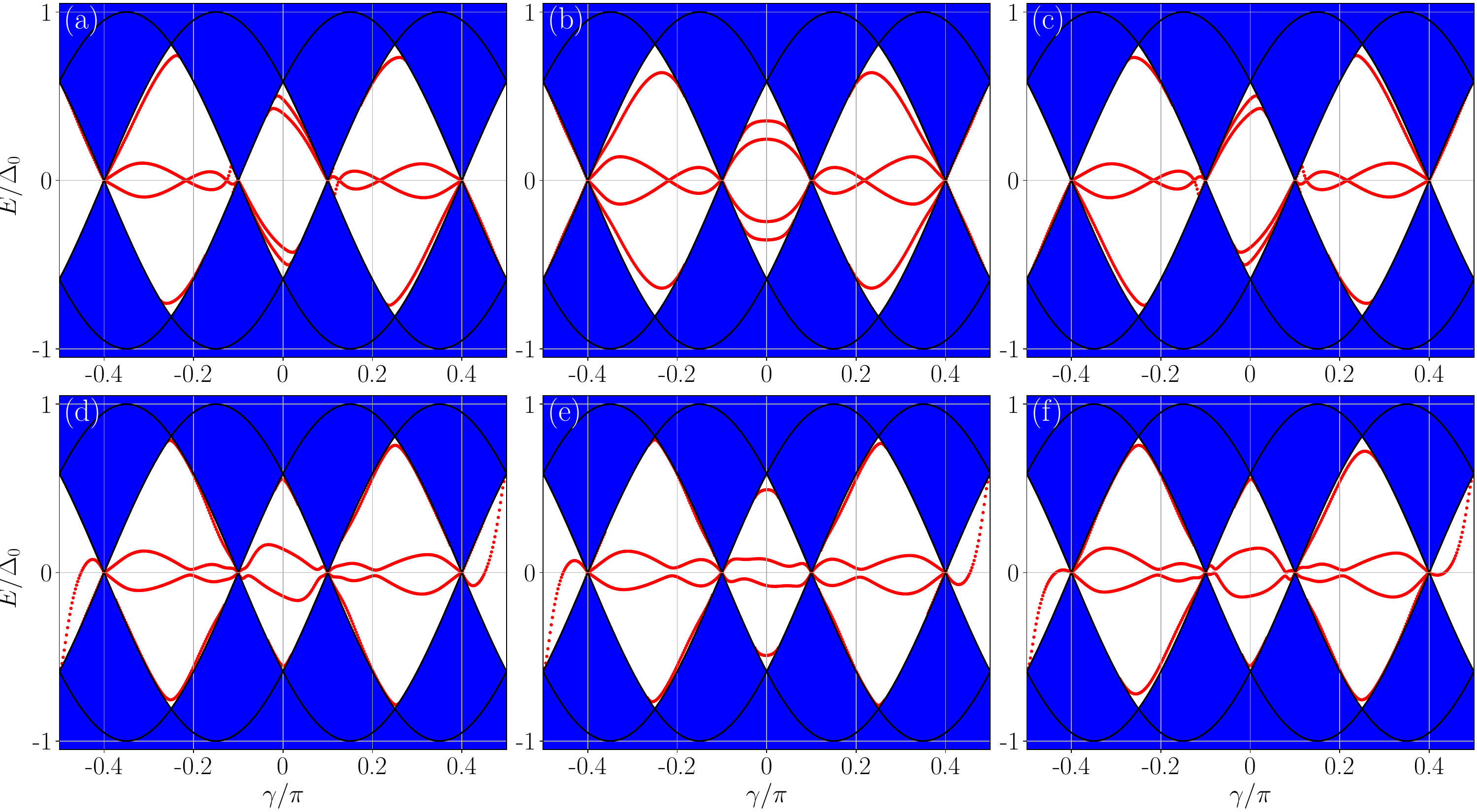}
	\caption{ABS energies as a function of incident angle $\gamma$ for a $(0.15\pi,0.15\pi)$ type $d$-wave Josephson junction for:  (a)-(c) homogeneous ferromagnetic barrier with $\theta_L=\theta_R=0.3\pi$ and $\phi_L=\phi_R=0.3\pi$, and (d)-(f) inhomogeneous ferromagnetic barrier with $\theta_L=0.3\pi$, $\phi_L=0.3\pi$, $\theta_R=0.13\pi$ and $\phi_R=1.16\pi$. Panels (a) and (d) are calculated for $\varphi=-0.5\pi$, (b) and (e) for $\varphi=0$, (c) and (f) for $\varphi=0.5\pi$. The remaining parameters are the same as in Fig. \ref{Fig2}.}\label{Fig16}		
\end{figure*}

Incidence angle dependent ABS energy bands for the  $(\beta_L,\beta_R)=(0.15\pi,0.15\pi)$ orientation of $d$-wave electrodes and  homogeneous ferromagnetic barrier, are shown in Figs. \ref{Fig16} (a)-(c) for phase differences $\varphi=-0.5\pi$, $0$ and $0.5\pi$, respectively. In this regime both AJE and JDE are not observed. The displayed bands confirm particle-hole symmetry, satisfying relation $E(\gamma, \varphi)=-E(-\gamma,\varphi)$. Since AJE is absent, the relation $E(\gamma,\varphi)=E(-\gamma,-\varphi)$ also holds, as can be verified by comparing Figs. \ref{Fig16} (a) and (c). Consequently, the relation $E(\gamma,\varphi)=-E(\gamma,-\varphi)$ holds as well. As a result of these relations, the phase dependent ABS spectra for $\gamma$ and $-\gamma$ exhibit mirror symmetry (see Fig. \ref{Fig11} (c1), (e1) and (g1) where AJE is absent), with intersections occuring at $\varphi=0$.

In Figs. \ref{Fig16} (d)-(e), $\gamma$ dependent ABS energy bands are shown for the case of an inhomogeneous ferromagnetic barrier where both AJE and JDE emerge. In this case, only the particle-hole symmetry relation remains preserved, while the others are broken. As a result, mirror symmetry in phase dependent bands for $\gamma$ and $-\gamma$ is lost, and any intersection between them becomes shifted awey from $\varphi=0$.

The white regions in the panels represent the superconducting gap as a function of quasiparticle direction. The directions where the gap closes are evident, coinciding with the nodal lines of $\Delta$ and $\tilde{\Delta}$. The presence of strongly reduced gap values results in bands being expelled from the superconducting gap. This is the main reason that, in the case of $d$-wave superconductors, the current carrying states are not dominated by ABS;  significant contributions also come from states in the continuum above the gap edge energies. Therefore, calculating the Josephson current using only ABS is insufficient, and the F-T technique provides a more complete description since it includes both ABS and states above the superconducting gap.

%\begin{figure*}[!h]
%	\centering
%	\includegraphics[width=17cm]{slika17}
%	\caption{EABS(g)0-0.25}\label{Fig17}		
%\end{figure*}
%\cite{Ji}

%\newpage

\twocolumngrid
\bibliography{reference1}

@article{Devoret2013,
	author  = {M. H. Devoret and R. J. Schoelkopf},
	title   = {Superconducting Circuits for Quantum Information: An Outlook},
	journal = {Science},
	year    = {2013},
	volume  = {339},
	number  = {6124},
	pages   = {1169--1174},
	doi     = {10.1126/science.1231930}
}

@article{Eschrig2015,
	author  = {Matthias Eschrig},
	title   = {Spin-polarized supercurrents for spintronics: a review of current progress},
	journal = {Reports on Progress in Physics},
	year    = {2015},
	volume  = {78},
	number  = {10},
	pages   = {104501},
	doi     = {10.1088/0034-4885/78/10/104501}
}

@article{Linder2015,
	author    = {Jacob Linder and Jason W. A. Robinson},
	title     = {Superconducting spintronics},
	journal   = {Nature Physics},
	year      = {2015},
	volume    = {11},
	number    = {4},
	pages     = {307--315},
	doi       = {10.1038/nphys3242},
	publisher = {Springer Nature}
}

@article{Danilo2025,
	title = {Quantum-geometric spin and charge {Josephson} diode effects},
	author = {Schulz, Niklas L. and Nikoli\ifmmode \acute{c}\else \'{c}\fi{}, Danilo and Eschrig, Matthias},
	journal = {Phys. Rev. B},
	volume = {112},
	issue = {10},
	pages = {104514},
	numpages = {6},
	year = {2025},
	month = {Sep},
	publisher = {American Physical Society},
	doi = {10.1103/nb38-v1jq},
	url = {https://link.aps.org/doi/10.1103/nb38-v1jq}
}

@article{Golod2022,
	author    = {Taras Golod and Vladimir M. Krasnov},
	title     = {Demonstration of a superconducting diode-with-memory, operational at zero magnetic field with switchable nonreciprocity},
	journal   = {Nature Communications},
	year      = {2022},
	volume    = {13},
	pages     = {3658},
	doi       = {10.1038/s41467-022-31256-w},
	publisher = {Springer Nature}
}

@article{Buzdin2005,
	author  = {A. I. Buzdin},
	title   = {Proximity effects in superconductor-ferromagnet heterostructures},
	journal = {Reviews of Modern Physics},
	year    = {2005},
	volume  = {77},
	number  = {3},
	pages   = {935--976},
	doi     = {10.1103/RevModPhys.77.935}
}

@article{Bergeret2005,
	author  = {F. S. Bergeret and A. F. Volkov and K. B. Efetov},
	title   = {Odd triplet superconductivity and related phenomena in superconductor-ferromagnet structures},
	journal = {Reviews of Modern Physics},
	year    = {2005},
	volume  = {77},
	number  = {4},
	pages   = {1321--1373},
	doi     = {10.1103/RevModPhys.77.1321}
}

@article{Josephson1962,
	author  = {B. D. Josephson},
	title   = {Possible new effects in superconductive tunnelling},
	journal = {Physics Letters},
	year    = {1962},
	volume  = {1},
	number  = {7},
	pages   = {251--253},
	doi     = {10.1016/0031-9163(62)91369-0}
}

@article{Golubov2004,
	author  = {A. A. Golubov and M. Yu. Kupriyanov and E. Il'ichev},
	title   = {The current-phase relation in {Josephson} junctions},
	journal = {Reviews of Modern Physics},
	year    = {2004},
	volume  = {76},
	pages   = {411--469},
	doi     = {10.1103/RevModPhys.76.411}
}

@article{Tanaka97,
	title = {Theory of {Josephson} effects in anisotropic superconductors},
	author = {Tanaka, Yukio and Kashiwaya, Satoshi},
	journal = {Phys. Rev. B},
	volume = {56},
	issue = {2},
	pages = {892--912},
	numpages = {0},
	year = {1997},
	month = {Jul},
	publisher = {American Physical Society},
	doi = {10.1103/PhysRevB.56.892},
	url = {https://link.aps.org/doi/10.1103/PhysRevB.56.892}
}

@article{Zutic2023,
	author = {Cai, Ranran and Žutić, Igor and Han, Wei},
	title = {Superconductor/Ferromagnet Heterostructures: A Platform for Superconducting Spintronics and Quantum Computation},
	journal = {Advanced Quantum Technologies},
	volume = {6},
	number = {1},
	pages = {2200080},
	doi = {https://doi.org/10.1002/qute.202200080},
	url = {https://advanced.onlinelibrary.wiley.com/doi/abs/10.1002/qute.202200080},
	year = {2023}
}

@article{Melnikov2022,
	author = {A. S. Mel’nikov and S. V. Mironov and A. V. Samokhvalov and A. I. Buzdin},
	title = {Superconducting spintronics: state of the art and prospects},
	publisher = {Physics-Uspekhi},
	year = {2022},
	journal = {Phys. Usp.},
	volume = {65},
	number = {12},
	pages = {1248-1289},
	url = {https://ufn.ru/en/articles/2022/12/f/},
	doi = {10.3367/UFNe.2021.07.039020}
}

@article{Eschrig2011,
	author    = {Matthias Eschrig},
	title     = {Spin-polarized supercurrents for spintronics},
	journal   = {Physics Today},
	volume    = {64},
	number    = {1},
	pages     = {43--49},
	year      = {2011},
	month     = {January},
	doi       = {10.1063/1.3541944}
}

@article{Buzdin1982,
	author       = {Buzdin, A I and Bulaevskii, L N and Panyukov, S V},
	title        = {Critical-current oscillations as a function of the exchange field and thickness of the ferromagnetic metal ({F}) in an {S}-{F}-{S} {Josephson} junction},
	url          = {https://www.osti.gov/biblio/5121669},
	journal      = {JETP Lett. (Engl. Transl.); (United States)},
	issn         = {ISSN JTPLA},
	volume       = {35:4},
	place        = {United States},
	year         = {1982},
	month        = {02}}

@article{Ryazanov2001,
	title = {Coupling of Two Superconductors through a Ferromagnet: Evidence for a $\ensuremath{\pi}$ Junction},
	author = {Ryazanov, V. V. and Oboznov, V. A. and Rusanov, A. Yu. and Veretennikov, A. V. and Golubov, A. A. and Aarts, J.},
	journal = {Phys. Rev. Lett.},
	volume = {86},
	issue = {11},
	pages = {2427--2430},
	numpages = {0},
	year = {2001},
	month = {Mar},
	publisher = {American Physical Society},
	doi = {10.1103/PhysRevLett.86.2427},
	url = {https://link.aps.org/doi/10.1103/PhysRevLett.86.2427}
}

@article{Kontos2002,
	title = {Josephson Junction through a Thin Ferromagnetic Layer: {Negative} Coupling},
	author = {Kontos, T. and Aprili, M. and Lesueur, J. and Gen\^et, F. and Stephanidis, B. and Boursier, R.},
	journal = {Phys. Rev. Lett.},
	volume = {89},
	issue = {13},
	pages = {137007},
	numpages = {4},
	year = {2002},
	month = {Sep},
	publisher = {American Physical Society},
	doi = {10.1103/PhysRevLett.89.137007},
	url = {https://link.aps.org/doi/10.1103/PhysRevLett.89.137007}
}

@article{Harlingen1995,
	title = {Phase-sensitive tests of the symmetry of the pairing state in the high-temperature superconductors---{Evidence} for ${d}_{{x}^{2}\ensuremath{-}{y}^{2}}$ symmetry},
	author = {Van Harlingen, D. J.},
	journal = {Rev. Mod. Phys.},
	volume = {67},
	issue = {2},
	pages = {515--535},
	numpages = {0},
	year = {1995},
	month = {Apr},
	publisher = {American Physical Society},
	doi = {10.1103/RevModPhys.67.515},
	url = {https://link.aps.org/doi/10.1103/RevModPhys.67.515}
}

@article{Nas21,
	author  = {Djurdjevi{\'c}, Stevan and Popovi{\'c}, Zorica},
	title   = {Influence of $d$-wave superconductor orientation on {Josephson} current and phase difference in junctions with inhomogeneous ferromagnet},
	journal = {Progress of Theoretical and Experimental Physics},
	volume  = {2021},
	number  = {8},
	pages   = {083I02},
	year    = {2021},
	doi     = {10.1093/ptep/ptab077}
}

@article{Yip1995,
	title = {Josephson current-phase relationships with unconventional superconductors},
	author = {Yip, Sungkit},
	journal = {Phys. Rev. B},
	volume = {52},
	issue = {5},
	pages = {3087--3090},
	numpages = {0},
	year = {1995},
	month = {Aug},
	publisher = {American Physical Society},
	doi = {10.1103/PhysRevB.52.3087},
	url = {https://link.aps.org/doi/10.1103/PhysRevB.52.3087}
}

@article{Sigrist1998,
	author = {Sigrist, Manfred},
	title = {Time-Reversal Symmetry Breaking States in High-Temperature Superconductors},
	journal = {Progress of Theoretical Physics},
	volume = {99},
	number = {6},
	pages = {899-929},
	year = {1998},
	month = {06},
	issn = {0033-068X},
	doi = {10.1143/PTP.99.899},
	url = {https://doi.org/10.1143/PTP.99.899}
}

@article{Kashivaya2000,
	doi = {10.1088/0034-4885/63/10/202},
	url = {https://doi.org/10.1088/0034-4885/63/10/202},
	year = {2000},
	month = {oct},
	publisher = {},
	volume = {63},
	number = {10},
	pages = {1641},
	author = {Satoshi Kashiwaya and Yukio Tanaka},
	title = {Tunnelling effects on surface bound states in
	unconventional superconductors},
	journal = {Reports on Progress in Physics}
}

@article{Yokoyama2014,
	title = {Anomalous {Josephson} effect induced by spin-orbit interaction and {Zeeman} effect in semiconductor nanowires},
	author = {Yokoyama, Tomohiro and Eto, Mikio and Nazarov, Yuli V.},
	journal = {Phys. Rev. B},
	volume = {89},
	issue = {19},
	pages = {195407},
	numpages = {14},
	year = {2014},
	month = {May},
	publisher = {American Physical Society},
	doi = {10.1103/PhysRevB.89.195407},
	url = {https://link.aps.org/doi/10.1103/PhysRevB.89.195407}
}

@article{Beenakker1991,
	title = {Universal limit of critical-current fluctuations in mesoscopic {Josephson} junctions},
	author = {Beenakker, C. W. J.},
	journal = {Phys. Rev. Lett.},
	volume = {67},
	issue = {27},
	pages = {3836--3839},
	numpages = {0},
	year = {1991},
	month = {Dec},
	publisher = {American Physical Society},
	doi = {10.1103/PhysRevLett.67.3836},
	url = {https://link.aps.org/doi/10.1103/PhysRevLett.67.3836}
}

@article{Bardeen1972,
	title = {Josephson Current Flow in Pure Superconducting-Normal-Superconducting Junctions},
	author = {Bardeen, John and Johnson, Jared L.},
	journal = {Phys. Rev. B},
	volume = {5},
	issue = {1},
	pages = {72--78},
	numpages = {0},
	year = {1972},
	month = {Jan},
	publisher = {American Physical Society},
	doi = {10.1103/PhysRevB.5.72},
	url = {https://link.aps.org/doi/10.1103/PhysRevB.5.72}
}

@article{Ando2020,
	author  = {Ando, Fumihiro and Miyasaka, Yuya and Li, Tianyi and Arakawa, Shingo and Shiota, Yoichi and Moriyama, Takahiro and Ono, Teruo and Yokoyama, Takehito and Nakamura, Yoshihisa and Saitoh, Eiji},
	title   = {Observation of superconducting diode effect},
	journal = {Nature},
	volume  = {584},
	number  = {7821},
	pages   = {373--376},
	year    = {2020},
	doi     = {10.1038/s41586-020-2590-4}
}

@article{Baumgartner2022,
	author  = {Baumgartner, Christian and Fuchs, Lorenz and Costa, Andreas and Reinhardt, Simon and Gronin, Sergei and Gardner, Geoffrey C. and Lindemann, Tyler and Manfra, Michael J. and Faria Junior, Paulo E. and Kochan, Denis and Fabian, Jaroslav and Paradiso, Nicola and Strunk, Christoph},
	title   = {Supercurrent rectification and magnetochiral effects in symmetric {Josephson} junctions},
	journal = {Nature Nanotechnology},
	volume  = {17},
	number  = {1},
	pages   = {39--44},
	year    = {2022},
	doi     = {10.1038/s41565-021-01009-9},
	url     = {https://doi.org/10.1038/s41565-021-01009-9}
}

@article{Costa2025,
	title = {Unconventional {Josephson} supercurrent diode effect induced by chiral spin-orbit coupling},
	author = {Costa, Andreas and Kanehira, Osamu and Matsueda, Hiroaki and Fabian, Jaroslav},
	journal = {Phys. Rev. B},
	volume = {111},
	issue = {14},
	pages = {L140506},
	numpages = {9},
	year = {2025},
	month = {Apr},
	publisher = {American Physical Society},
	doi = {10.1103/PhysRevB.111.L140506},
	url = {https://link.aps.org/doi/10.1103/PhysRevB.111.L140506}
}

@article{Asano2001,
	title = {Numerical method for dc {Josephson} current between $d$-wave superconductors},
	author = {Asano, Yasuhiro},
	journal = {Phys. Rev. B},
	volume = {63},
	issue = {5},
	pages = {052512},
	numpages = {4},
	year = {2001},
	month = {Jan},
	publisher = {American Physical Society},
	doi = {10.1103/PhysRevB.63.052512},
	url = {https://link.aps.org/doi/10.1103/PhysRevB.63.052512}
}

@article{Tanaka1996,
	title = {Theory of the {Josephson} effect in $d$-wave superconductors},
	author = {Tanaka, Yukio and Kashiwaya, Satoshi},
	journal = {Phys. Rev. B},
	volume = {53},
	issue = {18},
	pages = {R11957--R11960},
	numpages = {0},
	year = {1996},
	month = {May},
	publisher = {American Physical Society},
	doi = {10.1103/PhysRevB.53.R11957},
	url = {https://link.aps.org/doi/10.1103/PhysRevB.53.R11957}
}

@article{Ilichev2001,
	title = {Degenerate Ground State in a Mesoscopic $\mathrm{YBa}_{2}\mathrm{Cu}_{3}\mathrm{O}_{7\ensuremath{-}\mathit{x}}$ Grain Boundary {Josephson} Junction},
	author = {Il'ichev, E. and Grajcar, M. and Hlubina, R. and IJsselsteijn, R. P. J. and Hoenig, H. E. and Meyer, H.-G. and Golubov, A. and Amin, M. H. S. and Zagoskin, A. M. and Omelyanchouk, A. N. and Kupriyanov, M. Yu.},
	journal = {Phys. Rev. Lett.},
	volume = {86},
	issue = {23},
	pages = {5369--5372},
	numpages = {0},
	year = {2001},
	month = {Jun},
	publisher = {American Physical Society},
	doi = {10.1103/PhysRevLett.86.5369},
	url = {https://link.aps.org/doi/10.1103/PhysRevLett.86.5369}
}

@article{Testa2005,
	title = {Evidence of midgap-state-mediated transport in 45\ifmmode^\circ\else\textdegree\fi{} symmetric [001] tilt $\mathrm{Y}\mathrm{Ba}_{2}\mathrm{Cu}_{3}\mathrm{O}_{7\ensuremath{-}x}$ bicrystal grain-boundary junctions},
	author = {Testa, G. and Sarnelli, E. and Monaco, A. and Esposito, E. and Ejrnaes, M. and Kang, D.-J. and Mennema, S. H. and Tarte, E. J. and Blamire, M. G.},
	journal = {Phys. Rev. B},
	volume = {71},
	issue = {13},
	pages = {134520},
	numpages = {7},
	year = {2005},
	month = {Apr},
	publisher = {American Physical Society},
	doi = {10.1103/PhysRevB.71.134520},
	url = {https://link.aps.org/doi/10.1103/PhysRevB.71.134520}
}

@article{Buzdin2008,
	title = {Direct Coupling Between Magnetism and Superconducting Current in the {Josephson} ${\ensuremath{\varphi}}_{0}$ Junction},
	author = {Buzdin, A.},
	journal = {Phys. Rev. Lett.},
	volume = {101},
	issue = {10},
	pages = {107005},
	numpages = {4},
	year = {2008},
	month = {Sep},
	publisher = {American Physical Society},
	doi = {10.1103/PhysRevLett.101.107005},
	url = {https://link.aps.org/doi/10.1103/PhysRevLett.101.107005}
}

@article{Tanaka2009,
	title = {Manipulation of the {Majorana} Fermion, {Andreev} Reflection, and {Josephson} Current on Topological Insulators},
	author = {Tanaka, Yukio and Yokoyama, Takehito and Nagaosa, Naoto},
	journal = {Phys. Rev. Lett.},
	volume = {103},
	issue = {10},
	pages = {107002},
	numpages = {4},
	year = {2009},
	month = {Sep},
	publisher = {American Physical Society},
	doi = {10.1103/PhysRevLett.103.107002},
	url = {https://link.aps.org/doi/10.1103/PhysRevLett.103.107002}
}

@article{Brunetti2013,
	title = {Anomalous {Josephson} current, incipient time-reversal symmetry breaking, and {Majorana} bound states in interacting multilevel dots},
	author = {Brunetti, Aldo and Zazunov, Alex and Kundu, Arijit and Egger, Reinhold},
	journal = {Phys. Rev. B},
	volume = {88},
	issue = {14},
	pages = {144515},
	numpages = {13},
	year = {2013},
	month = {Oct},
	publisher = {American Physical Society},
	doi = {10.1103/PhysRevB.88.144515},
	url = {https://link.aps.org/doi/10.1103/PhysRevB.88.144515}
}

@article{Konschelle2015,
	title = {Theory of the spin-galvanic effect and the anomalous phase shift ${\ensuremath{\varphi}}_{0}$ in superconductors and {Josephson} junctions with intrinsic spin-orbit coupling},
	author = {Konschelle, Francois and Tokatly, Ilya V. and Bergeret, F. Sebasti\'an},
	journal = {Phys. Rev. B},
	volume = {92},
	issue = {12},
	pages = {125443},
	numpages = {16},
	year = {2015},
	month = {Sep},
	publisher = {American Physical Society},
	doi = {10.1103/PhysRevB.92.125443},
	url = {https://link.aps.org/doi/10.1103/PhysRevB.92.125443}
}

@article{Minutillo2018,
	title = {Anomalous {Josephson} effect in {S}/{S}{O}/{F}/{S} heterostructures},
	author = {Minutillo, M. and Giuliano, D. and Lucignano, P. and Tagliacozzo, A. and Campagnano, G.},
	journal = {Phys. Rev. B},
	volume = {98},
	issue = {14},
	pages = {144510},
	numpages = {9},
	year = {2018},
	month = {Oct},
	publisher = {American Physical Society},
	doi = {10.1103/PhysRevB.98.144510},
	url = {https://link.aps.org/doi/10.1103/PhysRevB.98.144510}
}

@article{Silaev2017,
	title = {Anomalous current in diffusive ferromagnetic {Josephson} junctions},
	author = {Silaev, M. A. and Tokatly, I. V. and Bergeret, F. S.},
	journal = {Phys. Rev. B},
	volume = {95},
	issue = {18},
	pages = {184508},
	numpages = {9},
	year = {2017},
	month = {May},
	publisher = {American Physical Society},
	doi = {10.1103/PhysRevB.95.184508},
	url = {https://link.aps.org/doi/10.1103/PhysRevB.95.184508}
}

@article{Meng2022,
	title = {Anomalous supercurrent modulated by interfacial magnetizations in {Josephson} junctions with ferromagnetic bilayers},
	author = {Meng, Hao and Wu, Xiuqiang and Ren, Yajie and Wu, Jiansheng},
	journal = {Phys. Rev. B},
	volume = {106},
	issue = {17},
	pages = {174502},
	numpages = {12},
	year = {2022},
	month = {Nov},
	publisher = {American Physical Society},
	doi = {10.1103/PhysRevB.106.174502},
	url = {https://link.aps.org/doi/10.1103/PhysRevB.106.174502}
}

@article{Lu2015,
	title = {Anomalous {Josephson} effect in $d$-wave superconductor junctions on a topological insulator surface},
	author = {Lu, Bo and Yada, Keiji and Golubov, A. A. and Tanaka, Yukio},
	journal = {Phys. Rev. B},
	volume = {92},
	issue = {10},
	pages = {100503},
	numpages = {5},
	year = {2015},
	month = {Sep},
	publisher = {American Physical Society},
	doi = {10.1103/PhysRevB.92.100503},
	url = {https://link.aps.org/doi/10.1103/PhysRevB.92.100503}
}

@article{Flensberg2016,
	title = {Effects of spin-orbit coupling and spatial symmetries on the {Josephson} current in {S}{N}{S} junctions},
	author = {Rasmussen, Asbj\o{}rn and Danon, Jeroen and Suominen, Henri and Nichele, Fabrizio and Kjaergaard, Morten and Flensberg, Karsten},
	journal = {Phys. Rev. B},
	volume = {93},
	issue = {15},
	pages = {155406},
	numpages = {5},
	year = {2016},
	month = {Apr},
	publisher = {American Physical Society},
	doi = {10.1103/PhysRevB.93.155406},
	url = {https://link.aps.org/doi/10.1103/PhysRevB.93.155406}
}

@article{Vakili2024,
	title = {Field-free {Josephson} diode effect in a $d$-wave superconductor heterostructure},
	author = {Vakili, Hamed and Ali, Moaz and Kovalev, Alexey A.},
	journal = {Phys. Rev. B},
	volume = {110},
	issue = {10},
	pages = {104518},
	numpages = {9},
	year = {2024},
	month = {Sep},
	publisher = {American Physical Society},
	doi = {10.1103/PhysRevB.110.104518},
	url = {https://link.aps.org/doi/10.1103/PhysRevB.110.104518}
}

@article{Osin2024,
	title = {Anomalous {Josephson} diode effect in superconducting multilayers},
	author = {Osin, A. S. and Levchenko, Alex and Khodas, Maxim},
	journal = {Phys. Rev. B},
	volume = {109},
	issue = {18},
	pages = {184512},
	numpages = {21},
	year = {2024},
	month = {May},
	publisher = {American Physical Society},
	doi = {10.1103/PhysRevB.109.184512},
	url = {https://link.aps.org/doi/10.1103/PhysRevB.109.184512}
}

@article{Tanaka2022,
	title = {Theory of giant diode effect in $d$-wave superconductor junctions on the surface of a topological insulator},
	author = {Tanaka, Yukio and Lu, Bo and Nagaosa, Naoto},
	journal = {Phys. Rev. B},
	volume = {106},
	issue = {21},
	pages = {214524},
	numpages = {13},
	year = {2022},
	month = {Dec},
	publisher = {American Physical Society},
	doi = {10.1103/PhysRevB.106.214524},
	url = {https://link.aps.org/doi/10.1103/PhysRevB.106.214524}
}

@article{Lu2023,
	title = {Tunable {Josephson} Diode Effect on the Surface of Topological Insulators},
	author = {Lu, Bo and Ikegaya, Satoshi and Burset, Pablo and Tanaka, Yukio and Nagaosa, Naoto},
	journal = {Phys. Rev. Lett.},
	volume = {131},
	issue = {9},
	pages = {096001},
	numpages = {7},
	year = {2023},
	month = {Aug},
	publisher = {American Physical Society},
	doi = {10.1103/PhysRevLett.131.096001},
	url = {https://link.aps.org/doi/10.1103/PhysRevLett.131.096001}
}

@article{Costa2023,
	title = {Microscopic study of the {Josephson} supercurrent diode effect in {Josephson} junctions based on two-dimensional electron gas},
	author = {Costa, Andreas and Fabian, Jaroslav and Kochan, Denis},
	journal = {Phys. Rev. B},
	volume = {108},
	issue = {5},
	pages = {054522},
	numpages = {13},
	year = {2023},
	month = {Aug},
	publisher = {American Physical Society},
	doi = {10.1103/PhysRevB.108.054522},
	url = {https://link.aps.org/doi/10.1103/PhysRevB.108.054522}
}

@article{Wang2024,
	title = {Efficient {Josephson} diode effect on a two-dimensional topological insulator with asymmetric magnetization},
	author = {Wang, J. and Jiang, Y. and Wang, Juan Juan and Liu, Jun-Feng},
	journal = {Phys. Rev. B},
	volume = {109},
	issue = {7},
	pages = {075412},
	numpages = {7},
	year = {2024},
	month = {Feb},
	publisher = {American Physical Society},
	doi = {10.1103/PhysRevB.109.075412},
	url = {https://link.aps.org/doi/10.1103/PhysRevB.109.075412}
}

@article{Ilic2024,
	title = {Superconducting diode effect in diffusive superconductors and {Josephson} junctions with {Rashba} spin-orbit coupling},
	author = {Ili\ifmmode \acute{c}\else \'{c}\fi{}, Stefan and Virtanen, Pauli and Crawford, Daniel and Heikkil\"a, Tero T. and Bergeret, F. Sebasti\'an},
	journal = {Phys. Rev. B},
	volume = {110},
	issue = {14},
	pages = {L140501},
	numpages = {6},
	year = {2024},
	month = {Oct},
	publisher = {American Physical Society},
	doi = {10.1103/PhysRevB.110.L140501},
	url = {https://link.aps.org/doi/10.1103/PhysRevB.110.L140501}
}

@article{Kokkeler2022,
	title = {Field-free anomalous junction and superconducting diode effect in spin-split superconductor/topological insulator junctions},
	author = {Kokkeler, T. H. and Golubov, A. A. and Bergeret, F. S.},
	journal = {Phys. Rev. B},
	volume = {106},
	issue = {21},
	pages = {214504},
	numpages = {8},
	year = {2022},
	month = {Dec},
	publisher = {American Physical Society},
	doi = {10.1103/PhysRevB.106.214504},
	url = {https://link.aps.org/doi/10.1103/PhysRevB.106.214504}
}

@article{Nas2025,
	author = {Djurdjević, Stevan and Popović, Zorica},
	title = {Josephson Diode Effect in d-Wave Superconductor/Ferromagnet/d-Wave Superconductor Junction with Interfacial {Rashba} Spin–Orbit Coupling},
	journal = {Progress of Theoretical and Experimental Physics},
	volume = {2025},
	number = {10},
	pages = {103I01},
	year = {2025},
	month = {10},
	issn = {2050-3911},
	doi = {10.1093/ptep/ptaf124},
	url = {https://doi.org/10.1093/ptep/ptaf124},
}

@article{Kulik1975,
	author = {{Kulik}, I.~O. and {Omel'Yanchuk}, A.~N.},
	title = "{Contribution to the microscopic theory of the {Josephson} effect in superconducting bridges}",
	journal = {Soviet Journal of Experimental and Theoretical Physics Letters},
	year = 1975,
	month = feb,
	volume = {21},
	pages = {96},
	adsurl = {https://ui.adsabs.harvard.edu/abs/1975JETPL..21...96K}
}

@article{Furusaki1991,
	title = {Dc {Josephson} effect and {Andreev} reflection},
	journal = {Solid State Communications},
	volume = {78},
	number = {4},
	pages = {299-302},
	year = {1991},
	issn = {0038-1098},
	doi = {https://doi.org/10.1016/0038-1098(91)90201-6},
	url = {https://www.sciencedirect.com/science/article/pii/0038109891902016},
	author = {Akira Furusaki and Masaru Tsukada}
}

@article{Furusaki1999,
	title = {Josephson current carried by {Andreev} levels in superconducting quantum point contacts},
	journal = {Superlattices and Microstructures},
	volume = {25},
	number = {5},
	pages = {809-818},
	year = {1999},
	issn = {0749-6036},
	doi = {https://doi.org/10.1006/spmi.1999.0730},
	url = {https://www.sciencedirect.com/science/article/pii/S0749603699907309},
	author = {Akira Furusaki}
}

@article{Sauls2018,
	author = {Sauls, J. A.},
	title = {Andreev bound states and their signatures},
	journal = {Philosophical Transactions of the Royal Society A: Mathematical, Physical and Engineering Sciences},
	volume = {376},
	number = {2125},
	pages = {20180140},
	year = {2018},
	month = {06},
	issn = {1364-503X},
	doi = {10.1098/rsta.2018.0140},
	url = {https://doi.org/10.1098/rsta.2018.0140},
}

@article{Mizushima2015,
	author = {Mizushima, T. and Machida, K.},
	title = {Multifaceted properties of {Andreev} bound states: interplay of symmetry and topology},
	journal = {Philosophical Transactions of the Royal Society A: Mathematical, Physical and Engineering Sciences},
	volume = {376},
	number = {2125},
	pages = {20150355},
	year = {2018},
	month = {06},
	issn = {1364-503X},
	doi = {10.1098/rsta.2015.0355},
	url = {https://doi.org/10.1098/rsta.2015.0355},
}

@article{Mondal2025,
	title = {Josephson diode effect with {Andreev} and {Majorana} bound states},
	author = {Mondal, Sayan and Fu, Pei-Hao and Cayao, Jorge},
	journal = {Phys. Rev. B},
	volume = {112},
	issue = {14},
	pages = {144506},
	numpages = {16},
	year = {2025},
	month = {Oct},
	publisher = {American Physical Society},
	doi = {10.1103/79tj-c3y4},
	url = {https://link.aps.org/doi/10.1103/79tj-c3y4}
}

@article{Wang2025a,
	title = {Theoretical study of superconducting diode effect in planar ${T}_{d}\text{\ensuremath{-}}\mathrm{{Mo}{Te}}_{2}$ {Josephson} junctions},
	author = {Wang, Gong-Qi and Miao, Jian-Jian and Chen, Wei-Qiang},
	journal = {Phys. Rev. B},
	volume = {112},
	issue = {1},
	pages = {014508},
	numpages = {12},
	year = {2025},
	month = {Jul},
	publisher = {American Physical Society},
	doi = {10.1103/kwz7-5stj},
	url = {https://link.aps.org/doi/10.1103/kwz7-5stj}
}

@article{Reynoso2008,
	title = {Anomalous {Josephson} Current in Junctions with Spin Polarizing Quantum Point Contacts},
	author = {Reynoso, A. A. and Usaj, Gonzalo and Balseiro, C. A. and Feinberg, D. and Avignon, M.},
	journal = {Phys. Rev. Lett.},
	volume = {101},
	issue = {10},
	pages = {107001},
	numpages = {4},
	year = {2008},
	month = {Sep},
	publisher = {American Physical Society},
	doi = {10.1103/PhysRevLett.101.107001},
	url = {https://link.aps.org/doi/10.1103/PhysRevLett.101.107001}
}

@article{Zazunov2008,
	title = {Anomalous {Josephson} Current through a Spin-Orbit Coupled Quantum Dot},
	author = {Zazunov, A. and Egger, R. and Jonckheere, T. and Martin, T.},
	journal = {Phys. Rev. Lett.},
	volume = {103},
	issue = {14},
	pages = {147004},
	numpages = {4},
	year = {2009},
	month = {Oct},
	publisher = {American Physical Society},
	doi = {10.1103/PhysRevLett.103.147004},
	url = {https://link.aps.org/doi/10.1103/PhysRevLett.103.147004}
}

@article{Blum2002,
	title = {Oscillations of the Superconducting Critical Current in {Nb}-{Cu}-{Ni}-{Cu}-{Nb} Junctions},
	author = {Blum, Y. and Tsukernik, A. and Karpovski, M. and Palevski, A.},
	journal = {Phys. Rev. Lett.},
	volume = {89},
	issue = {18},
	pages = {187004},
	numpages = {4},
	year = {2002},
	month = {Oct},
	publisher = {American Physical Society},
	doi = {10.1103/PhysRevLett.89.187004},
	url = {https://link.aps.org/doi/10.1103/PhysRevLett.89.187004}
}

@article{Nesterov2016,
	title = {Anomalous {Josephson} effect in semiconducting nanowires as a signature of the topologically nontrivial phase},
	author = {Nesterov, Konstantin N. and Houzet, Manuel and Meyer, Julia S.},
	journal = {Phys. Rev. B},
	volume = {93},
	issue = {17},
	pages = {174502},
	numpages = {9},
	year = {2016},
	month = {May},
	publisher = {American Physical Society},
	doi = {10.1103/PhysRevB.93.174502},
	url = {https://link.aps.org/doi/10.1103/PhysRevB.93.174502}
}

@article{Alidoust2021,
	title = {Cubic spin-orbit coupling and anomalous {Josephson} effect in planar junctions},
	author = {Alidoust, Mohammad and Shen, Chenghao and \ifmmode \check{Z}\else \v{Z}\fi{}uti\ifmmode \acute{c}\else \'{c}\fi{}, Igor},
	journal = {Phys. Rev. B},
	volume = {103},
	issue = {6},
	pages = {L060503},
	numpages = {8},
	year = {2021},
	month = {Feb},
	publisher = {American Physical Society},
	doi = {10.1103/PhysRevB.103.L060503},
	url = {https://link.aps.org/doi/10.1103/PhysRevB.103.L060503}
}

@article{Baumgartner2022c,
	doi = {10.1088/1361-648X/ac4d5e},
	url = {https://doi.org/10.1088/1361-648X/ac4d5e},
	year = {2022},
	month = {feb},
	publisher = {IOP Publishing},
	volume = {34},
	number = {15},
	pages = {154005},
	author = {Baumgartner, C and Fuchs, L and Costa, A and Picó-Cortés, Jordi and Reinhardt, S and Gronin, S and Gardner, G C and Lindemann, T and Manfra, M J and Faria Junior, P E and Kochan, D and Fabian, J and Paradiso, N and Strunk, C},
	title = {Effect of {Rashba} and {Dresselhaus} spin–orbit coupling on supercurrent rectification and magnetochiral anisotropy of ballistic {Josephson} junctions},
	journal = {Journal of Physics: Condensed Matter}
}

@article{Wu2022,
	author  = {Wu, H. and Wang, Y. and Xu, Y. and Liu, J. and Zhang, Q. and Zhao, S. and Chen, H. and Lin, Z. and Li, P. and Cao, T. and Zhang, X. and Jarillo-Herrero, P. and He, L.},
	title   = {The field-free {Josephson} diode in a van der {Waals} heterostructure},
	journal = {Nature},
	year    = {2022},
	volume  = {604},
	number  = {7907},
	pages   = {653--656},
	doi     = {10.1038/s41586-022-04504-8}
}

@article{Pal2022,
	author  = {Pal, Banabir and Chakraborty, Anirban and Sivakumar, Pranava K. and Davydova, Margarita and Gopi, Ajesh K. and Pandeya, Avanindra K. and Krieger, Jonas A. and Zhang, Yang and Date, Mihir and Ju, Sailong and Yuan, Noah and Schröter, Niels B. M. and Fu, Liang and Parkin, Stuart S. P.},
	title   = {Josephson diode effect from {Cooper} pair momentum in a topological semimetal},
	journal = {Nature Physics},
	year    = {2022},
	volume  = {18},
	number  = {10},
	pages   = {1228--1233},
	doi     = {10.1038/s41567-022-01699-5}
}

@article{Costa2023nat,
	author  = {Costa, A. and Baumgartner, C. and Reinhardt, S. and Fuchs, M. and Lu, L. and Watanabe, K. and Taniguchi, T. and Sch\"onenberger, C.},
	title   = {Sign reversal of the {Josephson} inductance magnetochiral anisotropy and 0--$\pi$-like transitions in supercurrent diodes},
	journal = {Nature Nanotechnology},
	year    = {2023},
	volume  = {18},
	number  = {12},
	pages   = {1266--1272},
	doi     = {10.1038/s41565-023-01451-x}
}

@article{Turini2022,
	author = {Turini, Bianca and Salimian, Sedighe and Carrega, Matteo and Iorio, Andrea and Strambini, Elia and Giazotto, Francesco and Zannier, Valentina and Sorba, Lucia and Heun, Stefan},
	title = {Josephson Diode Effect in High-Mobility {In}{Sb} Nanoflags},
	journal = {Nano Letters},
	volume = {22},
	number = {21},
	pages = {8502-8508},
	year = {2022},
	doi = {10.1021/acs.nanolett.2c02899},
	URL = {https://doi.org/10.1021/acs.nanolett.2c02899},
}

@article{Jeon2022,
	author  = {Jeon, K. R. and Kim, J. K. and Yoon, J. and Lee, G. H. and Song, K. M. and Kim, S. J. and Park, J. H. and Shin, Y. J. and Lee, S. and Choi, M. S. and Kim, T. Y. and Hwang, C. S.},
	title   = {Zero-field polarity-reversible {Josephson} supercurrent diodes enabled by a proximity-magnetized {Pt} barrier},
	journal = {Nature Materials},
	year    = {2022},
	volume  = {21},
	number  = {9},
	pages   = {1008--1013},
	doi     = {10.1038/s41563-022-01300-7}
}

@Article{Zhang2025,
	title={{Evidence of $\phi_0$-{Josephson} junction from skewed diffraction patterns in {Sn}-{In}{Sb} nanowires}},
	author={B. Zhang and Z. Li and V. Aguilar and P. Zhang and M. Pendharkar and C. Dempsey and J. S. Lee and S. D. Harrington and S. Tan and J. S. Meyer and M. Houzet and C. J. Palmstrom and S. M. Frolov},
	journal={SciPost Phys.},
	volume={18},
	pages={013},
	year={2025},
	publisher={SciPost},
	doi={10.21468/SciPostPhys.18.1.013},
	url={https://scipost.org/10.21468/SciPostPhys.18.1.013},
}

@misc{Kochan2023,
	title={Phenomenological Theory of the Supercurrent Diode Effect: The {Lifshitz} Invariant}, 
	author={Denis Kochan and Andreas Costa and Iaroslav Zhumagulov and Igor Žutić},
	year={2023},
	eprint={2303.11975},
	archivePrefix={arXiv},
	primaryClass={cond-mat.supr-con},
	url={https://arxiv.org/abs/2303.11975}, 
}

@article{Lotfizadeh2024,
	author  = {Lotfizadeh, N. and Schiela, W. F. and Pekerten, B. and Banerjee, A. and Dartiailh, M. C. and Matos-Abiague, A. and Shabani, J.},
	title   = {Superconducting diode effect sign change in epitaxial {Al}-{In}{As} {Josephson} junctions},
	journal = {Communications Physics},
	year    = {2024},
	volume  = {7},
	pages   = {120},
	doi     = {10.1038/s42005-024-01618-5}
}

@article{Hu2007,
	title = {Proposed Design of a {Josephson} Diode},
	author = {Hu, Jiangping and Wu, Congjun and Dai, Xi},
	journal = {Phys. Rev. Lett.},
	volume = {99},
	issue = {6},
	pages = {067004},
	numpages = {4},
	year = {2007},
	month = {Aug},
	publisher = {American Physical Society},
	doi = {10.1103/PhysRevLett.99.067004},
	url = {https://link.aps.org/doi/10.1103/PhysRevLett.99.067004}
}

@article{
	Davydova2022,
	author = {Margarita Davydova  and Saranesh Prembabu  and Liang Fu },
	title = {Universal {Josephson} diode effect},
	journal = {Science Advances},
	volume = {8},
	number = {23},
	pages = {eabo0309},
	year = {2022},
	doi = {10.1126/sciadv.abo0309},
	URL = {https://www.science.org/doi/abs/10.1126/sciadv.abo0309}}

@article{Zhang2022,
	title = {General Theory of {Josephson} Diodes},
	author = {Zhang, Yi and Gu, Yuhao and Li, Pengfei and Hu, Jiangping and Jiang, Kun},
	journal = {Phys. Rev. X},
	volume = {12},
	issue = {4},
	pages = {041013},
	numpages = {11},
	year = {2022},
	month = {Nov},
	publisher = {American Physical Society},
	doi = {10.1103/PhysRevX.12.041013},
	url = {https://link.aps.org/doi/10.1103/PhysRevX.12.041013}
}

@article{Nadeem2023,
	author  = {Nadeem, Muhammad and Fuhrer, Michael S. and Wang, Xiaolin},
	title   = {The superconducting diode effect},
	journal = {Nature Reviews Physics},
	year    = {2023},
	volume  = {5},
	pages   = {558--577},
	doi     = {10.1038/s42254-023-00632-w}
}

@article{He2022,
	doi = {10.1088/1367-2630/ac6766},
	url = {https://doi.org/10.1088/1367-2630/ac6766},
	year = {2022},
	month = {may},
	publisher = {IOP Publishing},
	volume = {24},
	number = {5},
	pages = {053014},
	author = {He, James Jun and Tanaka, Yukio and Nagaosa, Naoto},
	title = {A phenomenological theory of superconductor diodes},
	journal = {New Journal of Physics}
}

@article{Masaki2021,
	title = {Theory of the nonreciprocal {Josephson} effect},
	author = {Misaki, Kou and Nagaosa, Naoto},
	journal = {Phys. Rev. B},
	volume = {103},
	issue = {24},
	pages = {245302},
	numpages = {10},
	year = {2021},
	month = {Jun},
	publisher = {American Physical Society},
	doi = {10.1103/PhysRevB.103.245302},
	url = {https://link.aps.org/doi/10.1103/PhysRevB.103.245302}
}

@article{Jiang2022,
	author  = {Iang, K. and Hu, J.},
	title   = {Superconducting diode effects},
	journal = {Nature Physics},
	year    = {2022},
	volume  = {18},
	number  = {10},
	pages   = {1145--1146},
	doi     = {10.1038/s41567-022-01701-0}
}

@misc{Wang2025,
	title={Josephson diode effect: a phenomenological perspective}, 
	author={Da Wang and Qiang-Hua Wang and Congjun Wu},
	year={2025},
	eprint={2506.23200},
	archivePrefix={arXiv},
	primaryClass={cond-mat.supr-con},
	url={https://arxiv.org/abs/2506.23200}, 
}

@article{Zutic2004,
	title = {Spintronics: Fundamentals and applications},
	author = {\ifmmode \check{Z}\else \v{Z}\fi{}uti\ifmmode \acute{c}\else \'{c}\fi{}, Igor and Fabian, Jaroslav and Das Sarma, S.},
	journal = {Rev. Mod. Phys.},
	volume = {76},
	issue = {2},
	pages = {323--410},
	numpages = {0},
	year = {2004},
	month = {Apr},
	publisher = {American Physical Society},
	doi = {10.1103/RevModPhys.76.323},
	url = {https://link.aps.org/doi/10.1103/RevModPhys.76.323}
}

@article{DeFranceschi2010,
	author  = {De Franceschi, Silvano and Kouwenhoven, Leo and Schönenberger, Christian and Wernsdorfer, Wolfgang},
	title   = {Hybrid superconductor--quantum dot devices},
	journal = {Nature Nanotechnology},
	year    = {2010},
	volume  = {5},
	number  = {10},
	pages   = {703--711},
	doi     = {10.1038/nnano.2010.173},
	url     = {https://doi.org/10.1038/nnano.2010.173}
}

@article{Alam2023,
	author  = {Alam, Shamiul and Hossain, Md Shafayat and Srinivasa, Srivatsa Rangachar and Aziz, Ahmedullah},
	title   = {Cryogenic memory technologies},
	journal = {Nature Electronics},
	year    = {2023},
	volume  = {6},
	number  = {3},
	pages   = {185--198},
	doi     = {10.1038/s41928-023-00930-2},
	url     = {https://doi.org/10.1038/s41928-023-00930-2}
}

@article{Tanaka1997,
	title = {Theory of {Josephson} effect in superconductor-ferromagnetic-insulator-superconductor junction},
	journal = {Physica C: Superconductivity},
	volume = {274},
	number = {3},
	pages = {357-363},
	year = {1997},
	issn = {0921-4534},
	doi = {https://doi.org/10.1016/S0921-4534(97)80002-0},
	url = {https://www.sciencedirect.com/science/article/pii/S0921453497800020},
	author = {Yukio Tanaka and Satoshi Kashiwaya}
}

@article{Lofwander2001,
	doi = {10.1088/0953-2048/14/5/201},
	url = {https://doi.org/10.1088/0953-2048/14/5/201},
	year = {2001},
	month = {may},
	publisher = {},
	volume = {14},
	number = {5},
	pages = {R53},
	author = {T Löfwander and V S Shumeiko and G Wendin},
	title = {Andreev bound states in high-{Tc} superconducting
	junctions},
	journal = {Superconductor Science and Technology}
}

@article{Liu2010a,
	title = {Relation between symmetry breaking and the anomalous {Josephson} effect},
	author = {Liu, Jun-Feng and Chan, K. S.},
	journal = {Phys. Rev. B},
	volume = {82},
	issue = {12},
	pages = {125305},
	numpages = {5},
	year = {2010},
	month = {Sep},
	publisher = {American Physical Society},
	doi = {10.1103/PhysRevB.82.125305},
	url = {https://link.aps.org/doi/10.1103/PhysRevB.82.125305}
}

@article{Reynoso2012,
	title = {Spin-orbit-induced chirality of {Andreev} states in {Josephson} junctions},
	author = {Reynoso, Andres A. and Usaj, Gonzalo and Balseiro, C. A. and Feinberg, D. and Avignon, M.},
	journal = {Phys. Rev. B},
	volume = {86},
	issue = {21},
	pages = {214519},
	numpages = {18},
	year = {2012},
	month = {Dec},
	publisher = {American Physical Society},
	doi = {10.1103/PhysRevB.86.214519},
	url = {https://link.aps.org/doi/10.1103/PhysRevB.86.214519}
}

@article{Bergeret2015,
	doi = {10.1209/0295-5075/110/57005},
	url = {https://doi.org/10.1209/0295-5075/110/57005},
	year = {2015},
	month = {jun},
	publisher = {EDP Sciences, IOP Publishing and Società Italiana di Fisica},
	volume = {110},
	number = {5},
	pages = {57005},
	author = {Bergeret, F. S. and Tokatly, I. V.},
	title = {Theory of diffusive $\varphi_0$ {Josephson} junctions in the presence of spin-orbit coupling},
	journal = {Europhysics Letters}
}

@article{Tanaka1996a,
	title = {Local density of states of quasiparticles near the interface of nonuniform $d$-wave superconductors},
	author = {Tanaka, Yukio and Kashiwaya, Satoshi},
	journal = {Phys. Rev. B},
	volume = {53},
	issue = {14},
	pages = {9371--9381},
	numpages = {0},
	year = {1996},
	month = {Apr},
	publisher = {American Physical Society},
	doi = {10.1103/PhysRevB.53.9371},
	url = {https://link.aps.org/doi/10.1103/PhysRevB.53.9371}
}

@article{Cayao2024,
	title = {Enhancing the {Josephson} diode effect with {Majorana} bound states},
	author = {Cayao, Jorge and Nagaosa, Naoto and Tanaka, Yukio},
	journal = {Phys. Rev. B},
	volume = {109},
	issue = {8},
	pages = {L081405},
	numpages = {7},
	year = {2024},
	month = {Feb},
	publisher = {American Physical Society},
	doi = {10.1103/PhysRevB.109.L081405},
	url = {https://link.aps.org/doi/10.1103/PhysRevB.109.L081405}
}

@book{deGennes1966,
	address = {New York},
	author = {de Gennes, P. G.},
	comment = {ecuaciones de Bogoliubov-de Gennes},
	publisher = {Benjamin},
	title = {Superconductivity of Metals and Alloys},
	year = 1966
}

@article{Rashba1960,
	title={Properties of semiconductors with an extremum loop. {I}. {Cyclotron} and combinational resonance in a magnetic field perpendicular to the plane of the loop},
	author={Rashba, EI},
	journal={Soviet Physics, Solid State},
	volume={2},
	pages={1109--1122},
	year={1960},
	publisher={American Institute of Physics}
}

@article{Bychkov1984,
	title = {Properties of a 2{D} electron gas with lifted spectral degeneracy},
	author = {Bychkov, Yu. A. and Rashba, E. I.},
	journal = {JETP Lett.},
	volume = {39},
	issue = {2},
	pages = {66},
	year = {1984},
	doi = {},
	url = {http://jetpletters.ru/ps/0/article_19121.shtml},
}

@article{Muhlschlegel1959,
	author    = {B. M{\"u}hlschlegel},
	title     = {Die thermodynamischen Funktionen des Supraleiters},
	journal   = {Zeitschrift f{\"u}r Physik},
	year      = {1959},
	volume    = {155},
	pages     = {313--327},
	doi       = {10.1007/BF01332932}
}

@article{McMillan1968,
	title = {Transition Temperature of Strong-Coupled Superconductors},
	author = {McMillan, W. L.},
	journal = {Phys. Rev.},
	volume = {167},
	issue = {2},
	pages = {331--344},
	numpages = {0},
	year = {1968},
	month = {Mar},
	publisher = {American Physical Society},
	doi = {10.1103/PhysRev.167.331},
	url = {https://link.aps.org/doi/10.1103/PhysRev.167.331}
}

@article{Pajovic2006,
	title = {Josephson coupling through ferromagnetic heterojunctions with noncollinear magnetizations},
	author = {Pajovi\ifmmode \acute{c}\else \'{c}\fi{}, Z. and Bo\ifmmode \check{z}\else \v{z}\fi{}ovi\ifmmode \acute{c}\else \'{c}\fi{}, M. and Radovi\ifmmode \acute{c}\else \'{c}\fi{}, Z. and Cayssol, J. and Buzdin, A.},
	journal = {Phys. Rev. B},
	volume = {74},
	issue = {18},
	pages = {184509},
	numpages = {7},
	year = {2006},
	month = {Nov},
	publisher = {American Physical Society},
	doi = {10.1103/PhysRevB.74.184509},
	url = {https://link.aps.org/doi/10.1103/PhysRevB.74.184509}
}

@article{Costa2017,
	title = {Magnetoanisotropic {Josephson} effect due to interfacial spin-orbit fields in superconductor/ferromagnet/superconductor junctions},
	author = {Costa, Andreas and H\"ogl, Petra and Fabian, Jaroslav},
	journal = {Phys. Rev. B},
	volume = {95},
	issue = {2},
	pages = {024514},
	numpages = {18},
	year = {2017},
	month = {Jan},
	publisher = {American Physical Society},
	doi = {10.1103/PhysRevB.95.024514},
	url = {https://link.aps.org/doi/10.1103/PhysRevB.95.024514}
}

@article{Bagwell1992,
	title = {Suppression of the {Josephson} current through a narrow, mesoscopic, semiconductor channel by a single impurity},
	author = {Bagwell, Philip F.},
	journal = {Phys. Rev. B},
	volume = {46},
	issue = {19},
	pages = {12573--12586},
	numpages = {0},
	year = {1992},
	month = {Nov},
	publisher = {American Physical Society},
	doi = {10.1103/PhysRevB.46.12573},
	url = {https://link.aps.org/doi/10.1103/PhysRevB.46.12573}
}

@InProceedings{Beenaker1992,
	author="Beenakker, C. W. J.",
	editor="Fukuyama, Hidetoshi
	and Ando, Tsuneya",
	title="Three ``Universal'' Mesoscopic {Josephson} Effects",
	booktitle="Transport Phenomena in Mesoscopic Systems",
	year="1992",
	publisher="Springer Berlin Heidelberg",
	address="Berlin, Heidelberg",
	pages="235--253",
	isbn="978-3-642-84818-6"
}

@article{Beenaker2023,
	title = {Phase-shifted {Andreev levels} in an altermagnet {Josephson} junction},
	author = {Beenakker, C. W. J. and Vakhtel, T.},
	journal = {Phys. Rev. B},
	volume = {108},
	issue = {7},
	pages = {075425},
	numpages = {7},
	year = {2023},
	month = {Aug},
	publisher = {American Physical Society},
	doi = {10.1103/PhysRevB.108.075425},
	url = {https://link.aps.org/doi/10.1103/PhysRevB.108.075425}
}

@article{Alipourzadeh2025,
	title = {Andreev bound states and supercurrent in an unconventional superconductor-altermagnet {Josephson} junction},
	author = {Alipourzadeh, Mohammad and Hajati, Yaser},
	journal = {Phys. Rev. B},
	volume = {111},
	issue = {21},
	pages = {214515},
	numpages = {12},
	year = {2025},
	month = {Jun},
	publisher = {American Physical Society},
	doi = {10.1103/mj4b-2fnr},
	url = {https://link.aps.org/doi/10.1103/mj4b-2fnr}
}

@article{Behner2026,
	title = {Superconducting diode effect in selectively grown topological insulator based {Josephson} junctions},
	author = {Behner, Gerrit and Jalil, Abdur Rehman and Gr\"utzmacher, Detlev and Sch\"apers, Thomas},
	journal = {Phys. Rev. B},
	volume = {113},
	issue = {3},
	pages = {035440},
	numpages = {8},
	year = {2026},
	month = {Jan},
	publisher = {American Physical Society},
	doi = {10.1103/gc7k-rn8q},
	url = {https://link.aps.org/doi/10.1103/gc7k-rn8q}
}

@article{Cayao2026,
	title = {Nonlocal {Josephson} diode effect in minimal {Kitaev} chains},
	author = {Cayao, Jorge and Sato, Masatoshi},
	journal = {Phys. Rev. Res.},
	volume = {8},
	issue = {1},
	pages = {013326},
	numpages = {10},
	year = {2026},
	month = {Mar},
	publisher = {American Physical Society},
	doi = {10.1103/hf7s-f7tj},
	url = {https://link.aps.org/doi/10.1103/hf7s-f7tj}
}

@article{Zikic2007,
	title = {Superharmonic Josephson relations in unconventional superconductor junctions with a ferromagnetic barrier},
	author = {Zikic, R. and Dobrosavljevi\ifmmode \acute{c}\else \'{c}\fi{}-Gruji\ifmmode \acute{c}\else \'{c}\fi{}, L.},
	journal = {Phys. Rev. B},
	volume = {75},
	issue = {10},
	pages = {100502(R)},
	numpages = {4},
	year = {2007},
	month = {Mar},
	publisher = {American Physical Society},
	doi = {10.1103/PhysRevB.75.100502},
	url = {https://link.aps.org/doi/10.1103/PhysRevB.75.100502}
}
%\printbibliography
%\cite{}

\end{document}